\documentclass[aps,prd,preprint,floatfix,superscriptaddress,nofootinbib,longbibliography,a4paper]{revtex4-1}
\pdfoutput=1
\usepackage{xcolor}
\usepackage{graphicx}
\usepackage{textcomp}
\usepackage{subfigure}
\usepackage{feynmp}
\usepackage{amsmath,amssymb,array,mathtools}
\allowdisplaybreaks
\usepackage{tikz}
\usepackage{natbib}
\usepackage{enumerate}
\usepackage{hhline}
\usepackage{float}
\usepackage{siunitx}
\usepackage{hyperref}
\usepackage{orcidlink}
\usepackage{comment}
\usepackage{multirow}
\hypersetup{
    citecolor= red,
    colorlinks=true,
    linkcolor=blue,
    filecolor=magenta,      
    urlcolor=blue,}

\usepackage{epsfig}
\usepackage{xfrac}
\usepackage{slashed}
\usepackage{pifont}

\newcommand{\nudobe}{\texttt{$\nu$DoBe} }
\newcommand{\beqa}{\begin{eqnarray}}
\newcommand{\eeqa}{\end{eqnarray}}
\newcommand{\be}{\begin{equation}}
\newcommand{\ee}{\end{equation}}
\newcommand{\ba}{\begin{array}} 
\newcommand{\ea}{\end{array}}

\newcommand{\nl}{\nonumber \\}
\newcommand{\cc}{\, {\mathcal{C}}^{-1}\,}
\newcommand{\hc}{\;+\; \mathrm{H.c.}}
\newcommand{\hcn}{& + & \mathrm{H.c.} }

\newcommand{\ft}{\mathbf{10}}
\newcommand{\ff}{\mathbf{\overline{5}}}

\newcommand{\hh}{\mathrm{H}}
\newcommand{\eq}{\,&=&\,}
\newcommand{\ad}{\,&+&\,}
\newcommand{\mi}{\,&-&\,}
\newcommand{\nubb}{$0\nu\beta\beta$~}
\newcommand{\Or}{\mathcal{O}}

\newcommand{\Change}[1]{{#1}}

\begin{document} 
\preprint{CETUP-2025-003}
\title{Scalar-induced Neutrinoless Double Beta Decay in $SU(5)$}
\author{P. S. Bhupal Dev}
\email{bdev@wustl.edu}
\affiliation{Department of Physics and McDonnell Center for the Space Sciences, Washington University, St. Louis, MO 63130, USA}
\affiliation{PRISMA$^+$ Cluster of Excellence \& Mainz Institute for Theoretical Physics, 
Johannes Gutenberg-Universit\"{a}t Mainz, 55099 Mainz, Germany}
\author{Srubabati Goswami}
\email{sruba@prl.res.in}
\affiliation{Theoretical Physics Division, Physical Research Laboratory,\\ Navrangpura, Ahmedabad-380 009, India}
\author{Debashis Pachhar}
\email{debashispachhar@prl.res.in}
\affiliation{Theoretical Physics Division, Physical Research Laboratory,\\ Navrangpura, Ahmedabad-380 009, India}
\affiliation{Indian Institute of Technology Gandhinagar, Palaj-382 055, India}
\author{Saurabh K. Shukla}
\email{krshuklassaurabh@gmail.com}
\affiliation{Theoretical Physics Division, Physical Research Laboratory,\\ Navrangpura, Ahmedabad-380 009, India}


\begin{abstract}
We discuss the role of heavy scalar fields in mediating  neutrinoless double beta decay $(0\nu\beta\beta)$ within the $SU(5)$ Grand Unified Theory framework, extended suitably to include neutrino mass. In such a minimal realistic $SU(5)$ setup for fermion masses, the scalar  contributions to $0\nu\beta\beta$ are extremely suppressed as a consequence of the proton decay bound. We circumvent this problem by imposing a discrete ${\cal Z}_3$ symmetry. However, the scalar contributions to $0\nu\beta\beta$ remain  suppressed in this $SU(5) \times {\cal Z}_3$ model due to the neutrino mass constraint. We find that the $0\nu\beta\beta$ contribution can be enhanced by extending the scalar sector with an additional $\mathbf{15}$-dimensional scalar representation with suitable ${\cal Z}_3$ charge. Such an extension not only yields realistic fermion mass spectra but also leads to experimentally testable predictions in upcoming ton-scale $0\nu\beta\beta$ searches, which can be used as a sensitive probe of the new scalars across a broad range, from LHC-accessible scales up to $\sim  10^{10}\,\text{GeV}$. 
\end{abstract}
\maketitle
\tableofcontents
\section{Introduction}\label{sec:bg}
Baryon number ($B$) and lepton number ($L$) are accidental global symmetries of the Standard Model~(SM), broken only via non-perturbative sphaleron effects at high temperatures. However, in many beyond-the-SM (BSM) scenarios, where the SM particle spectrum is augmented with additional degrees of freedom --- new gauge bosons, fermions, and/or scalars --- $B$ or $L$-violating processes can be naturally induced with potentially observable rates. Therefore, these rare processes such as proton decay and neutrinoless double beta decay ($0\nu\beta\beta$) serve as  powerful probes of BSM physics.  

In a bottom-up approach, the strength of such new $B/L$-violating interactions remains largely unconstrained, restricted only by phenomenological considerations and experimental non-observations so far. However, this freedom is  significantly reduced when we adopt a top-down approach in the context of Grand Unified Theories (GUTs)~\cite{Pati:1974yy, Georgi:1974sy,Fritzsch:1974nn, Langacker:1980js, Mohapatra:1986uf,  
Croon:2019kpe}. GUTs represent an elegant
group-theoretical framework that naturally incorporates the  extended degrees of freedom and the associated rare processes into a self-consistent ultraviolet (UV)-completion of the SM. 
GUTs are based on the novel idea of unifying the strong, electromagnetic and weak interactions within a single, larger symmetry group. In these theories, new fields emerge as components of enlarged representations that couple collectively to quark-lepton multiplets. The Yukawa sector proves especially significant in the context of  GUT, as the SM Higgs field usually co-exists with additional scalar fields in these extended frameworks. Consequently, the strength of new interactions becomes intimately connected to low-energy observables --- particularly the fermion mass spectrum. This characteristic predictive power can distinguish GUTs from other simplified extensions of the SM. 

Since the SM gauge group $SU(3)_c\times SU(2)_L\times U(1)_Y$ has rank four (two for $SU(3)_c$ and one each for $SU(2)_L$ and $U(1)_Y$), the minimal choice for unification in a simple group is the rank-four $SU(5)$. However, the original $SU(5)$ model proposed in Ref.~\cite{Georgi:1974sy} suffers from a number of issues, including (i) inability to account for the observed charged fermion mass ratios, (ii) failure to achieve gauge coupling unification, (iii) rapid proton decay , (iv) massless neutrinos, and (v) doublet-triplet splitting problem. Extensions to the minimal $SU(5)$ model~\cite{Georgi:1974sy} to address these issues typically involve additional scalar/fermion multiplets~\cite{Georgi:1979df, Ellis:1979fg, Dorsner:2005fq,Bajc:2006ia, Saad:2019vjo, Klein:2019jgb,  Dorsner:2019vgf}. In this paper, we discuss the role of the heavy $SU(5)$ scalar multiplets in mediating the $L$-violating process of $0\nu\beta\beta$, as well as in generating correct neutrino masses and mixing, while being consistent with the observed charged fermion masses. 

To this end, we develop a realistic yet minimally extended $SU(5)$ framework in which the leptoquark contribution to \nubb   is enhanced relative to the standard light neutrino exchange mechanism.  The role of scalar leptoquarks in mediating $0\nu\beta\beta$ has been investigated before; see e.g., Refs.~\cite{Hirsch:1996ye, Bonnet:2012kh,Helo:2013ika,Helo:2015ffa,Pas:2015hca,Gonzalez:2016ztm, Graesser:2022nkv,Graf:2022lhj, Scholer:2023bnn,Li:2023wfi,Fajfer:2024uut,Dev:2024tto}. Within the context of $SU(5)$ grand unification, the \nubb mechanism has been further examined in Refs.~\cite{Fonseca:2015ena,Parida:2018apw,Dorsner:2025rzj}. 
However, there exists no minimal realistic $SU(5)$ GUT framework discussing {\it both} \nubb and neutrino mass arising from the heavy scalars.  
Here we propose such a scenario which ensures that the same set of scalar fields is responsible for generating neutrino masses, inducing $0\nu\beta\beta$, and contributing to fermion masses via one-loop corrections---thereby yielding a realistic and predictive $SU(5)$ framework.

In this work we focus our attention on the canonical $SU(5)$ scenario extended by a triplet scalar $\Delta$ belonging to the $\mathbf{15}$-dimensional representation. This allows generation of neutrino mass at tree level via the type-II seesaw mechanism~\cite{Magg:1980ut,Schechter:1980gr,Mohapatra:1980yp,Lazarides:1980nt} as well as radiatively at one loop induced by the  pair of  scalar leptoquarks -- $S_3$ 
and $R_2$~\cite{Dorsner:2016wpm,Dorsner:2017wwn, Babu:2019mfe, Babu:2020hun}. In this model, the scalar leptoquark fields $S_3$ and $\tilde{R}_2$ (see Tab.~\ref{tab:scalars}) inducing the $L$-violating process of $0\nu\beta\beta$ are also the ones inducing the $B$-violating process of proton decay. The stringent requirement of satisfying the proton decay constraints suppresses the leptoquark induced $0\nu\beta\beta$ rate to extremely small values. In order to evade this, we construct an $SU(5)\times {\cal Z}_3$ model which forbids the diquark interactions of $S_3$ capable of inducing proton decay. Although it yields inconsistent tree-level Yukawa relations in the down-quark and charged lepton sectors at the GUT scale, this inconsistency is resolved by incorporating radiative corrections from the heavy degrees of freedom, achieving realistic charged and neutral fermion mass spectra and mixing angles~\cite{Shukla:2024bwf}. We then compute the \nubb in this model, after taking into account other constraints such as those coming from the charged lepton-flavor-violating (cLFV) process of $\mu\to e$ conversion which restrict the mass of the $\tilde{R}_2$ leptoquark.  However, we find that  the set of scalars~$\left(\Delta\,,\,\tilde{R}_2,\,S_3\right)$ inducing the \nubb process, predicts a much suppressed rate as compared to the standard decay rate induced by light neutrinos. 
This suppression arises because, to achieve viable neutral fermion mass spectra, the scalar mass $M_\Delta$ is required to be around $10^{16}$ GeV, making the scalar-induced contribution negligible. 
In the allowed parameter space obtained by fitting the fermion mass spectrum, the  canonical light neutrino-mediated effective neutrino mass governing $0\nu\beta\beta$ rejects the inverted ordering from the existing  KamLAND-Zen bound ~\cite{KamLAND-Zen:2024eml}.

In order to enhance the scalar-induced $0\nu\beta\beta$ rate, an additional triplet scalar~$\left(\Delta_2\right)$ is then introduced in such a way that it is decoupled from matter multiplet interactions.  In this extended scenario, cancellations between standard and non-standard contributions to $0\nu\beta\beta$ allow the inverted mass ordering in certain parameter regions. Furthermore, it is also shown that \nubb can be used as a sensitive probe of the new scalar mass $M_{\Delta_2}$ across a broad range, from collider-accessible TeV scale all the way up to $\sim 10^{10}\,\text{GeV}$. In particular, future ton-scale experiments like nEXO~\cite{nEXO:2021ujk} and LEGEND-1000~\cite{LEGEND:2021bnm} with half-life sensitivities up to $10^{28}$ years can probe a wide range of the allowed parameter space in this $SU(5)$ GUT construction.

The rest of the paper is organised as follows: We review the generic $SU(5)$ framework in Section~\ref{sec:allmodels} and discuss the contribution of the $SU(5)$ scalars to \nubb  process. In Section~\ref{sec:SU5Z3}, in order to evade the proton decay constraint, an $SU(5) \times {\cal Z}_3$ scenario is constructed where some leptoquarks can remain light. This particular framework prohibits the diquark coupling of scalars contributing to $0\nu\beta\beta$. Subsequently, radiative corrections to the Yukawa relations are considered, ensuring a realistic scenario. In section~\ref{sec:nubb1}, the scalar contributions to the \nubb process is studied in the considered $SU(5) \times {\cal Z}_3$ model. The parameter fitting procedure is described in Section~\ref{sec:benchmark}. In Section~\ref{sec:0nubb}, we present the model predictions for \nubb and find that the scalar contributions are suppressed compared to the canonical light neutrino contribution. In Section~\ref{sec:0nubb2}, a viable scenario is presented where the scalar contribution to \nubb process can be significantly enhanced. Using the current experimental limit on the \nubb half-life, we then derive stringent constraints on the new scalar mass. Section~\ref{sec:summary} summarizes our main findings. A set of Appendices~\ref{app:sub-amplitudes}--\ref{app:bestfit}  are provided to support the discussion in the main text: Appendix~\ref{app:sub-amplitudes} gives the \nubb sub-amplitudes, nuclear matrix elements and phase space factors; Appendix~\ref{app:nubbSU5} discusses the \nubb contribution in the canonical $SU(5)$; Appendix~\ref{app:LF} gives the expressions of the loop integration factors used in the main text; Appendix~\ref{app:clfvs} discusses the cLFV constraints; and Appendix~\ref{app:bestfit} gives the best-fit solutions for two benchmark points. 

\section{Basic $SU(5)$ framework}
\label{sec:allmodels}

In the $SU(5)$ framework, the SM Weyl fermions are embedded in the $\ff$ and $\ft$ dimensional irreps, as follows~\cite{Langacker:1980js}:
\beqa{\label{eq:ff&ft}}
\ff_{a}\eq \varepsilon_{ab}\,\ell^b,\hspace{0.5cm}\ff_{\alpha}\;=\;d^C_{\alpha}\,,\nl
\ft^{a\alpha} \eq \frac{1}{\sqrt{2}}\,q^{a\,\alpha},\hspace{0.5cm}\ft^{\alpha\beta}\;=\;\frac{\varepsilon^{\alpha\beta\gamma}}{\sqrt{2}}\,u^C_{\gamma},\hspace{0.5cm}\ft^{ab}\;=\;\frac{\varepsilon^{ab}}{\sqrt{2}}\,e^C\,,
\eeqa
where Greek letters $\left(1\leq\,\,\alpha,\, \beta,\,\gamma...\,\,\leq 3 \right)$ denote $SU(3)_c$ indices while $SU(2)_L$ labels are depicted by the lowercase Latin alphabets $(4\leq a,b,c... \leq 5 )$. The convention of two-indexed Levi-Civita tensor is as follows: $\varepsilon_{45}\,=1\,=\,\varepsilon^{54}\,=\,-\varepsilon_{54}\,=\,-\varepsilon^{45}$. The three-indexed Levi-Civita follows the convention where $\varepsilon_{123}\,=\,1$ and for other cyclic permutations. The superscript $C$ stands for the charge-conjugated spinor, i.e., $\psi^C=i\sigma_2\psi^*$, where $\sigma_2$ is the second Pauli matrix.  

\begin{table}[t]
    \centering
    \begin{tabular}{c|cc}
    \hline\hline
    
        $SU(5)$ Multiplet & ~~~~~~~Notation~~~~~~~ & ~~~~~~~SM Charge~~~~~~~ \\\hline
        
      \multirow{2}{*}{$\mathbf{5_{\hh}}$}    & $H_1$& $\left(1,2,\frac{1}{2}\right)$ \\
    &    $S_1$ & $\left(3,1,-\frac{1}{3}\right)$\\\hline
    &  $\Delta$ & $\left(1,3,1\right)$\\
    $\mathbf{15_\hh}$ & $\tilde{R}_2$ & $\left(3,2,\frac{1}{6}\right)$ \\
    & $\Sigma$ & $\left(6,1,-\frac{2}{3}\right)$ \\ 
    \hline

    & $H_2$& $\left(1,2,\frac{1}{2}\right)$ \\
     &$S'_1$ & $\left(3,1,-\frac{1}{3}\right)$\\
     & $\tilde{S}_1$ & $\left(\overline{3},1,\frac{4}{3}\right)$ \\
   $\mathbf{45_{\hh}}$ & $R_2$ & $\left(\overline{3},2,-\frac{7}{6}\right)$ \\
    &  $S_3$ & $\left(3,3,-\frac{1}{3}\right)$ \\
    &  $\mathbb{S}$ & $\left(\overline{6},1,-\frac{1}{3}\right)$ \\
    &  $O$ & $\left(8,2,\frac{1}{2}\right)$ \\
\hline\hline
    \end{tabular}
    \caption{Scalar multiplets  residing inside the $\mathbf {5}_\hh$, $\mathbf{15}_\hh$ and $\mathbf{45}_{\hh}$-dimensional irreps of $SU(5)$ which participate in the Yukawa interactions at renormalizable level.  Their charges under the SM gauge group $SU(3)_c\times SU(2)_L\times U(1)_Y$ are also shown.}
    \label{tab:scalars}
\end{table}

The Higgs sector of the minimal $SU(5)$ consists of a $\mathbf{5}_{\hh}$-dimensional irrep and can be augmented with a $\mathbf{45}_\hh$-dimensional irrep to have viable tree-level Yukawa relations in the charged fermion sector~\cite{Georgi:1979df}. In order to also account for the neutrino masses and mixing, the Yukawa sector can be extended by a $\mathbf{15}_\hh$ scalar irrep. The different scalar multiplets residing in $\mathbf{5}_\hh$, $\mathbf{15}_\hh$ and $\mathbf{45}_\hh$ can be inferred from Tab.~\ref{tab:scalars}. The $SU(5)$-invariant Yukawa Lagrangian with these irreps participating in the Yukawa sector is given as follows:  
\beqa{\label{eq:SU5Tree}}
-{\cal L}_{\mathrm{Y}} \eq \frac{1}{4}\,\left(Y_{5}\right)_{AB}\,\ft_{A}^T\,\cc\ft_B\,\mathbf{5}_{\hh} \,+\,\sqrt{2}\, \left(\tilde{Y}_{5}\right)_{AB}\,\ft_{A}^T\,\cc\,\ff_B\,\mathbf{5}_{\hh}^{\dagger}\, \nl
\ad \frac{1}{2}\,\left(\tilde{Y}_{45}\right)_{AB}\,\ft_{A}^T\,\cc\ft_B\,\mathbf{45}_{\hh} \,+\,\sqrt{2}\, \left(Y_{45}\right)_{AB}\,\ft_{A}^T\,\cc\,\ff_B\,\mathbf{45}_{\hh}^{\dagger}\,\nl
\ad \left(Y_{15}\right)_{AB}\,\ff_{A}^T\,\cc\,\ff_B\,\mathbf{15}_{\hh}\,\hc
\eeqa
Here $\left(A,B=1,2,3\right)$ are the generation labels and ${\cal C}$ is the charge conjugation operator. 
In the above Yukawa Lagrangian, $Y_5$~$(\tilde{Y}_{45})$ is symmetric~(antisymmetric) in flavor indices and $Y_{15}$ is symmetric in generation labels while the remaining matrices have complex entries.  Using the embedding of $\ft$ and $\ff$ from Eq.~(\ref{eq:ff&ft}), it is straightforward to decompose the interaction terms written above in Eq.~(\ref{eq:SU5Tree}) and compute the interactions of different scalars with SM fermions. The Yukawa Lagrangian written in Eq.~(\ref{eq:SU5Tree}) has enough free parameters to yield the observed values of charged and neutral fermion mass spectra, even at the tree level. \Change{Note that the role of  $\mathbf{15}_\hh$ in facilitating unification has been discussed in Ref.~\cite{Dorsner:2005fq}.}

The phenomenology of different scalars stemming from the $\mathbf{5}_\hh$, $\mathbf{15}_\hh$ and $\mathbf{45}_\hh$, including contributions to neutrino mass, $B$ and $L$- violating interactions, cLFV and lepton-flavor-universality-violating interactions,  has been extensively studied ~\cite{Dorsner:2016wpm}.  
Particularly, the scalar fields $S_{1}\in\mathbf{5}_\hh$ and $S_{1'},S_3\in \mathbf{45}_\hh$ are known to induce tree-level proton decays while $\tilde{S}_{1}\in\mathbf{45}_\hh$ induces proton decay at one-loop~\cite{Dorsner:2012nq,Patel:2022wya}. The scalar field $\Delta\in \mathbf{15}_\hh$ induces tree-level neutrino mass via type-II seesaw~\cite{Magg:1980ut,Schechter:1980gr,Mohapatra:1980yp,Lazarides:1980nt} and the pair of scalar fields $S_{1,1'}-\tilde{R}_{2}$, $S_3-\tilde{R}_2$ and $S_3-R_2$ can also contribute to neutrino masses at one-loop~\cite{Dorsner:2016wpm,Dorsner:2017wwn, Babu:2019mfe, Babu:2020hun}. 

In general, the pair of scalar leptoquarks $S_{1,1'}-\tilde{R}_{2}$ and $S_3-\tilde{R}_2$ are also known to contribute to \( 0\nu\beta\beta \) decay~\cite{Dev:2024tto,Fajfer:2024uut}. However, in the $SU(5)$ embedding, the proton decay constraint on the mass of \( S_{1,1',3} - \tilde{R}_2 \) suppresses their contribution to \( 0\nu\beta\beta \), as elaborated below. A distinctive feature of scalar-induced proton decay is that the proton preferentially decays into the $\nu \sf{m}^+$, where $\sf{m}^+$ is a meson composed of either first or second-generation  quarks~\cite{Patel:2022wya}. This process violates \( B-L \) by two units~\cite{Wilczek:1979hc}~\footnote{Here, $B-L$ is a global quantum number which is violated with additional interaction terms like $\mathbf{45}_\hh\mathbf{45}_\hh\mathbf{15}^{\dagger}_\hh$ in $SU(5)$~\cite{Mohapatra:1986uf}. }, and therefore, we also expect a contribution to $0\nu\beta\beta$. The leading proton decay mode induced by the pair \( S_{1',3} - \tilde{R}_2 \) that would be relevant for \nubb is \( p \to \nu\, \pi^+ \),  whose decay width can be computed as follows~\cite{Nath:2006ut}:\footnote{Here, we provide the expression for the $(S_3-\tilde{R}_2)$-mediated contribution. Similar expression holds for the $(S_{1,1'}-\tilde{R}_2)$ pair for which $M_{S_3}$ is replaced by $M_{S_{1,1'}}$.}
\beqa\label{eq:naiveproton}
\Gamma\left(p\to \nu\,\pi^+\right) \simeq \sum_i \left( \frac{\left(Y_{15}\right)_{1i}\,\left(Y_{45}\right)_{11}}{M^2_{S_3}\,M^2_{\tilde{R}_2}}\right)^2\,\left(\eta\,v\right)^2\, \frac{m_p}{32\pi\,f^2_{\pi}}\,\alpha^2\,A^2 \, ,
\eeqa
where $\eta$ is the strength of the  $\mathbf{45}_{\hh}\mathbf{45}_{\hh}\mathbf{15}^{\dagger}_\hh$ vertex with positive mass dimension, $v$ is the vacuum expectation value (vev) of the SM Higgs boson, \( \alpha \sim 0.01 \) GeV\(^3\) is the hadronic matrix element, \( A \sim 1.4 \) is the long-distance renormalization factor, and \( f_\pi\simeq 130 \) MeV is the pion decay constant. Putting the different factors mentioned above and using the current lower bound of $3.9\times 10^{32}$ yr on the lifetime of $p\to \nu\,\pi^+$~\cite{Super-Kamiokande:2013rwg} gives the following estimate:
\beqa\label{eq:estimate}
\frac{\left(Y_{15}\right)_{11}\,\left(Y_{45}\right)_{11}\,\eta\,v}{M^2_{S_3}\,M^2_{\tilde{R}_2}} & \lesssim & 10^{-28}\, \text{GeV}^{-2} .
\eeqa
The above relation severely constrains the product of the Yukawa couplings and leptoquark masses from proton decay. 

The same combination goes into the \nubb amplitude mediated by leptoquark pairs (see Appendices~\ref{app:sub-amplitudes} and \ref{app:nubbSU5} for details): 
\beqa
{\cal A}_{\rm LQ} &=& V_{ud}\, M_{PS}\,\left(\frac{m_N}{m_e} \right) \left( \frac{\left(Y_{15}\right)_{11}\,\left(Y_{45}\right)_{11}\,\eta\,v}{M^2_{S_3}\,M^2_{\tilde{R}_2}} \right)\,v^2 
\lesssim  10^{-21} ,\label{eq:nubb_tree}
\eeqa
where $M_{PS}$ is the relevant nuclear matrix element (NME), $m_N$ and $m_e$ are the nucleon and electron masses  respectively, and $V_{ud}$ is the (1,1) element of the CKM matrix.
To arrive at the above estimation, we have used the upper bound given in Eq.~\eqref{eq:estimate}. On the other hand, the canonical light neutrino-mediated amplitude, ${\cal A}_{st} \simeq m_{ee}^{\rm std}/m_e \sim 10^{-8}$, where $m_{ee}^{\rm std} = \sum_{i} U_{ei}^2\, m_i$, with $U$ being the PMNS mixing matrix and $m_i$ the masses of active neutrinos. Thus, we find that the leptoquark-mediated \nubb amplitude is more than $\sim 13$ orders of magnitude smaller than the canonical light neutrino-mediated amplitude, and thus, \nubb cannot place any meaningful constraints on the leptoquarks in this case. It also shows that the $B$-violating process of proton decay occurring at dimension-7 is more constraining than the $L$-violating process of \nubb occurring at dimension-9.

In the next section, we explore if the situation can be remedied by forbidding the diquark coupling of $S_3$, thereby removing the constraint on its mass from proton decay and thus allowing a larger contribution to $0\nu\beta\beta$. In order to achieve this, we impose a discrete ${\cal Z}_3$ symmetry, in addition to the $SU(5)$ gauge symmetry, which forbids the diquark interactions of $S_3$.

\section{An $SU(5)\times {\cal Z}_3$ model }
\label{sec:SU5Z3}
The scalar sector of the $SU(5)$ model, considered in the earlier section, is comprised of $\mathbf{5}_{\hh}$, $\mathbf{15}_{\hh}$ and $\mathbf{45}_{\hh}$-dimensional irreps along with $\mathbf{24}_\hh$ which breaks the $SU(5)$ gauge symmetry into the SM. The $\mathbf{5}_\hh$-dimensional scalar irrep can be decomposed into different sub-multiplets as follows:
\beqa\label{eq:5Hdeco}
\mathbf{5}^a_\hh \eq H_{1}^{a}\,, \hspace{1cm} \mathbf{5}^\alpha_\hh\,\,=\, S_1^{\alpha}\,.
\eeqa
The decomposition of $\mathbf{15}_{\hh}$ into its constituent scalars, such that one obtains canonically normalized kinetic term, is as follows:
\beqa \label{eq:15hdeco}
\mathbf{15}^{ab}_{\hh} \eq \Delta^{ab}, \hspace{1cm} \mathbf{15}^{a\alpha}_\hh\,\,=\,\,\frac{1}{\sqrt{2}}\tilde{R}^{a\alpha}_2,\;\;\text{and} \hspace{1cm}\mathbf{15}^{\alpha\beta}_\hh\;=\; \Sigma^{\alpha\beta}. 
\eeqa
Similarly, the decomposition of $\mathbf{45}_{\hh}$-plet into the constituent scalars can be inferred from Ref.~\cite{Patel:2022wya} and is written below for convenience:
\beqa \label{eq:45dec}
\mathbf{45}_{\hh\,\gamma}^{\alpha \beta} &\equiv& {\mathbb{S}}_{\gamma}^{\alpha \beta} + \frac{1}{2 \sqrt{2}} \left( \delta_\gamma^{\alpha} {S_1'}^\beta - \delta_\gamma^\beta {S'_1}^\alpha \right)\,,~~~~\mathbf{45}_{\hh\,a}^{\alpha \beta}  \equiv   {R_2}_a^{\alpha \beta}\,\,, \nonumber \\
\mathbf{45}_{\hh\,\beta}^{\alpha a} &\equiv& \frac{1}{\sqrt{2}} O_\beta^{\alpha a} + \frac{1}{2\sqrt{6}} \delta_\beta^\alpha H_2^a\,,~~~~\mathbf{45}_{\hh\,\beta}^{ab} \equiv \frac{1}{\sqrt{2}} \varepsilon^{ab} {\tilde{S_1}}_\beta\,,\nonumber \\
\mathbf{45}_{\hh\,a}^{b \alpha} & \equiv & \frac{1}{\sqrt{2}} {S_3}_a^{b \alpha} - \frac{1}{2 \sqrt{2}} \delta_a^b {S'_1}^\alpha\,,~~~~ \mathbf{45}_{\hh\,c}^{ab} \equiv -\frac{\sqrt{3}}{2\sqrt{2}} \left( \delta_c^a H_2^b - \delta_c^b H_2^a\right)\,. \eeqa

\begin{table}[t]
    \centering
    \begin{tabular}{cc}
    \hline\hline
    
        ~~~~$SU(5)$ Multiplet~~~~ & ~~~~${\cal Z}_3$ Charge~~~~ \\\hline
        
          $ \ff_{A}$ & $\omega^2 $\\
          $\ft_{A}$ & $\omega $\\
          $\mathbf{5}_{\hh}$ & $\omega $\\
          $\mathbf{15}_{\hh}$ & $\omega^2$  \\
          $\mathbf{24}_{\hh}$ & $1$\\
          $\mathbf{45}_{\hh}$ & $1$ \\
          \hline
            $\widehat{\mathbf{15}}_{\hh}$ & $1 $ \\
\hline\hline
    \end{tabular}
    \caption{Assignment of ${\cal Z}_3$ charges to different fermion and scalar multiplets in our $SU(5)$ model. The subscript $A$ denotes family labels, with the ${\cal Z}_3$ charge being the same for all generations for the given fermionic multiplet. Here $\omega$ is the cube-root of unity. The last row shows the ${\cal Z}_3$ charge of an additional $\widehat{\mathbf{15}}_\hh$ which could significantly enhance the \nubb  rate in the considered framework (see Section~\ref{sec:0nubb2}).}
    \label{tab:Z3charges}
\end{table}

The assignment of ${\cal Z}_3$ charges to various (scalar and fermion) multiplets\footnote{\Change{Different options of imposing flavor symmetry in the context of minimal $SU(5)$ can be found in Ref.~\cite{Lindestam:2021dyk}.}} is depicted in Tab.~\ref{tab:Z3charges}.  The $\mathbf{24}_\hh$-dimensional scalar irrep transforms as a singlet under the imposed ${\cal Z}_3$ symmetry. Consequently, the ${\cal Z}_3$ symmetry remains intact even after the breaking of $SU(5)$. Moreover, the assigned ${\cal Z}_3$ charges are such that they prohibit any mixing term between $\mathbf{5}_\hh$ and $\mathbf{45}_\hh$. As a result, the scalar fields $H_{1,2}$ and $S_{1},\,S_{1'}$, residing in $\mathbf{5}_\hh$ and $\mathbf{45}_\hh$ respectively, cannot mix as long as ${\cal Z}_3$ remains unbroken. One immediate consequence of the non-mixing of $H_{1,2}$ is that the considered model effectively reduces to a Type-II Two Higgs Doublet Model (THDM)~\cite{Branco:2011iw}. The ${\cal Z}_3$ symmetry is broken when any of the SM Higgs fields residing in $\mathbf{5}_\hh$ or $\mathbf{45}_\hh$ acquires a vev, thus mixing $H_{1,2}$ and one of the linear combinations will be identified as the SM Higgs boson. Note that the ${\cal Z}_3$-breaking can potentially regenerate the unwanted interactions (e.g. $S_1-S_3$ mixing) inducing proton decay. However, since ${\cal Z}_3$ is broken at a  lower scale~(electroweak scale), the strength of these undesired couplings is suppressed by $(v/M_{\rm GUT})^2$, which ensures that the proton decay rate remains well below the experimental limit.\footnote{{It is a commonly employed strategy to overcome the proton decay problem using discrete symmetries, e.g., using $R$-parity~\cite{Farrar:1978xj} or matter parity~\cite{Dimopoulos:1981zb, Dimopoulos:1981dw}, baryon triality~\cite{Ibanez:1991hv, Ibanez:1991pr}, $Z_4$~\cite{Emmanuel-Costa:2011xdu, Emmanuel-Costa:2013gia}, proton hexality~\cite{Dreiner:2005rd, Forste:2010pf}, gauged discrete symmetries~\cite{Mohapatra:2007vd, Lee:2007qx, Hur:2008sy, Azatov:2008vu, Berasaluce-Gonzalez:2011gos} or discrete flavor symmetries~\cite{Dutta:2004zh, Kajiyama:2005rk, Dev:2012nm}.}}    

Note that spontaneous breaking of a discrete symmetry in the early Universe generates degenerate vacua. These vacua
are disconnected in the three-dimensional space, thus leading to the formation of domain walls between them~\cite{Kibble:1976sj}. This can be a  problem, because once they form after inflation, they may soon
dominate the energy density and overclose the Universe
during the Hubble expansion~\cite{Zeldovich:1974uw, Vilenkin:1984ib}. However, there are various ways to solve this, e.g. by diluting them away during/after inflation for a suitable choice of the reheating temperature~\cite{Kolb:1990vq}, by introducing bias terms in the potential~\cite{Larsson:1996sp}, perforating them by fast primordial black holes~\cite{Stojkovic:2005zh}, assuming that the discrete symmetry arises as a low-energy remnant symmetry after the spontaneous breaking of some continuous gauge symmetry~\cite{King:2018fke}, suppressing the thermal production of domain walls~\cite{Dvali:1995cc}, etc. \Change{In particular, the ${\cal Z}_3$ domain wall collapse has recently been studied in Refs.~\cite{Wu:2022stu, Wu:2022tpe}. For EW-scale ${\cal Z}_3$ breaking, a bias term $\varepsilon\gtrsim \varepsilon_{\rm min} (T_{\rm dec})\simeq \sigma T_{\rm dec}^2/(0.301 g_*^{-1/2}M_{\rm Pl})\sim  10^{-16}~{\rm GeV}^4$ is enough to make the domain walls collapse  before the BBN epoch, i.e. $T_{\rm dec}\sim 1$ MeV (see also Refs.~\cite{Larsson:1996sp, Gelmini:1988sf}). Here, $g_*$ denotes the effective relativistic degrees of freedom, $M_{\rm Pl}$ is the reduced Planck mass, and $\sigma\sim c_\sigma v^3/\lambda_{\rm eff}^{-1/2}$ is the surface energy tension of the domain wall, $\lambda_{\rm eff}$ includes the scalar self-coupling and thermal couplings to other fields, $c_\sigma$ is an ${\cal O}(1)$ number depending on the exact potential/profile. Such tiny bias terms ($\ll v^4$) can be easily generated by Planck-suppressed operators. }

The assignment of the ${\cal Z}_3$ charges in Tab.~\ref{tab:Z3charges} also forbids the $\tilde{Y}_{5,45}$ Yukawa couplings in Eq.~(\ref{eq:SU5Tree}) and the only allowed terms in the Yukawa Lagrangian are shown below:
\beqa{\label{eq:SU5Z3Tree}}
-{\cal L}_{\mathrm{Y}} \eq \frac{1}{4}\,\left(Y_{5}\right)_{AB}\,\ft_{A}^T\,\cc\ft_B\,\mathbf{5_{\hh}} \,+\,\sqrt{2}\, \left(Y_{45}\right)_{AB}\,\ft_{A}^T\,\cc\,\ff_B\,\mathbf{45_{\hh}}^{\dagger}\, \nl
\ad \left(Y_{15}\right)_{AB}\,\ff_{A}^T\,\cc\,\ff_B\,\mathbf{15_{\hh}}\,\hc\,,
\eeqa
where $Y_5$ and $Y_{15}$ are symmetric in flavor indices. Using the notations given in Eq.~(\ref{eq:ff&ft}) together with the canonically normalized decomposition of $\mathbf{5}_\hh$ and $\mathbf{45}_{\hh}$ provided in Eqs.~(\ref{eq:5Hdeco}) and \eqref{eq:45dec}, the vertex involving the SM Higgs and fermions can be written as follows:
\beqa {\label{45H-SM}}
-{\cal L}_{\mathrm{Y}} &\supset&  - \left(Y_{5}\right)_{AB}\,\varepsilon_{ab}\,q^{a\alpha\,T}_A\,\cc\,u^C_{\alpha\,B}\,H_1^b \nl \mi \left(Y_{45}\right)_{AB}\,\left(\frac{1}{\sqrt{6}}\,q^{a\alpha\,T}_A\,\cc\,d^C_{\alpha\,B}\,H^{\dagger}_{2a} - \sqrt{\frac{3}{2}}\,e^{C\,T}_A\,\cc\,\ell_B^a\,H^{\dagger}_{2\,a} \right)
 \hc\eeqa
The interaction terms in Eq.~(\ref{45H-SM}) lead to the following tree-level Yukawa relations for the charged fermions valid at the GUT scale:
\beqa{\label{eq:tree-yuk}}
\left(Y_{u}\right)_{AB} \eq \,\left(Y_{5}\right)_{AB}\,,\nl
\left(Y_{d}\right)_{AB} \eq \frac{1}{\sqrt{6}}\,\left(Y_{45}\right)_{AB}\,,\nl
\left(Y_{e}\right)_{AB} \eq -\sqrt{\frac{3}{2}}\,\left(Y_{45}^T\right)_{AB}\,.
\eeqa
The tree-level Yukawa relations written in Eq.~(\ref{eq:tree-yuk}) have enough freedom to reproduce the correct up-type Yukawa sector. However, in the down quark and charged lepton sectors, it leads to $3\,Y_d\,=-\,Y_e^T$ which results in $\frac{y_d}{y_e}\,=\,\frac{y_c}{y_{\mu}}\,=\,\frac{y_b}{y_{\tau}}\,=\,\frac{1}{3}$. In contrast, at the traditional GUT scale~$(M_{\rm{GUT}} \sim 10^{16} \, \text{GeV})$, the renormalization group evolution (RGE)-extrapolated SM values predict $\frac{y_d}{y_e} = 2$, $\frac{y_c}{y_{\mu}} = \frac{1}{5}$, and $\frac{y_b}{y_{\tau}} = \frac{2}{3}$~\cite{Das:2000uk}. Thus,  the tree-level Yukawa relations obtained here are not viable in the down quark and charged lepton sectors, unlike in the $SU(5)$ model without the ${\cal Z}_3$ [cf.~Eq.~\eqref{eq:SU5Tree}].  However, this issue is resolved by switching on the one-loop correction imparted by various heavier degrees of freedom~(scalar and gauge bosons), which are already present in the model, as discussed in the next section.

\subsection{Charged fermion masses}
\label{ssec:SU5NLO}
The inconsistency in the fermion mass relations at the tree-level [c.f. Eq.~\eqref{eq:tree-yuk}] can be addressed once the heavy scalar and gauge boson-mediated one-loop corrections to the Yukawa vertices are considered~\cite{Patel:2023gwt}.
The one-loop matching condition for the Yukawa couplings at a given renormalization scale $\mu$, following Refs.~\cite{Weinberg:1980wa,Hall:1980kf,Oliensis:1982sd,Kane:1993td,Hempfling:1993kv,Wright:1994qb}, was derived in Ref.~\cite{Patel:2023gwt} and is given as follows:
\beqa{\label{eq:matchcond}}
Y_f\left(\mu\right) \eq Y_f^0\left(1-\frac{K_H\left(\mu\right)}{2}\right) + \delta Y_f\left(\mu\right) - \frac{1}{2}\left(K_f^T\,\left(\mu\right)\,Y_f^0 + Y_f^0\,K_f^C\,\left(\mu\right)\right), 
\eeqa
where $Y_f^0$ are the tree-level Yukawa couplings of fermions $f\,\supset\,\big\{q,\,u^C,\,d^C,\,\ell,\,e^C\big\} $ with the SM Higgs boson. The one-loop corrected Yukawa coupling at the scale $\mu$, i.e. $Y_f\left(\mu\right)$ is deterministic and calculable in terms of the finite part of the vertex corrections $\left( \delta\,Y_f\right)$ and wave-function renormalization factors $\left( K_{H}, K_{f}\right)$. Vertex corrections to various Yukawa couplings and wave function renormalization for different fields are induced by the heavy fields inherent to this scenario. 

To compute the one-loop corrections, the interactions of the heavy degrees of freedom with the SM fermions are required. 
The interactions between various scalars with SM fermions stemming from $\mathbf{5}_\hh$ and $\mathbf{45}_\hh$ are obtained from the decomposition of Eq.~(\ref{eq:SU5Tree}) using Eqs.~\eqref{eq:ff&ft}, \eqref{eq:5Hdeco}, \eqref{eq:15hdeco} and \eqref{eq:45dec}, as shown below: 
\beqa{\label{eq:45scalarscoupling}}
-{\cal L}_{\mathrm{Y}} & \supset & -Y_{5\,AB}\,\left(u^{C\,T}_{\gamma\,A}\,\cc\,e^C_{B}\,+\, \frac{1}{2}\,\varepsilon_{\alpha\beta\gamma}\,\varepsilon_{ab}\,q^{a\alpha\,T}_{A}\,\cc\,q^{b\beta}_B\right)\,S_{1}^{\gamma}\nl
\ad \frac{Y_{45\,AB}}{\sqrt{2}}\,\left( \frac{1}{\sqrt{2}}\varepsilon^{\alpha\beta\rho}\,u^{C\,T}_{\rho\,A}\,\cc\,d^C_{\gamma\,B}\,{\mathbb S}^{\dagger\gamma}_{\alpha\beta} + \frac{1}{\sqrt{2}}\,\varepsilon^{\alpha\beta\gamma}\,u^{C\,T}_{\alpha\,A}\,\cc\,d^C_{\beta\,B}\,S^{'\dagger}_{1\,\gamma} \right. \nl
\mi \left. \frac{1}{\sqrt{2}}\,\varepsilon^{\alpha\beta\rho}\,\varepsilon_{ab}\,u^{C\,T}_{\rho\,A}\,\cc\,\ell^a_B\,R^b_{2\,\alpha\beta} - \sqrt{2} q^{a\alpha\,T}_A\,\cc\,d^C_{\beta\,B}\,O^{\dagger\beta}_{\alpha\,a}  \right. \nl 
\mi \left. \sqrt{2}\,\varepsilon_{ab}\,q^{p\alpha\,T}_A\,\cc\,\ell^a_B\,S^{\dagger\,b}_{3\,p\alpha}\,-\, \sqrt{2}\, e^{C\,T}_A\,\cc\,d^C_{\beta\,B}\,\tilde{S}_1^{\beta}\, - \frac{1}{\sqrt{2}}\,\varepsilon_{ab}\,q^{a\alpha\,T}_A\,\cc\,\ell^b_B\,S^{'\,\dagger}_{1\,\alpha}\right)\nl
\ad Y_{15\,AB}\,\left( \varepsilon_{am}\,\varepsilon_{bn}\,\ell^{m\,T}_A\,\cc\,\ell^n_B\,\Delta^{ab} \,+\, \sqrt{2}\,\varepsilon_{ab}\,\ell^{b\,T}_A\,\cc\,d^C_{\alpha\,B}\,\tilde{R}_2^{a\alpha}\,+\,d^{C\,T}_{\alpha\,A}\,\cc\,d^{C}_{\beta\,B}\,\Sigma^{\alpha\beta}\right)\nl
\hcn
\eeqa
It is to be noted that due to the imposed ${\cal Z}_3$ symmetry, the diquark coupling of $S_3$ is forbidden and hence it cannot induce nucleon decay here. The one-loop matching condition also requires the contribution from heavy gauge bosons, whose Lagrangian is shown below: 
\beqa \label{eq:L_gauge}
{\cal L}_{G} \eq \overline{\ff}\,\overline{\sigma}_{\mu}\,{\mathsf{D}}^{\mu}\,\ff + \overline{\ft}\,\overline{\sigma}_{\mu}\,{\mathsf{D}}^{\mu}\,\ft\,, \eeqa
where $D_\mu$ is the covariant derivative and $\bar{\sigma}_{\mu}= \left(1, \vec{\sigma}\right)$, with  $\vec{\sigma}$ being the Pauli matrices. The couplings of heavy gauge boson~$X_{\mu}\left(3,2,-\frac{5}{6}~\right)$ with the SM fermions is shown below:
\beqa\label{eq:L_X}
-{\cal L}_{\rm G}^{(X)} &\supset& \frac{g_5}{\sqrt{2}} \overline{X}_\mu \left( \overline{d^C}_i \overline{\sigma}^\mu \ell_i - \overline{q}_i \overline{\sigma}^\mu u^C_i - \overline{e^C}_i \overline{\sigma}^\mu q_i \right) \hc\,, 
\eeqa
where  $g_5$ is the $SU(5)$ gauge coupling.\\

 The vertex corrections to the various Yukawa vertices are induced by heavier scalars and gauge bosons propagating inside the loop. These can be computed from Eqs.~\eqref{eq:45scalarscoupling} and \eqref{eq:L_X}, as shown below:
\beqa{\label{eq:deltaf}}
\left(\delta Y_{u}\right)_{AB} \eq 4\, g_5^2 \left(Y_5\right)_{AB} f[M_X^2,0] 
\;,\nl
\left(\delta Y_d\right)_{AB} \eq \frac{2}{\sqrt{6}}\, g_5^2 \left(Y_{45}\right)_{AB} f[M_X^2,0]
\;,\nl
\left(\delta Y_e\right)_{AB} \eq -6\,\sqrt{\frac{3}{2}}\, g_5^2 \left(Y_{45}^T\right)_{AB} f[M_X^2,0]\,,
\eeqa
where $(\delta Y_f)$ represents the finite part of the correction to the Yukawa interaction of fermion $f$ with the SM Higgs. The loop integration factor $f[M_i^2,0]$ is given in Appendix~\ref{app:LF}, where $M_i$ is the mass of the scalar or gauge boson propagating inside the loop. These corrections depend on the renormalization scale $\mu$ and are fully determined in terms of the tree-level Yukawa couplings $Y_{5,45}$, along with a function that involves the masses of heavy particles. While calculating these corrections, all SM fields are considered massless. Due to the assigned ${\cal Z}_3$ charges, scalar fields from $\mathbf{5}_\hh$ and $\mathbf{45}_\hh$ do not mix;  consequently, the vertex corrections receive contribution only from the heavy gauge bosons~$(X_{\mu})$, not from the scalar fields. Moreover, the vertex corrections to $Y_d$ and $Y_e$ are unable to break the tree-level inconsistency, $3\,Y_d = Y_e^T$, as $\delta Y_d$ and $\delta Y_e$ also follow the same behavior. However, the wave-function renormalization to different fermions will break the generational degeneracy, as shown below.   

The finite part of the wave function renormalization factor of the (scalar or fermion) field \( f \) is computed by taking the derivative of the self-energy correction of the field \( f \) with respect to the outgoing momentum and then setting the momentum to zero. The contribution to the wave function renormalization due to various scalars and gauge bosons to the external leg of SM fermions and Higgs are shown below:
\beqa{\label{eq:kfs}}
\left(K_q\right)_{AB} \eq 3\,g_5^2\,\delta_{AB}\,h[M_X^2,0] - \frac{1}{2}\,\left(Y_5^*Y_5^T\right)_{AB}\,h[M_{S_1}^2,0] \nl \mi  \left( 6 \,h[M_O^2,0] + 4 \,h[M_{S_3}^2,0] + 0.5 \, h[M_{S^{'}_1}^2,0] \right)\, \left(Y_{45}\,Y_{45}^{\dagger}\right)_{AB}
\;,\nl
\left(K_{u^C}\right)_{AB} \eq 4\,g_5^2\,\delta_{AB}\,h[M_X^2,0] - h[M_{S_1}^2,0]\left(Y_5^*\,Y_5^T\right)
\nl \mi \left( 3 h[M_{\mathbb S}^2,0] + 1.5 h[M_{S^{'}_1}^2, 0] + 2 h[M_{R_2}^2,0] \right)\, \left(Y_{45}\,Y_{45}^{\dagger}\right)_{AB}\;, \nl
\left(K_{d^C}\right)_{AB} \eq 2\,g_5^2\,\delta_{AB}\,h[M_X^2,0] \nl\mi \left( 6 h [M^2_{\mathbb{S}},0] + h[M_{S^{'}_1}^2,0] + 12 h[M^2_O,0] + 2\, h[M_{{\tilde{S}_1}}^2,0]\right)\,\left( Y_{45}^T\,Y_{45}^*\right)_{AB} \nl
\mi 4\,h[M_{\Sigma}^2,0]\,\left(Y_{15}\,Y_{15}^{\dagger}\right)_{AB} - 4\,h[M_{R_2}^2,0]\,\left(Y_{15}^T\,Y_{15}^*\right)_{AB}\nl
\left(K_\ell\right)_{AB} \eq 3\,g_5^2\,\delta_{AB}\,h[M_X^2,0] -\left( 6 h[M^2_{R_2},0] + 6 h[M_{S_3}^2,0] + 1.5 h[M_{S_1^{'}}^2,0] \right)\, \left(Y_{45}^T\,Y_{45}^*\right)_{AB}\,, \nl
\mi 3\,h[M_{\Delta}^2,0]\,\left(Y_{15}\,Y_{15}^{\dagger}\right)_{AB} - 6\,h[M_{\tilde{R}_2}^2,0]\,\left(Y_{15}\,Y_{15}^{\dagger}\right)_{AB}\,,\nl
\left(K_{e^C}\right)_{AB} \eq 6\,g_5^2\,\delta_{AB}\,h[M_X^2,0] - 3\,h[M_{S_1}^2,0]\,Y_5^{\dagger}\,Y_5\,\nl
 K_{H_1} \eq \frac{g_5^2}{2}\Big[2\,\left(f[M_X^2,M_{S_1}^2] + g[M_X^2,M_{S_1}^2]\right),\nl
  K_{H_{2}} \eq \frac{g_5^2}{2}\Big[ 2\,\left(f[M_X^2,M_{S^{'}_1}^2] + g[M_X^2,M_{S^{'}_1}^2]\right)   \nl \ad  4\,\left(f[M_X^2,M_{S_3}^2] + g[M_X^2,M_{S_3}^2]\right) + 4\,\left(f[M_X^2,M_{\tilde{S}}^2] + g[M_X^2,M_{\tilde{S}}^2]\right)\Big].
\eeqa
Here $K_f$ characterizes the finite parts of the wave function renormalization factor corresponding to the field $f$. The loop integrating factors $g[M_i^2,0]$ and $h[M_i^2,0]$ are given in Appendix~\ref{app:LF}. In contrast to the vertex corrections, wave function renormalization factors receive corrections from  both heavy scalars and gauge bosons. The interaction of each scalar with SM fermions determines its contribution to different wave-function renormalization factors. Notably, $K_\ell$ and $K_{d^C}$ are different because the scalar fields ${\mathbb{S}}$, $O$ and $\Sigma$ contribute only to $K_{d^C}$, while $R_2$ and $\Delta$ contribute to $K_\ell$, as the former only have diquark interactions. As a result, these two factors are not the same. Thus, including the wave-function renormalization effects can help in resolving the inconsistency in the tree-level Yukawa relation in the down quark and charged lepton sectors. Moreover, the  SM Higgs residing in $\mathbf{5}_{\hh}$ and $\mathbf{45}_{\hh}$ irreps also receives different contributions from scalars, which is due to the imposition of the ${\cal Z}_3$ symmetry. 

Substituting Eqs.~\eqref{eq:deltaf} and \eqref{eq:kfs} into Eq.~(\ref{eq:matchcond}), the effective Yukawa relations at one loop valid at a given renormalization scale $\mu$ are as follows:
\beqa{\label{eq:oneloopsu5}}
Y_{u}\left(\mu\right) &\simeq & Y_5\,\left(1 - \frac{1}{2} K_{H_1} \right) + \delta Y_{u} - \frac 12\, \left(K_q^T\,Y_5 + Y_5\,K_{u^C}\right)\, , \nl
Y_{d}\left(\mu\right) & \simeq & \frac{Y_{45}}{\sqrt{6}}\,\left(1 - \frac 12 K_{H_{1}} \right) + \delta Y_{d} - \frac{1}{2\sqrt{6}}  \times \left(K_q^T\,Y_{45} + Y_{45}\,K_{d^C}\right) \, , \nl
Y_{e}\left(\mu\right) &\simeq & -\sqrt{\frac{3}{2}}\,Y_{45}^T\left(1 - \frac 12\, K_{H_{2}} \right) + \delta Y_{e} - \frac 12 \left( -\sqrt{\frac{3}{2}}\right) \times \left(K_\ell^T\,Y_{45}^T + Y_{45}^T\,K_{e^C}\right),
\eeqa
where $Y_{5,45}$ are the tree-level Yukawa couplings depicted in Eq.~(\ref{eq:SU5Tree}).

\subsection{Neutrino mass}
The capability of the considered $SU(5)$ framework to generate neutrino mass is now evaluated. This model includes a $\mathbf{15}_\hh$-dimensional scalar irrep, which is traditionally associated with neutrino mass generation through the Type-II seesaw mechanism~\cite{Dorsner:2007fy}. As the charged fermion mass spectrum has been computed up to the one-loop level, consistency requires extending the neutrino mass calculation to the same order. The neutrino mass receives tree level contribution from $\Delta$~(c.f. Tab.~\ref{tab:scalars}) with the Yukawa coupling $Y_{15}$ and also receives contribution from $R_{2}-S_3$ at one loop. The resulting expression is given by,
\beqa{\label{eq:SU5NM}}
M_\nu \eq -2\,\eta\,\frac{\langle H_1 \rangle^2}{M_{\Delta}^2}\,Y_{15}
\nl
\ad 6\,\rho\,\eta\,\frac{\langle H_1 \rangle^4}{M_\Delta^2}\,\Big[\left(Y_{45}^T\,Y_u^*\,Y_{45}\right) + \left(Y_{45}^T\,Y_u^*\,Y_{45}\right)^T\Big] p[M_{S_3}^2,M_{R_2}^2] \, , 
\eeqa
 where the first term is the tree-level Type-II seesaw contribution, and the second term is the one-loop contribution.
 Here $\langle H_1 \rangle $ is the vev of the SM Higgs field  residing in the $\mathbf{5}_\hh$-dimensional Higgs, $\eta$ is a dimension-full trilinear coupling and is assumed to be close to the GUT scale, while $\rho$ is an ${\cal{O}}\left(1\right)$ quartic coupling and has been set to unity. The definition of the loop integration factor $p[m_1^2,m_2^2]$ is provided in Appendix~\ref{app:LF}.  It is to be noted that the trilinear vertex $\mathbf{45}_\hh\,\mathbf{45}_\hh\,\mathbf{15}_\hh^{\dagger}$ is forbidden due to the imposed ${\cal Z}_3$ symmetry; as a result, $H_2$ cannot contribute to the neutrino masses. It is also imperative to note that the neutrino mass receives contributions from the one-loop corrected Yukawa couplings $Y_u$.  Moreover, for a hierarchical neutrino mass spectrum far away from the quasi-degenerate regime, as strongly suggested by the current cosmological limits on the sum of neutrino masses~\cite{Planck:2018vyg, DESI:2025zgx}, the RGE running of neutrino parameters are known to be small~\cite{Chankowski:1993tx,Babu:1993qv,Antusch:2005gp,Mei:2005qp} and will not be considered here.   

\section{Contribution to \nubb }
\label{sec:nubb1}
The inclusion of $\mathbf{15}_\hh$ in $SU(5)$ framework renders neutrinos to be of Majorana type by virtue of Type-II seesaw, which gives rise to the smoking gun signal of \nubb~\cite{Schechter:1981bd}. In addition to the long-range contribution mediated by left-handed neutrinos, we also have other long-range contributions mediated by the scalars in $SU(5)$. Since \nubb is a low energy process, they can be described by low-energy Effective Field Theory (LEFT)~\cite{Kotila:2021xgw,Graf:2018ozy}.

The most general LEFT  Lagrangian for long range mechanisms can be written as~\cite{Pas:1997fx,Ali:2007ec,Kotila:2021xgw}
\beqa
-{\cal L}_{\rm eff} \supset \frac{G_F}{\sqrt{2}} \, \Big[ j^{\mu}_{V-A}J_{V-A,\mu} + \sum_{\dot{\alpha},\dot{\beta}} \epsilon_{\dot{\alpha}}^{\dot{\beta}} \, j_{\dot{\beta}} \, J_{\dot{\alpha}} + \hc \Big], \label{eq:LEFT}
\eeqa
where $G_F$ is the Fermi constant, and the leptonic and hadronic currents are defined as $j_{\dot{\beta}} = \bar e {\cal O}_{\dot{\beta}} \nu^C_e$ and $J_{\dot{\alpha}} = \bar u {\cal O}_{\dot{\alpha}} d$. The Greek indices $\dot{\beta} ,\dot{\alpha}$ can be $V\mp A, \, S\mp P,\, T\mp T_5$, where $V,A,S,P,T,T_5$ correspond to vector, axial-vector, scalar, pseudo-scalar, tensor, and axial-tensor respectively. In Eq.~(\ref{eq:LEFT}), the standard canonical long range mechanism (first term) has been separated from the non-standard contributions, with ${\epsilon}_{\dot{\alpha}}^{\dot{\beta}} $ being the corresponding Wilson coefficients of the non-standard operators.

\begin{figure}[t!]
    \centering    \includegraphics[width=0.5\textwidth]{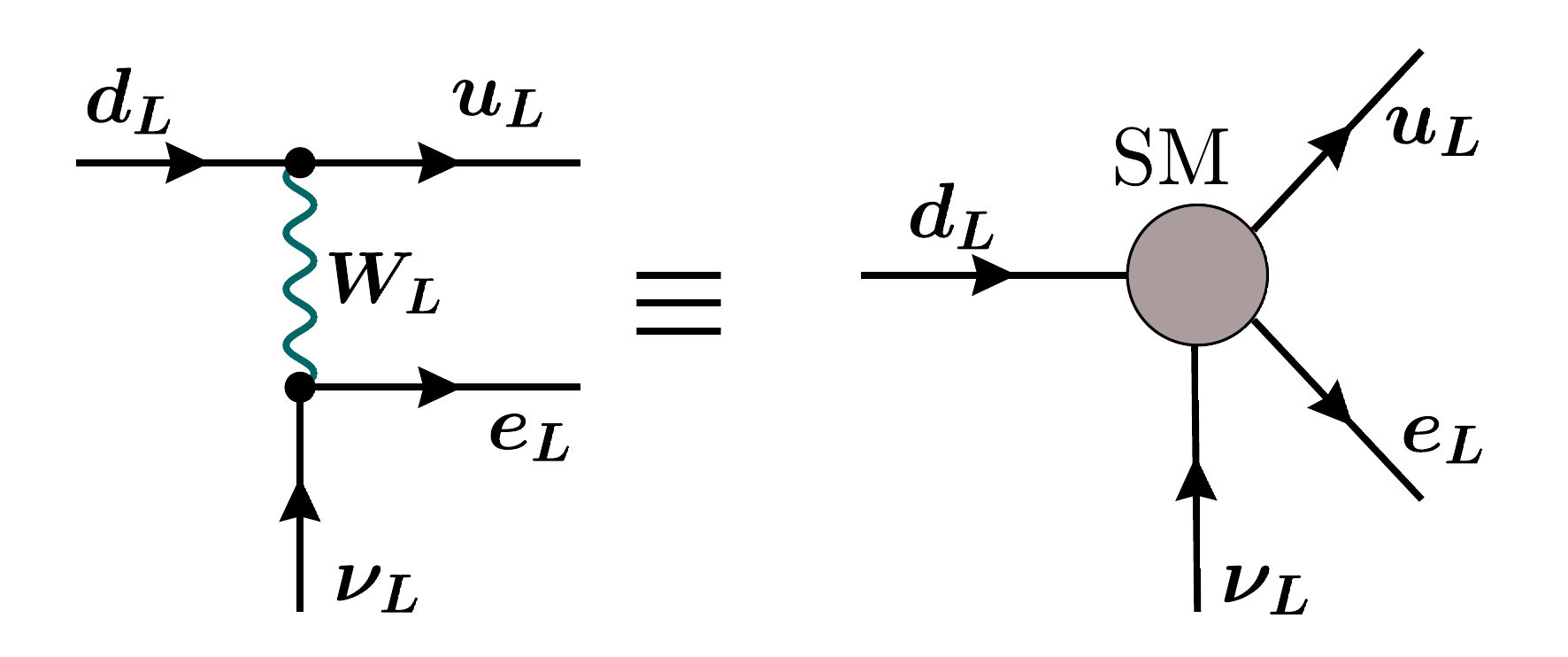}
    \caption{The canonical effective operator that contributes to \nubb. \textbf{Left:}  SM interactions. \textbf{Right:} LEFT operator after integrating out the $W_L$ boson.} 
    \label{fig:SM_operator}
    \end{figure}
    
\begin{figure}[t!]
    \centering
   \includegraphics[width=0.7\textwidth]{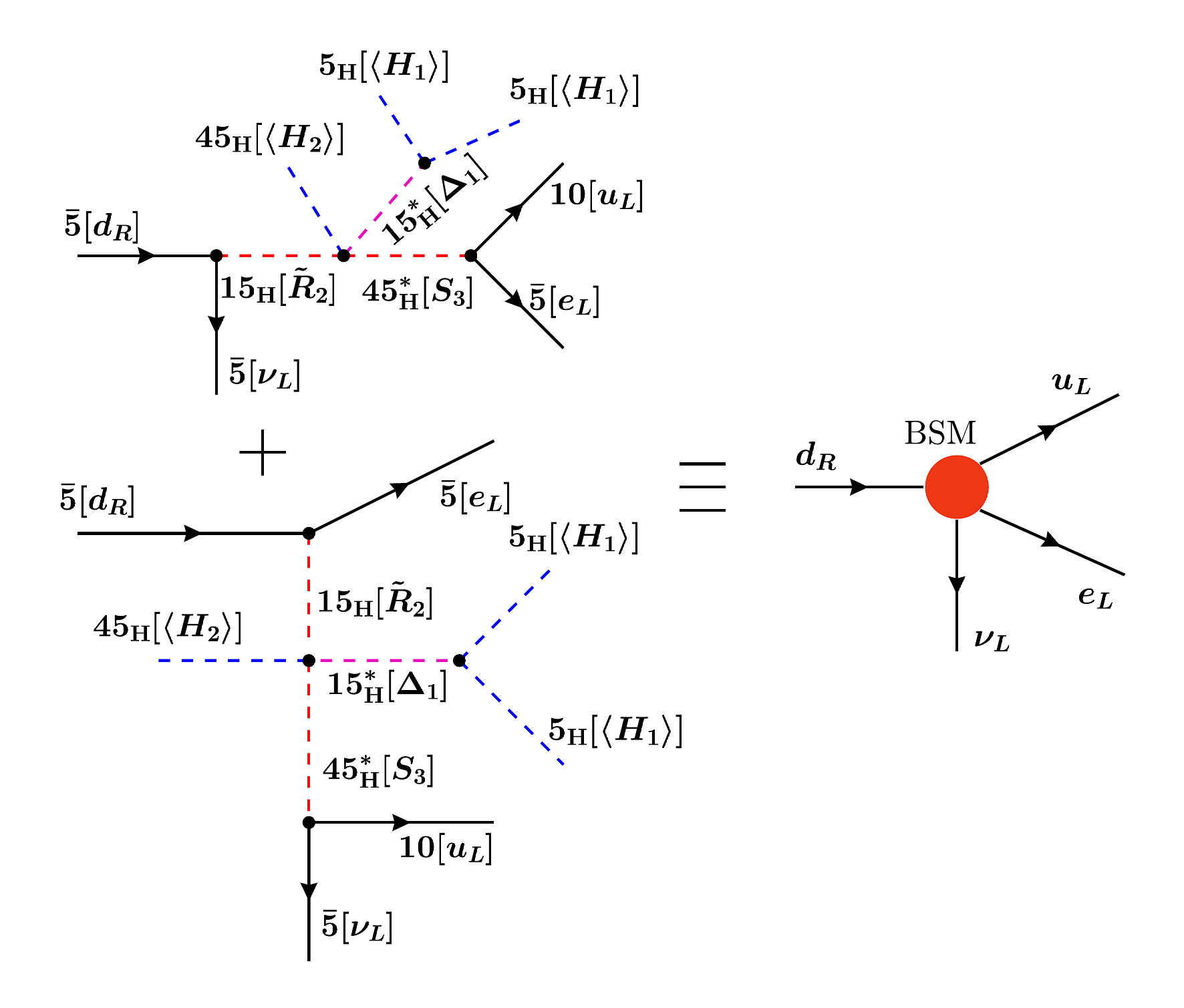}
    \caption{Effective operators that contribute to \nubb in the considered $SU(5)\times {\cal Z}_3$ model. \textbf{Left:} Diagrams contributing to \nubb generated by the $SU(5)$ scalars. The number outside the bracket denote the $SU(5)$ representation while the labels in the bracket denote the field under SM charges. \textbf{Right:} The LEFT operator corresponding to the left diagrams after integrating out the heavy scalar fields. } 
    \label{fig:0nbb_operator}
    \end{figure}

The operator contributing to the canonical long-range mechanism is shown in Fig.~\ref{fig:SM_operator}. The non-standard operators contributing to the other long-range mechanisms can be inferred from diagrams given in Fig.~\ref{fig:0nbb_operator} and are given by
\beqa
\mathcal{O}^{(6)}_{\rm BSM} =  \left( \frac{G_F}{\sqrt{2}} \right)\, \Big[\epsilon_{S+P}^{S+P}  \left( \bar{u} P_R d \right) \left( \bar{e} P_R \nu_e^C \right) + \epsilon_{T+T_5}^{T+T_5}   \left( \bar{u} \sigma^{\alpha\beta}P_R d \right) \left(\bar{e} \sigma_{\alpha\beta}P_R \nu_e^C \right) \Big] \, .
\label{eq:OBSM}
\eeqa
Both the diagrams given in Fig.~\ref{fig:0nbb_operator} originate from the same kind of interactions which stems from the $SU(5)$ Lagrangian. They differ only in the electric charges of the mixing partners: the first diagram involves the mixing of leptoquarks with $Q_{\rm em} = 1/3$ components of $M_{S_3}$ and $M_{\tilde{R}_2}$, while the second appears when the leptoquarks with $Q_{\rm em}=2/3$ mix. 

The dimensionless parameters $\epsilon_{S+P}^{S+P}$ and $\epsilon_{T+T_5}^{T+T_5}$ appearing in Eq.~\eqref{eq:OBSM} at the LEFT scale are given by
\beqa \label{eq:nubb}
\epsilon_{S+P}^{S+P} &\approx & \frac{\rho \, \eta \, v^5}{M_{\Delta}^2} \, \frac{\left(\sqrt{2}\, Y_{\tilde{R}_2} \right)_{11}}{M_{\tilde{R}_2}^{2}}\, \frac{\left(\sqrt{2}\, Y_{S_3}\right)_{11}}{M_{S_3}^2}  \, ,\label{eq:S+P}\\
\epsilon_{T+T_5}^{T+T_5} &=& - \frac 14 \,\epsilon_{S+P}^{S+P} \, ,\label{eq:TR}
\eeqa
corresponding to the $S+P$ and $T+T_5$ contributions, respectively.

Incorporating the standard and non-standard contributions, the LEFT Lagrangian in Eq.~\eqref{eq:LEFT} becomes the following: 
\beqa
-{\cal L}_{\rm eff} \supset \frac{G_F }{\sqrt{2}} \, \Big[ j^{\mu}_{V-A}J_{V-A,\mu} + \epsilon_{S+P}^{S+P} \, j_{S+P} \, J_{S+P} + \epsilon_{T+T_5}^{T+T_5} \, j_{T+T_5} \, J_{T+T_5}  \hc \Big] . \label{eq:LEFT_SU5}
\eeqa
For the \nubb process, one has to take two terms from the effective Lagrangian in Eq.~\eqref{eq:LEFT_SU5} which can be written as 
\beqa
\left({\cal L}_{1} {\cal L}_{2} \right) & =& \frac{G_F^2 }{2} \left[ j^{\mu}_{V-A}J_{V-A,\mu} \, j^{\nu}_{V-A}J_{V-A,\nu} + \sum_{\substack{i = S + P, \\ \phantom{i = {}} T + T_5}} \, \epsilon_{i}^{i} \, j^{\mu}_{V-A} \,J_{V-A,\mu}  \, j_{i} \, J_{i} + \,{\cal O} (\epsilon^2)\right]. \label{eq:onbb_amplitude}
\eeqa
\begin{figure}[t!]
    \centering
    \includegraphics[scale=0.4]{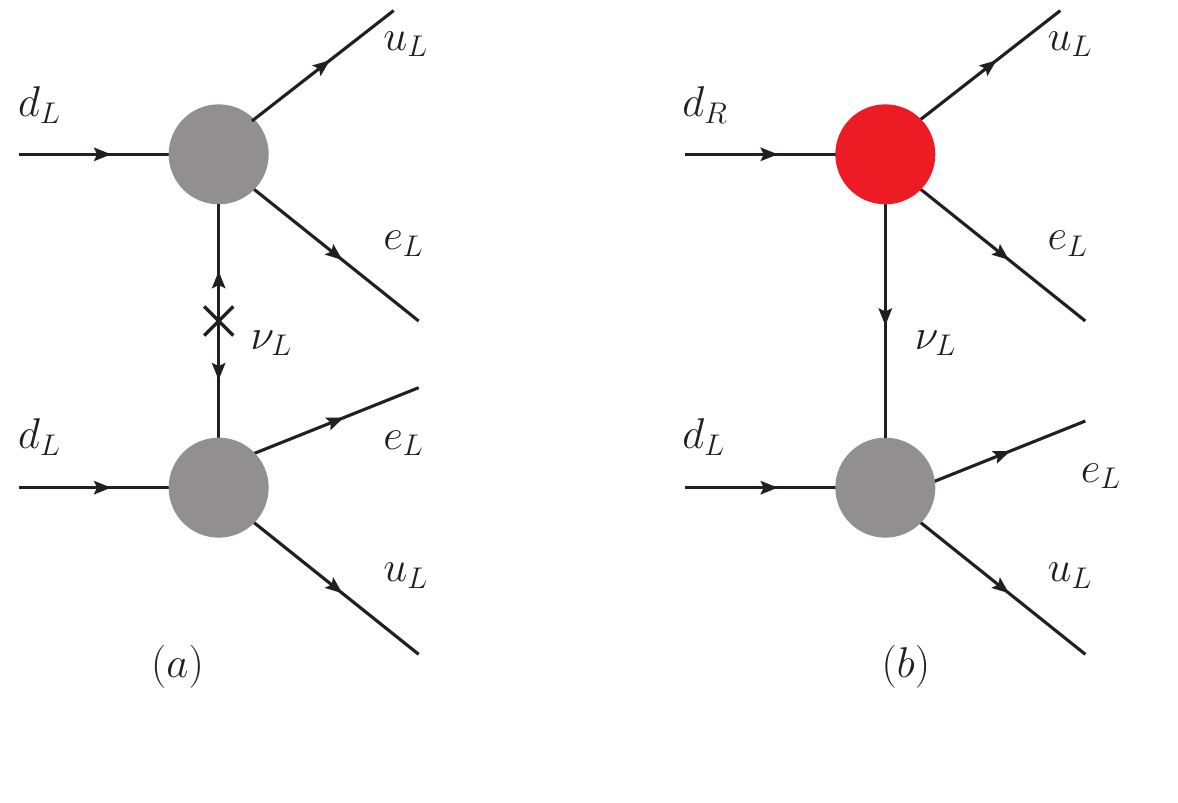}
    \caption{\textbf{Left:} \nubb process where both vertices contain SM interaction terms $\left( j^\mu_{V-A}\,J_{V-A\;\mu}\right)$. \textbf{Right:} \nubb process where one of the SM interaction vertices in the left diagram  is replaced by the non standard operators ($j_{i}\,J^{i}$) generated from $SU(5)$ scalars. }
    \label{fig:0nbb_diag}
    \end{figure}
The standard long-range $W_L$-$W_L$ contribution to \nubb is represented by the first term in Eq.~\eqref{eq:onbb_amplitude}, while the second term corresponds to the product of the standard and non-standard contribution, suppressed by a factor of $\epsilon$. The third term,  proportional to $\, {\cal O}(\epsilon^2)$, arises purely from non-standard contributions.
In Fig.~\ref{fig:0nbb_diag}, \nubb contributions are shown up to ${\cal O}(\epsilon)$, as the ${\cal O}(\epsilon^2)$ terms are typically negligible.

Having identified the relevant LEFT operators and their corresponding Wilson coefficients, the half-life can now be computed using the master formula presented in Ref.~\cite{Cirigliano:2018yza}:
\beqa
\left(T^{0\nu}_{1/2}\right)_{\rm total}^{-1} &=& g_A^4 \, \sum_{k} G_{0k} \, \left| {\cal A}_k\left( \left\{ C_i\right\}\right)\right|^2,
\label{eq:half-life}
\eeqa
where $G_{0k}$ denotes the atomic phase space factors (PSFs) and ${\cal A}_k\left(\left\{ C_i\right\} \right)$ are the sub-amplitudes which depend on the NMEs, low-energy constants (LECs) and the Wilson coefficients of the relevant operators; see Appendix~\ref{app:sub-amplitudes} for details. In this work, we use the python package $\tt \nu DoBe$~\cite{Scholer:2023bnn} to compute the half-life of \nubb for $^{76}$Ge and $^{136}$Xe nuclei, \Change{incorporating different NME schemes and taking into account the uncertainties in the low-energy constants (LECs).} 
The $\tt \nu DoBe $ package handles the RG running of the Wilson coefficients—particularly $\epsilon_{S+P}^{S+P}$ and $\epsilon_{T+T_5}^{T+T_5}$—from the electroweak scale ($\mu=\mu_{\rm ew}$) down to $\mu =  2$ GeV. It is to be noted that the decay-rate formula in $\tt \nu DoBe $ is expressed in terms of the Wilson coefficients evaluated at the chiral symmetry breaking scale ($\mu\sim 2$ GeV)~\cite{Cirigliano:2018yza}. The effective Majorana mass can be extracted from Eq.~(\ref{eq:half-life}) as [see Eq.~\eqref{eq:app_meff}]:
\beqa
\left(T_{1/2}^{0\nu}\right)_{\rm total}^{-1} & = & g_A^4 G_{01}\left| {\cal M}_{\nu}^{(3)} \right|^2\, \frac{ \left| m_{ee}^{\rm eff}\right|^2}{m^2_e} \, .
\label{eq:thalf1}
\eeqa
The effective Majorana mass is defined as 
\beqa
 m_{ee}^{\rm eff} =  m_{ee}^{\rm std} +  m_{ee}^{\rm nstd}, \label{eq:meff}
\eeqa
where $m_{ee}^{\rm std} = \sum_{i}U_{ei}^2\,m_i$ denotes the canonical light neutrino contribution, and $m_{ee}^{\rm nstd}$ denotes the non-standard contribution. 
By employing this parameterization, it is ensured that in the limit where non-standard contributions become negligible ($m_{ee}^{\rm nstd} \rightarrow 0$), the effective mass reduces to the standard result, $m_{ee}^{\rm eff}\to m_{ee}^{\rm std}$. Consequently, the deviation of $m_{ee}^{\rm eff}$ from  $m_{ee}^{\rm std}$ serves as a measure of the non-standard contributions, including interference with the SM contribution.  

\section{Parameter fitting} \label{sec:benchmark}
The viability of the expressions given in Eqs.~(\ref{eq:oneloopsu5}) and \eqref{eq:SU5NM} in yielding realistic charged and neutral fermion mass spectra observed at low energies is done via a $\chi^2$ optimization procedure, similar to that in Ref.~\cite{Patel:2023gwt}. The \(\chi^2\) function is defined as 
\beqa\label{eq:chisqdef}
\chi^2 \eq  \sum_{i} \left(\frac{O_{i,\,\rm{theo}} - O_{i,\,\rm{exp}}}{\sigma_{i,\,\rm{exp}}}\right)^2\,,
\eeqa
where  \( O_{i,\mathrm{theo}} \) is the theoretical prediction for each observable, and \( O_{i,\mathrm{exp}} \) is the corresponding experimentally measured value, where index \( i \) runs over all observables in the set. The experimental uncertainty for each observable is represented by \( \sigma_i \). The $\chi^2$ function in Eq.~\eqref{eq:chisqdef} includes nine charged fermion Yukawa couplings, four CKM parameters, two neutrino mass-squared differences and three PMNS mixing angles. Since all the \( O_{i\,\rm{theo}} \) values computed in Eqs.~\eqref{eq:oneloopsu5} and \eqref{eq:SU5NM} are at the GUT scale, the corresponding \( O_{i\,\rm{exp}} \) values are also evolved to the GUT scale. The GUT scale values of the experimental observables are obtained after extrapolating the low energy observables using two-loop THDM RGE equations from $M_t=173$ GeV (pole mass of the top quark~\cite{ParticleDataGroup:2024cfk}) to  $M_{\rm{GUT}}$, chosen here to be $10^{16}$ GeV. To compute the observables at the GUT scale, the analysis starts at \( M_t \) in a basis where \( Y_{u,e} = {\rm Diag}(M_{u,e})/v \), \( Y_d = V_{\rm{CKM}} {\rm Diag}(M_d)/v \),  with \( M_{u,d,e} \) as diagonal fermion mass matrices at \( M_t \). These serve as inputs for two-loop THDM RG running, with the corresponding $\beta$-functions evaluated using the {\tt PyR@TE} package~\cite{Sartore:2020gou}. As there are two Higgs vevs involved in the THDM framework, the ratio of their vevs, i.e. $\tan\beta\,\equiv\,\frac{v_1}{v_2}$, is fixed at $1.5$~\cite{Mummidi:2018myd} for our numerical purpose.  After diagonalization of the Yukawa matrices, the GUT-scale values of the various observables are obtained (as shown in Tab.~\ref{tab:sol}). The low-energy input values of the charged fermion masses and CKM mixing parameters are taken from PDG~\cite{ParticleDataGroup:2024cfk}, while the neutrino mass and mixing parameters are taken from a recent {\tt NuFIT} update~\cite{Esteban:2024eli}.  

Since uncertainties in the experimental observables at the GUT scale are not precisely known, conservative estimates are adopted. A \(\pm\SI{10}{\percent}\) uncertainty is used for all the observables. Despite knowing the leptonic parameters with high precision at low energies~\cite{ParticleDataGroup:2024cfk}, such a large uncertainty is assumed at the GUT scale due to several factors: (i) unknown scalar effects in the RGEs, (ii) uncertainties in scalar masses, (iii) uncertainty in the matching scale, and (iv) higher-order threshold corrections. The limited knowledge of GUT-scale physics justifies the increased uncertainty~(see e.g. Refs.~\cite{Dueck:2013gca, Babu:2016bmy} for similar fits with different uncertainty assumptions for the leptonic spectrum).

The allowed Yukawa values have been taken to be in the perturbative limit, i.e $\big |Y_{5,15,45}\big|\,\leq\,\sqrt{4\pi}$ for obtaining a numerical solution minimizing the $\chi^2$ function. This perturbative limit on the magnitude of Yukawa coupling is obtained from $2\to\,2 $ tree-level scattering at the high energy limit~\cite{Allwicher:2021rtd}. The mass of the heavy gauge boson~$(M_X)$ appearing in Eq.~(\ref{eq:L_X}) has been set to be equal to the matching scale~$(\mu)$ which is same as the conventional GUT scale $(\mu\,=\,M_{\rm GUT}\,=\,M_{X}\,=\,10^{16}$ GeV) and the gauge coupling  $g_5$ is taken to be $0.524$, which is the mean value of the SM gauge couplings at the GUT scale. We do not make an attempt to achieve exact unification of the gauge couplings in this scenario, which can be done by extending the model by adding more degrees of freedom that do not couple to SM fermions thereby not affecting our analysis. The trilinear coupling $\eta$ appearing in the expression of neutrino mass [cf.~Eq.~(\ref{eq:SU5NM})] is also varied around the GUT scale in the range of $(0.1,10)M_{\rm GUT}$, whereas the quartic coupling $\rho$ appearing in the same expression has been set to unity.  

In Eq.~\eqref{eq:oneloopsu5}, $Y_{45}$ can be chosen as a real-diagonal matrix, $Y_5$ a symmetric matrix with complex entries and  two cases can be considered for $Y_{15}$, i.e., {\bf Case-I} where $Y_{15}$ is  real symmetric, and {\bf Case-II} where $Y_{15}$ is a complex-symmetric matrix. This yields a total of 21~(27) Yukawa couplings in Case-I~(Case-II). Moreover, there are one, three and six BSM scalar fields inside the $\mathbf{5}_\hh$, $\mathbf{15}_\hh$ and $\mathbf{45}_\hh$-dimensional irreps, whose masses can take any value between roughly 1 TeV (to satisfy the LHC constraints) and GUT scale. Additionally, the cubic coupling $\eta$ can in principle take any value; however, a value much smaller than the GUT scale would bring a fine-tuning problem. Therefore, we vary $\eta$ near the GUT scale for our analysis. Altogether, there are 32 (38) unknown parameters in Case-I (Case-II) which go into the theoretical predictions for the 18 observables in Eq.~\eqref{eq:chisqdef} to be fitted to their corresponding experimental values.  Moreover, the neutrino masses can have either normal ordering~(NO) or inverted ordering~(IO); and we shall analyze both possibilities.  The fitting of low-energy observables require splitting in the masses of scalars residing in a particular irrep $\left(\mathbf{15}_\hh, \, \mathbf{45}_\hh\right)$. Some of the scalar masses are required to significantly deviate from the matching scale in order to have substantial threshold corrections, which has also been observed in Ref.~\cite{Shukla:2024bwf}. This scenario goes beyond the Extended Survival Hypothesis~\cite{delAguila:1980qag,Mohapatra:1982aq,Dimopoulos:1984ha}, which states that all the scalars except the one breaking the symmetry must be heavier than the symmetry-breaking scale. This type of framework calls for a mechanism that generates splitting in the masses of scalars residing in the common multiplet, reminiscent of the doublet-triplet splitting problem generic to all GUTs~\cite{Dimopoulos:1981zb, Sakai:1981gr}. This could be addressed, e.g. by invoking higher-dimensional operators~\cite{Dorsner:2024seb}, the details of which do not really matter for our phenomenological analysis.

From Eq.~(\ref{eq:nubb}), it is evident that the \nubb rate, mediated by the $SU(5)$ scalars, depends on two main factors: (i) the Yukawa couplings of the leptoquarks with first-generation fermions,  and (ii) the leptoquark masses. This section explores the possibility of obtaining a viable fermion mass fit together with maximizing the \nubb rate, given in Eq.~(\ref{eq:app_meff}). 

As discussed above, the model with Case-I~(II) contains 32~(38) free parameters, including the scalar masses. It is evident from earlier discussions that the pair of scalars $S_3-R_2$ contribute to neutrino masses at one-loop while $S_3-\tilde{R}_2$ contribute to $0\nu\beta\beta$. Therefore, to relate these two effects and to maximize the $0\nu\beta\beta$, we fix the leptoquark masses as follows: $M_{S_3} \sim M_{R_2} \sim  2.0\,\text{TeV}$ to satisfy the LHC constraints~\cite{CMS:2018lab,CMS:2018iye,CMS:2018ncu, ATLAS:2019ebv,ATLAS:2020dsk,CMS:2020wzx, ATLAS:2021oiz,CMS:2022nty,ATLAS:2023vxj, ATLAS:2023prb}\footnote{The current LHC constraint on scalar leptoquark masses is $1.58~(1.59)$ TeV at 95\% confidence level (C.L) considering its decay to a top quark and electron~(muon) with 100\% branching ratio~\cite{ATLAS:2023prb}. To satisfy the LHC bounds, the leptoquark masses are fixed around $2$ TeV. Additionally, since the neutrino mass loop function in Appendix~\ref{app:LF} diverges for $M_{R_2} = M_{S_3}$, a small mass splitting is introduced to avoid this issue.  
}, while $M_{\tilde{R}_2}  \sim 10^3 $ TeV to satisfy the cLFV constraint from $\mu \to e$ conversion (see Appendix~\ref{app:clfvs}). The other scalar masses are varied within one order of magnitude from the GUT scale. Some of these scalars also induce proton decay; thus, setting their masses close to $M_{\rm{GUT}}$ automatically satisfies the proton decay constraints. Additionally, the masses of the extra scalars $\Sigma$ and ${\mathbb{S}}$ are allowed to vary between 1 TeV and the GUT scale, which is necessary to generate a mass splitting between the charged lepton and down quark sectors. These particular choices reduce the number of free parameters contributing to the different observables and maximize the \nubb contribution.

\begin{figure}[t!]
    \centering
    \includegraphics[scale=0.48]{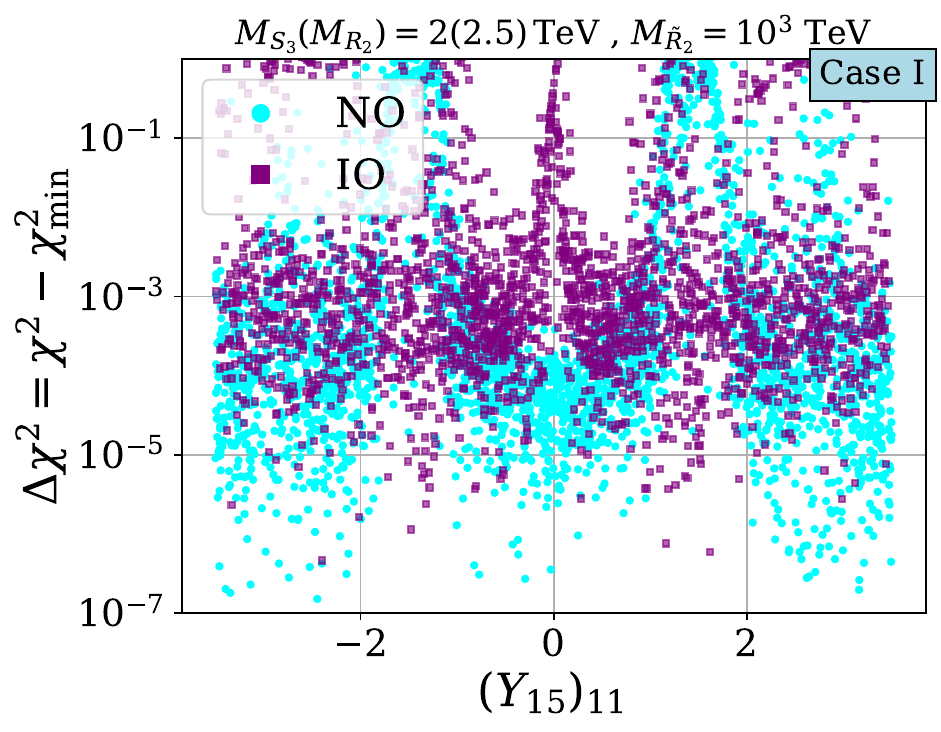}
    \includegraphics[scale=0.48]{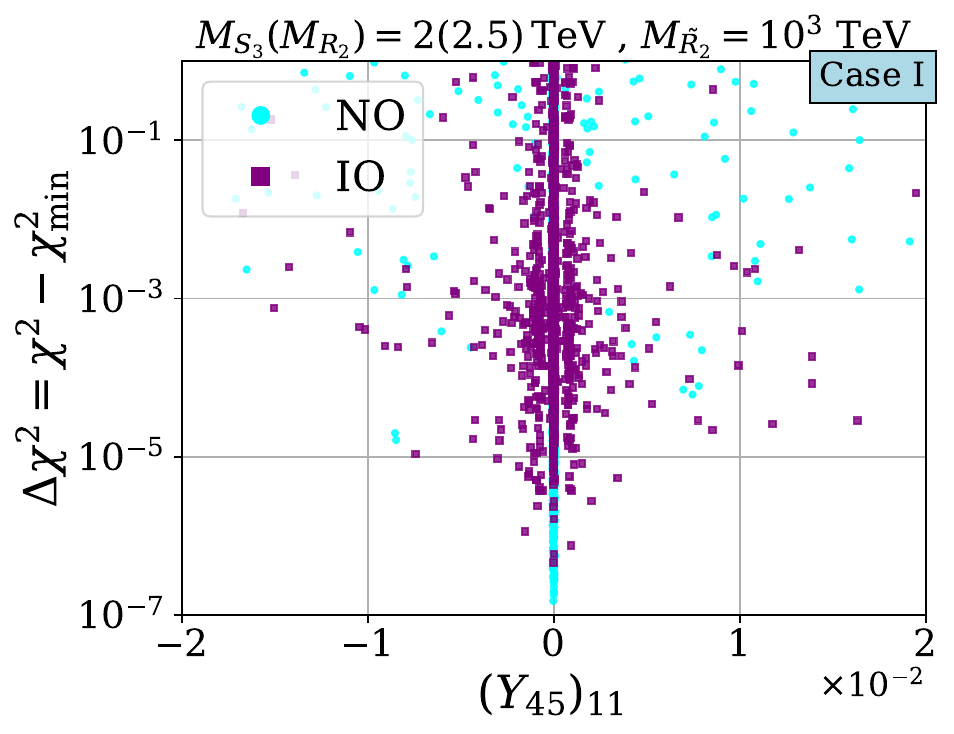}\\
        \includegraphics[scale=0.48]{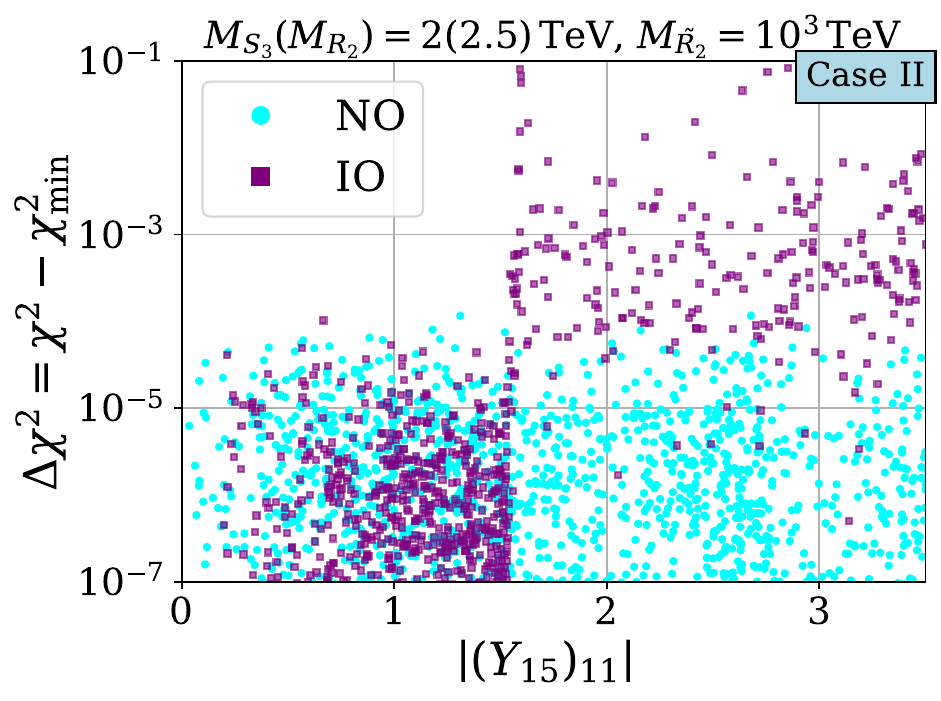}
    \includegraphics[scale=0.48]{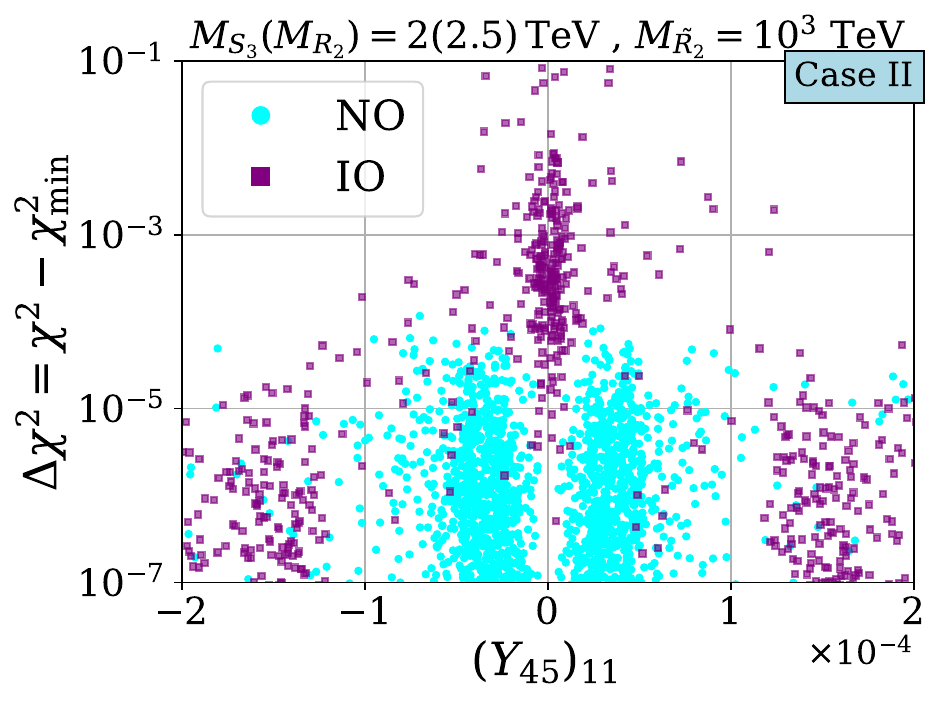}\\
\caption{Variation of $\Delta\chi^2$ with $\left(Y_{15}\right)_{11}$ (left panels) and $\left(Y_{45}\right)_{11}$ (right panels) in the range of $(-3.5, 3.5)$.
The top panel is for Case-I with real $Y_{15}$, while the bottom panel is for Case-II with complex $Y_{15}$. The \textcolor{cyan}{cyan} (\textcolor{purple}{purple})-colored points represent the normal (inverted) mass ordering.}
    \label{fig:chisquare&Yukawa}
\end{figure}

Having fixed the masses of scalar leptoquarks contributing to $0\nu\beta\beta$, the allowed range of Yukawa couplings need to be explored. As our aim is to maximize the leptoquark contribution to \nubb together with the viable fermion mass fit, the Yukawa coupling $(Y_{15})_{11}$ for both Cases I and II are varied in steps between  $(-3.5, 3.5)$ under the aforementioned constraints. The value of 3.5 is chosen so that the Yukawa coupling remains within its perturbative limit of $\sqrt{4\pi}$. With this choice, the algorithm is configured to fit the theoretical observables. The variation of $\Delta\chi^2\,\equiv\,\chi^2 - \chi^2_{\rm{min}}$ as a function of $(Y_{15})_{11}$ and $\left|\left(Y_{15}\right)_{11}\right|$ is shown in Fig.~\ref{fig:chisquare&Yukawa} left panels for Case-I (top) and Case-II (bottom). The right panels show the range of preferred values for $(Y_{45})_{11}$. The  \textcolor{cyan}{cyan} (\textcolor{purple}{purple}) dots in Fig.~\ref{fig:chisquare&Yukawa} represent normal~(inverted) neutrino mass ordering.  

\begin{figure}[t!]
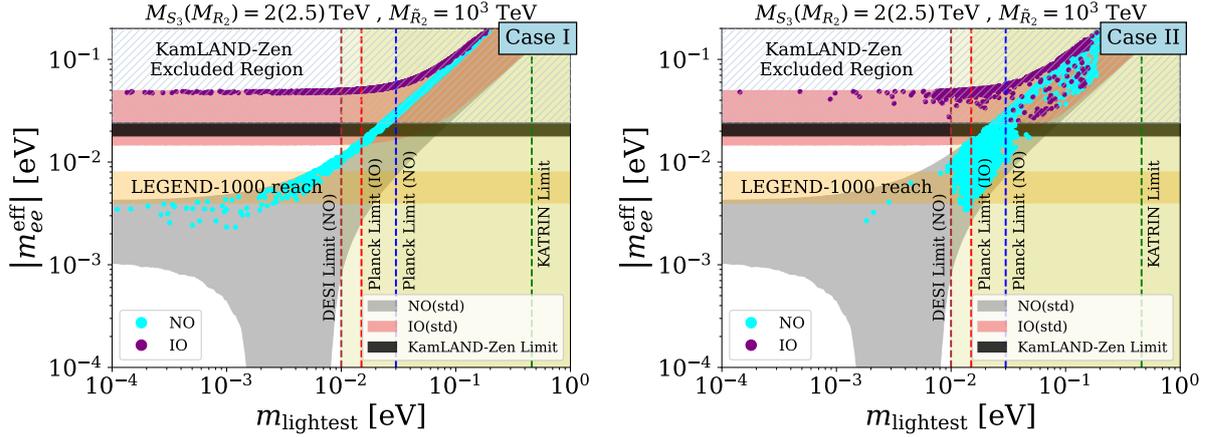

    \centering
    \includegraphics[width=0.48\linewidth]{replot/mlightest-mbb-NO_uncertainity.pdf}
    \includegraphics[width=0.48\linewidth]{replot/mlightest-mbb-NO_complex_uncertainity.pdf}
    \caption{Variation of $m_{ee}^{\rm eff}$ as a function of lightest neutrino mass obtained from the fermion mass fitting in our $SU(5)$ model with fixed leptoquark masses $M_{S_3}=2$ TeV, $M_{R_2}=2.5$ TeV, $M_{\tilde{R}_2}=10^3$ TeV. The left (right) panel is for Case-I (II). The gray and pink regions show the allowed ranges for NO and IO in the standard \nubb mechanism. The teal and orange bands show the current KamLAND-Zen limit and future LEGEND-1000 sensitivity, respectively. The vertical lines show the direct (KATRIN) and indirect (Planck, DESI) limits on the absolute neutrino mass.} 
    \label{fig:mbb}
\end{figure}

It is clear from Fig.~\ref{fig:chisquare&Yukawa} that all fermion masses and mixing angles can be fitted in this scenario considering the above-mentioned choices. As evident from the plot, the magnitude of the first element of the Yukawa matrix $|\left(Y_{15}\right)_{11}|$ can take any value in the entire range of $(0, 3.5)$ for yielding a good numerical fit. On the other hand, the preferred value of the first element satisfies $\big|\left(Y_{45}\right)_{11}\big| \lesssim 2 \times 10^{-2}$ in Case-I and $\lesssim 2 \times 10^{-4}$ in Case-II, for both normal and inverted orderings. 

\section{Model predictions for \nubb}
\label{sec:0nubb}
Now we compute the model predictions for \nubb. Any solution with acceptable $\chi^2$ must respect the current experimental upper bound on $m^{\rm eff}_{ee}$. We use the result from KamLAND-Zen experiment using $^{136}$Xe which is quoted as $m^{\rm eff}_{ee} < \left(0.015- 0.024\right)$\, eV~\cite{KamLAND-Zen:2024eml}, where the range is due to \Change{NME uncertainties and the variation of unknown  LECs. The values of the LECs used are given in Tab.~\ref{Tab:LECs}.  As for the NMEs, we have used all the NME schemes already implemented in $\tt \nu DoBE$, namely, IBM-2~\cite{Deppisch:2020ztt}, QRPA~\cite{Hyvarinen:2015bda}, Shell Model~\cite{Menendez:2017fdf} and CDFT~\cite{Ding:2024obt}, as tabulated in Tab.~\ref{tab:NMEs}. 
} Also, note that the experimental limits for $m^{\rm eff}_{ee}$ are obtained using $\tt \nu DoBE$~\cite{Scholer:2023bnn}, which includes short-range effects. This makes these constraints slightly different from direct experimental quoted results. 

Fig.~\ref{fig:mbb} shows the variation of $m_{ee}^{\rm eff}$ as a function of the lightest neutrino mass for Case-I (left panel) and Case-II~(right panel). The \textcolor{cyan}{cyan} (\textcolor{purple}{purple}) points are the solutions reproducing the correct fermion mass spectra for NO (IO). The gray (pink) shaded regions denote the standard NO~(IO) region, whereas the black and orange bands show the current KamLAND-Zen limit~\cite{KamLAND-Zen:2024eml} and future LEGEND-1000 sensitivity~\cite{LEGEND:2021bnm} calculated using $\tt \nu DoBE$~\cite{Scholer:2023bnn}. The vertical dashed lines (from right to left) indicate the direct limit on the absolute neutrino mass $m_\nu<0.45$ eV (at 90\% CL) from KATRIN~\cite{KATRIN:2024cdt}, and the indirect limits derived from the sum of neutrino mass constraint of $\sum_i m_i<0.12$ eV from Planck~\cite{Planck:2018vyg} and $\sum m_\nu<0.064$ eV from DESI~\cite{DESI:2025zgx}. Note that the DESI limit disfavors the IO; therefore, there is no DESI line corresponding to the IO case. From Fig.~\ref{fig:mbb}, it is evident that the values of $m_{ee}^{\rm eff}$ obtained here are all within the allowed range of standard $m_{ee}$ values, implying that the non-standard scalar contributions are small. This can also be seen from Eq.~\eqref{eq:app_meff}: 
\beqa
m_{ee}^{\rm nstd} & \simeq & 2 \, m_N \times \frac{\rho \, \eta \, v^5}{M_{\Delta}^2} \, \frac{\left( Y_{\tilde{R}_2} \right)_{11}}{M_{\tilde{R}_2}^{2}}\, \frac{\left( Y_{S_3}\right)_{11}}{M_{S_3}^2}  \lesssim  3.5 \times 10^{-17}~\rm eV \label{eq:meenstd-01}
\eeqa  
In the above equation, the considered values for the scalar masses and the trilinear scalar coupling contributing to \nubb are obtained from the $\chi^2$ analysis, which are $\eta \sim 10^{16}$ GeV, $M_\Delta \sim 10^{16}$ GeV, and   $M_{S_3} = 2 \times 10^3$ GeV, $ M_{R_2} = 2.5 \times 10^3$ GeV and $M_{\tilde{R}_2} = 10^6$ GeV. The Yukawa couplings are taken as $\left(Y_{\tilde{R}_2
}\right)_{11}\approx  \left(Y_{15}\right)_{11} = 3.5 $ and $\left(Y_{S_3}\right)_{11} \approx \left(Y_{45}\right)_{11} = 2\times 10^{-2}$ which are the maximum allowed Yukawa couplings, required for the fermion mass fitting, as evident from Fig.~\ref{fig:chisquare&Yukawa}. Additionally, in Eq.~\eqref{eq:meenstd-01}, a factor of two is multiplied for the two contributions yielding the same LEFT operator, as shown in Fig.~\ref{fig:0nbb_operator}. With these assumptions, the value of maximum possible $m_{ee}^{\rm nstd}$ is negligible compared to the standard contribution, $m_{ee}^{\rm std} \sim 10^{-2}$ eV. The suppression of $m_{ee}^{\rm nstd}$ is due to the high mass scale of the triplet scalar ($M_{\Delta}$), which is required to generate the observed small neutrino masses [cf.~Eq.~\eqref{eq:SU5NM}]. Consequently, within the considered settings of the $SU(5)$ framework, the effective Majorana mass is dominated by the standard light-neutrino exchange contribution, yielding $m_{ee}^{\rm eff} \approx m_{ee}^{\rm std.}$. It is also clear from Fig.~\ref{fig:mbb} that in both cases, for the considered choice of parameters, IO is ruled out as the prediction for $m^{\rm eff}_{ee}$ is larger than the KamLAND-Zen limit.

In the left panel of Fig.~\ref{fig:mbb}, since $Y_{15}$ is real~(Case-I), the Majorana phases are zero, and $m_{ee}^{\rm eff}$ values stay close to the upper limit of the standard contribution for NO. On the other hand, in the right panel, the Majorana phases are non-zero due to complex $Y_{15}$. However, it is evident from the right panel that the fermion mass fits prefer the higher values of $m_1$ (through higher values of $Y_{15}$) and the cancellation regime of the standard $m^{\rm eff}_{ee}$ is not reached. 

\section{Enhancing the \nubb contribution}
\label{sec:0nubb2}
\begin{figure}[t!]
    \centering
    \includegraphics[width=0.7\textwidth]{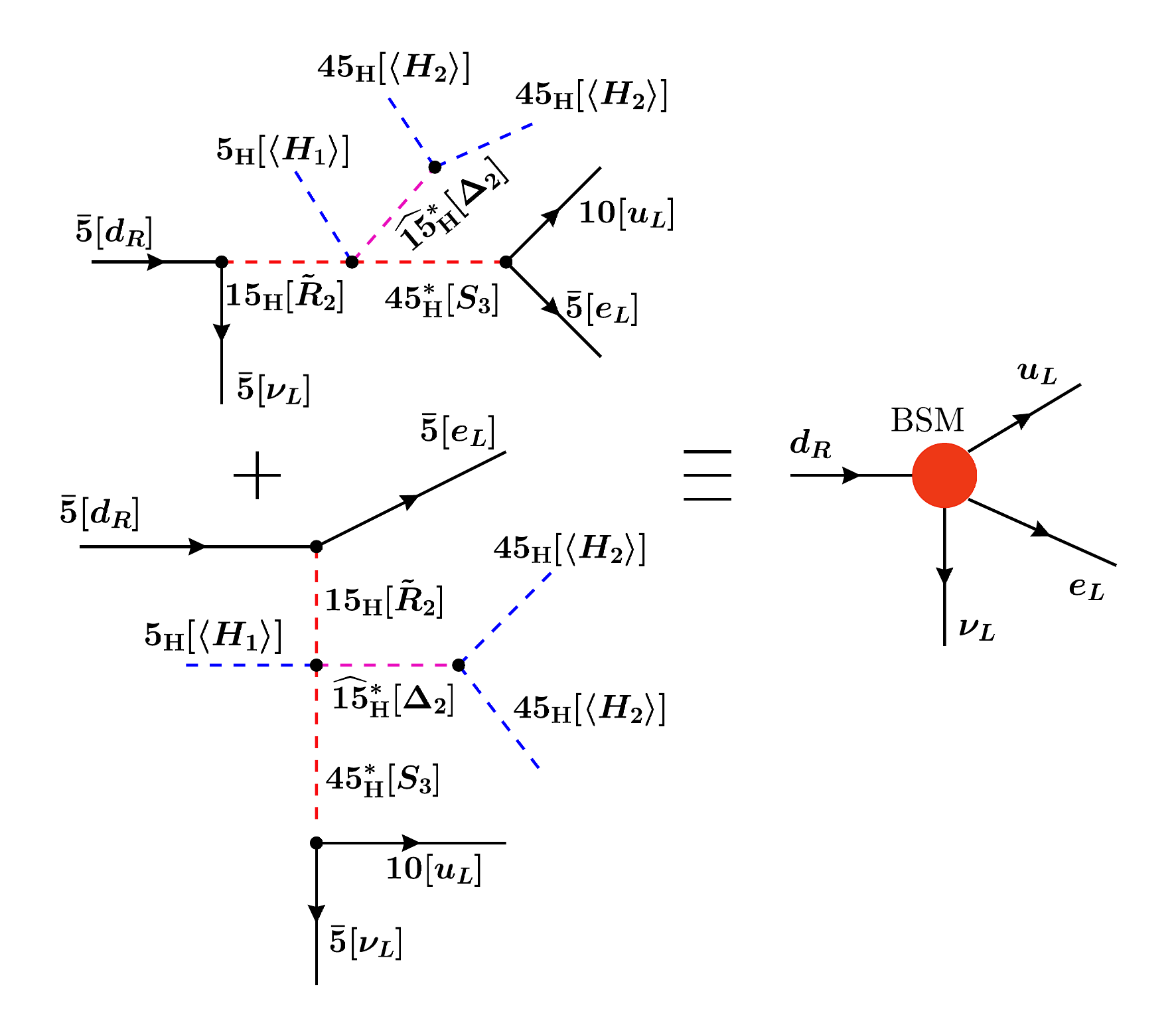}
    \caption{Effective operator diagrams that contribute to \nubb in the $SU(5)\times {\cal Z}_3$ model in the presence of $\mathbf{\widehat{15}_\hh}$. The diagram in the left panel shows the scalar-induced \nubb in the considered scenario, where the irreps outside the parenthesis depict the parent multiplets. The right panel shows the equivalent LEFT operator.}
    \label{fig:onbb_extended}

    \end{figure}
\begin{figure}[t!]
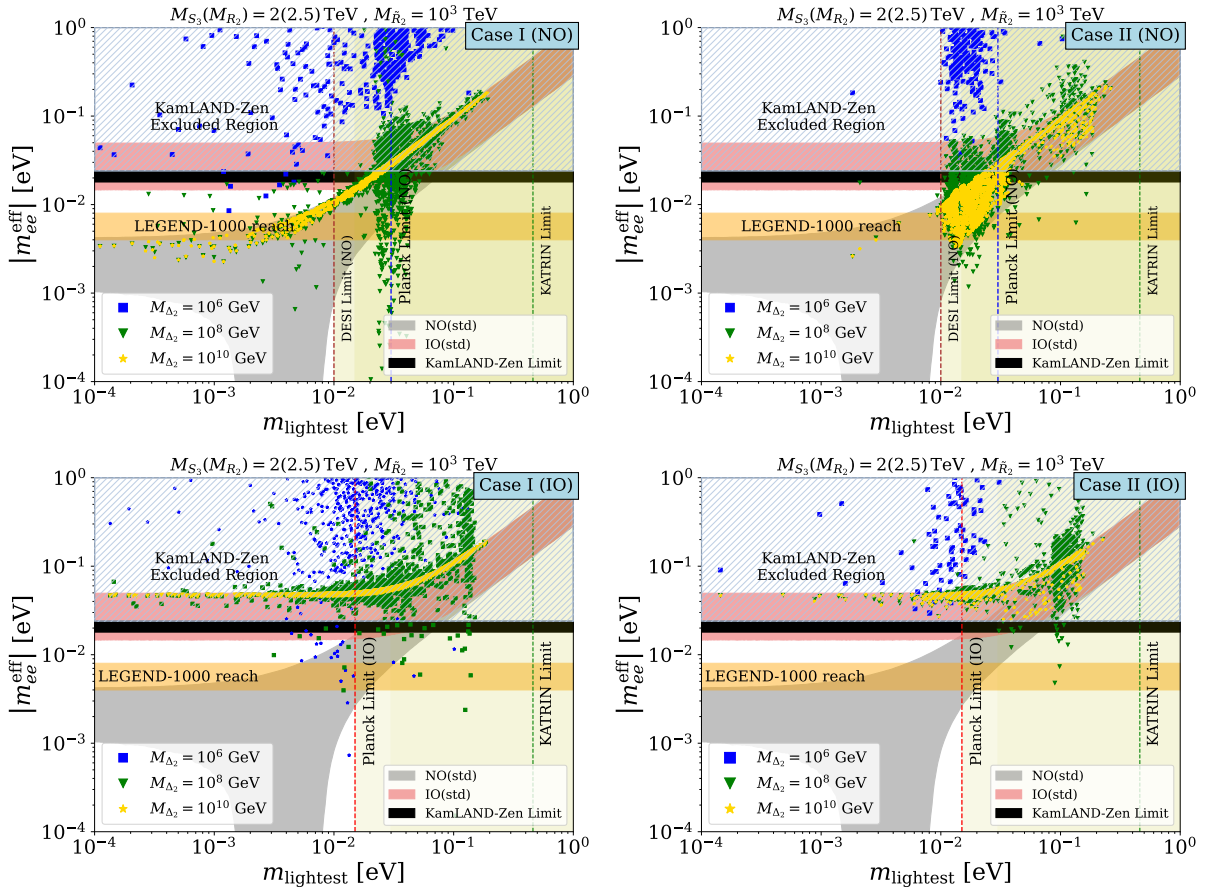

    \centering
    \includegraphics[width=0.48\linewidth]{replot/mlightest-mbb-NO_mdelta_uncertain.pdf}
     \includegraphics[width=0.48\linewidth]{replot/mlightest-mbb-NO_mdelta_complex_uncertain.pdf}

     \includegraphics[width=0.48\linewidth]{replot/mlightest-mbb-IO_mdelta_uncertain.pdf}
     \includegraphics[width=0.48\linewidth]{replot/mlightest-mbb-IO_mdelta_complex_uncertain.pdf}
    \caption{Same as in Fig.~\ref{fig:mbb} but with an additional scalar contribution from $\Delta_2$ arising from an extra $\mathbf{15_{\hh}}$. 
    The results are shown for different benchmark values of $M_{\Delta_2}$. For clarity of presentation, we have separated the NO and IO results for Cases-I and II.}  
    \label{fig:enhanced_mbb}
    \end{figure}

As concluded in the previous section, the \nubb contribution arising from leptoquarks in the considered $SU(5)$ framework is suppressed due to the high mass scale of $\Delta$, which is required to accommodate the observed smallness of neutrino masses. In this section, we consider an alternative scenario by extending the scalar sector with a new $\mathbf{15_\hh}$ irrep, where the scalar-induced contributions to the \nubb can be enhanced. This additional $\mathbf{15_\hh}$ is denoted as $\widehat{\mathbf{15}}_\hh$ in Tab.~\ref{tab:Z3charges} \Change{and can, in principle, be motivated by a broader theoretical consideration, e.g. by  embedding the considered framework within a higher gauge group such as $SO(10)$. In such a scenario, the $\mathbf{15}_\hh$ can reside in a $\mathbf{54}_\hh$ dimensional irrep of $SO(10)$, which is often used to break the $SO(10)$ gauge symmetry~\cite{Fritzsch:1974nn, Slansky:1981yr}. Furthermore, the $\mathbf{54}_\hh$ also does not participate in Yukawa interactions at the renormalisable level~\cite{Buccella:1984ft}, and therefore, does not alter the fermion mass fits.}

In terms of the scalar content, $\widehat{\mathbf{15}}_\hh$ resembles with the $\mathbf{15_\hh}$ irrep shown in Tab.~\ref{tab:scalars} and the new scalar stemming from this extra $\mathbf{15_\hh}$ which contributes to \nubb  is denoted as $\Delta_2\left(1,3,1\right)$. Here, $\widehat{\mathbf{15}}_\hh$ irrep is assigned a ${\cal Z}_3 = 1$ charge that forbids its coupling to the $SU(5)$ fermionic multiplets i.e. $\ff$ and $\ft$.  As a result, the scalars in $\widehat{\mathbf{15}}_\hh$ do not contribute to the fermion mass fitting discussed earlier. 

Since it does not contribute to neutrino masses, the triplet scalar~$(\Delta_2)$ residing in $\mathbf{\widehat{15}_\hh}$ can be as light as possible, enabling it to generate substantial contributions to the \nubb amplitude.\footnote{\Change{There may exist other possibilities to enhance \nubb in $SU(5)$, e.g. by adding higher multiplets such as ${\bf 50}_\hh$ or ${\bf 70}_\hh$.}} The corresponding diagrams are shown in Fig.~\ref{fig:onbb_extended}. The current non-observation of \nubb can therefore be utilized to establish a lower bound on the mass of this triplet scalar.  The role of $\Delta_2$ in enhancing the \nubb rate is illustrated in Fig.~{\ref{fig:enhanced_mbb}}, by considering different values of $M_{\Delta_2}\,\supset\{ \,10^{6},\,10^{8},\,10^{10}\}$ GeV. The value of $m_{ee}^{\rm std}$ as a function of the lightest neutrino mass is obtained from the solution set shown in Fig.~\ref{fig:mbb}, whereas the non-standard \nubb contribution for $M_{\Delta_2}$ is given below in Eq.~\eqref{eq:nstd}. It is found that for $M_{\Delta_2} \sim 10^{8}$~GeV the interplay between the standard and non-standard contribution can lead to the cancellation among these contributions depending upon the sign of the first matrix elements of $Y_{15,\, 45}$, which will be discussed subsequently by choosing two suitable benchmark values~(BP1, BP2) from Case-I~(given in Appendix~\ref{app:bestfit}). 
The destructive interference between the standard and non-standard contributions reduces the effective Majorana mass to levels compatible with current experimental bounds, thereby allowing the IO scenario in Case-I, as can be seen from the lower left panel of Fig.~\ref{fig:enhanced_mbb}. 
  \begin{figure}[t!]
    \centering
    \includegraphics[scale=0.45]{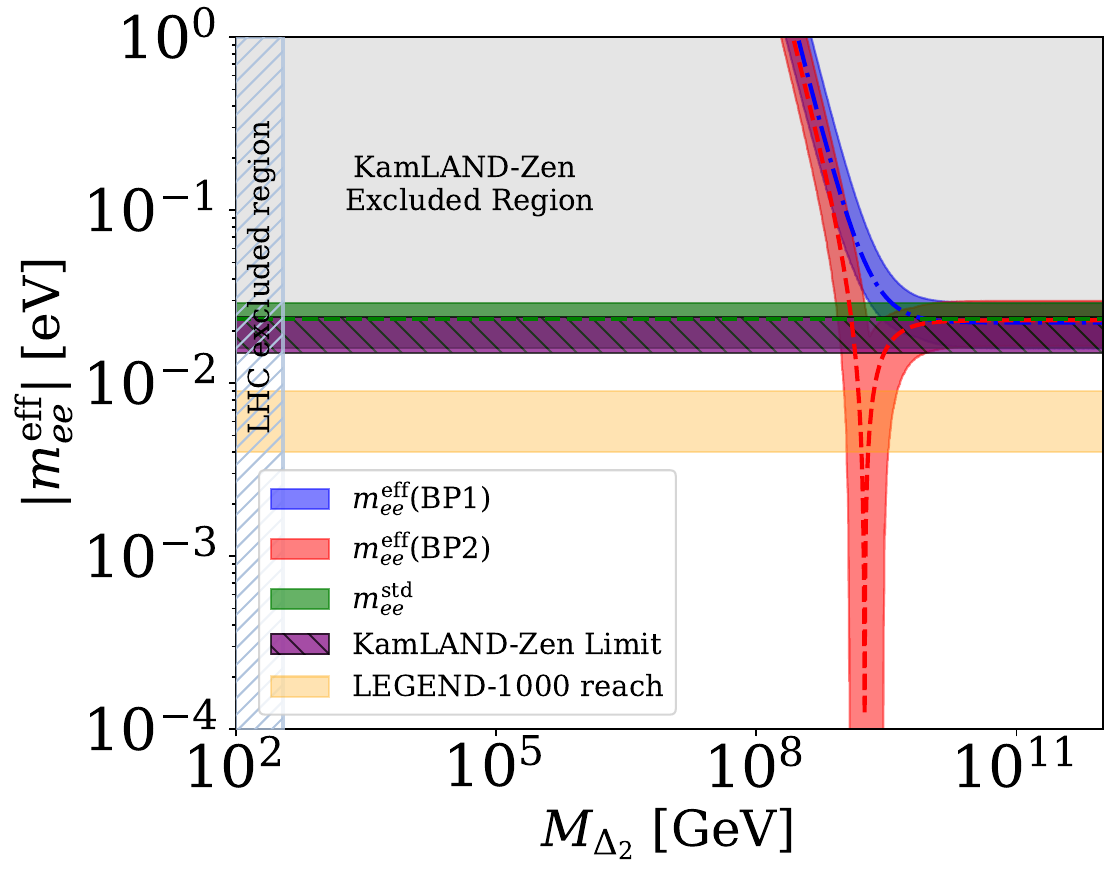}
    \caption{Variation of $m_{ee}^{\rm eff}$ for BP1~(blue) and BP2~(red) as function of $M_{\Delta_2}$. The standard contributions for the two BPs are calculated as $0.0229$ eV and $0.0236$ eV, shown by the green dashed line. The purple-shaded region corresponds to the current KamLAND-Zen limit,  while the orange band shows the LEGEND-1000 sensitivity. The vertical hatched region is disfavored by LHC data. The red, blue and green bands denote the uncertainty in the corresponding $m_{ee}^{\rm eff}$ values due to various NME schemes and variation in the unknown LECs. }
    \label{fig:onbb_enhanced}
    \end{figure}

The expression relating $T_{1/2}^{0\nu}$ and $m_{ee}^{\rm eff}$ is given in Eq.~\eqref{eq:app_meff}. 
The standard and non-standard contributions can interfere either constructively or destructively depending on the sign of $\epsilon_{S+P}^{S+P}$. For BP1, as $\left(Y_{15}\right)_{11}\times\left(Y_{45}\right)_{11} < 0 $, it implies $\epsilon_{S+P}^{S+P} <0$~[cf. Eq.~\eqref{eq:S+P}] and ${\cal M}_{\nu}^{\prime (6)}/{\cal M}_{\nu}^{(3)}<0$. Consequently, the non-standard term in Eq.~\eqref{eq:app_meff} is positive and it constructively interferes with $m_{ee}^{\rm std}$. On the other hand, the converse happens for the BP2, where the non-standard term is negative and it destructively interferes with $m_{ee}^{\rm std}$. This constructive or destructive interference effect between the standard and non-standard contributions is shown in Fig.~\ref{fig:onbb_enhanced} as a function of $M_{\Delta_2}$, where the cancellation is evident for BP2 (red band). The figure also shows that for BP1, $M_{\Delta_2}$ values below $ 4.7\times 10^{9}$ GeV are ruled out by KamLAND-Zen whereas for BP2, this bound gets relaxed to $1.2\times 10^9$ GeV due to the cancellation effect.  Both Figs.~\ref{fig:enhanced_mbb} and ~\ref{fig:onbb_enhanced} indicate that the non-standard contribution becomes sub-dominant for $M_{\Delta_2} \gtrsim 10^{10}$ GeV. It implies that \nubb can provide more stringent limits on $M_{\Delta_2}$ well beyond the reach of colliders. For comparison, the current LHC bound on $M_{\Delta_2}$ is around 350 GeV from the diboson decay channel~\cite{ATLAS:2021jol}, as shown by the vertical hatched region in Fig.~\ref{fig:onbb_enhanced}. Note that since $\Delta_2$ does not couple to SM fermions, the more stringent limits on triplet scalars from (di)leptonic decay modes do not apply in our case.

We can derive a lower bound on $M_{\Delta_2}$ from \nubb using the expression
\beqa\label{eq:mdel2bound}
m_{ee}^{\rm nstd } & \simeq &  2\, m_N \times \frac{\rho \, \eta \, v^5}{M_{\Delta_2}^2} \, \frac{\left( \sqrt{2}\,Y_{15} \right)_{11}}{M_{\tilde{R}_2}^{2}}\, \frac{\left(\sqrt{2}\,Y_{45}\right)_{11}}{M_{S_3}^2} \, .  \label{eq:nstd}
\eeqa 
By applying the current KamLAND-Zen upper limit of $m_{ee} < 0.02 $ eV, a conservative upper bound on $M_{\Delta_2}$ can be derived as:
\beqa \label{eq:mdelta2no}
M_{\Delta_2} & \gtrsim & 10^{10} {\rm GeV} \left[ \frac{m_N}{1\,\rm GeV}  \, \frac{10^{-2} \,\rm eV}{m_{ee}}\,  \frac{\rho}{{\cal O}(1)}\, \frac{\eta}{10^{16} \,\rm GeV}\,\frac{\left( Y_{15} \right)_{11}}{{\cal O}(1)} \, \frac{\left( Y_{45}\right)_{11}}{10^{-2}} \right. \nonumber \\
& & \hspace{3cm}\left. \left( \frac{v}{10^2\,\rm GeV} \right)^5 \left( \frac{10^6 \,\rm GeV}{M_{\tilde{R}_2}} \right)^2 \, \left(\frac{2 \times 10^3 \,\rm GeV}{M_{S_{3}}} \right)^2 \right]^{-1/2}  \, .\label{eq:boundmt2}
\eeqa
The bound mentioned in Eq.~\eqref{eq:boundmt2} is subjected to the minimum masses of the leptoquarks and maximum values of the Yukawa couplings yielding a viable fermion mass spectrum and evading the cLFVs. For larger leptoquark mass and smaller Yukawa couplings, the bound on $M_{\Delta_2}$ can be relaxed down to the LHC exclusion limit. The point is that the \nubb bound on $M_{\Delta_2}$ can be much stronger than the collider bounds, depending on the leptoquark masses and Yukawa couplings.


\section{Conclusions}
\label{sec:summary} 
Processes violating the Baryon and/or Lepton number provide `smoking gun' signatures of new physics beyond the Standard Model. Models based on Grand Unified Theories provide a viable setup in which such rare processes arise naturally. This work investigated a realistic $SU(5)$ scenario with  contributions to neutrinoless double beta decay resulting from Majorana neutrinos as well as from $SU(5)$ scalars, while satisfying the proton decay constraints, as well as the observed fermion mass spectrum. GUT models are known to  strongly constrain the Yukawa parameters by unifying quarks and leptons in the same multiplet. This study intended to analyze the possibility of enhancing the \nubb contribution from leptoquarks embedded in $SU(5)$ multiplets. It is found that in the minimal $SU(5)$ setup extended by a triplet scalar belonging to the 15 dimensional representation to incorporate neutrino mass,  the same set of scalars inducing \nubb also induce proton decay.  Consequently, compliance with the proton decay constraints precludes any observable effect of the scalars on the \nubb process. 

To avoid this problem, we then constructed an $SU(5)\times {\cal Z}_3$ model which forbids the diquark couplings of the $(S_3)$ leptoquark, thus removing the proton decay constraint. However, imposing the discrete symmetry makes the Yukawa relations unrealistic at the tree level. Nevertheless, switching on radiative corrections we were able to generate the observed  fermion mass spectra (within the assumed uncertainties). On the other hand, in order to yield the observed tiny neutrino masses, the desired value of the mass of triplet scalar field $\Delta$ was found to be close to the GUT scale. Such a high value of the mass of $\Delta$  suppresses the \nubb amplitude orders of magnitude below the standard light-neutrino exchange contribution. Moreover, within the region of parameter space consistent with the fermion mass spectrum, the light-neutrino-induced effective neutrino mass in neutrinoless double beta decay for the inverted ordering exceeds the KamLAND-ZEN limits, thereby disfavoring this scenario.

Finally, we showed that a significant enhancement in the \nubb rate can be achieved by introducing an additional scalar irrep~$(\mathbf{\widehat{15}_\hh})$ consisting of another triplet ($\Delta_2$) with ${\cal Z}_3 = 1$. This charge assignment disables it from having fermion interactions at the GUT scale. This ensures that $\Delta_2$ does not contribute to neutrino masses~(or any other fermion mass), allowing its mass to be much smaller than the GUT scale. By optimizing the Yukawa couplings and minimizing  the leptoquark masses within experimental limits, we derive a conservative constraints on $M_{\Delta_2} \geq 10^{10}$ GeV from non-observation of $0\nu\beta\beta$. This scenario demonstrates that while the minimal $SU(5)$ framework faces severe phenomenological constraints, extended symmetry structures with carefully chosen scalar sectors can enhance the \nubb process beyond the standard neutrino contribution, thus making such GUT models testable in future \nubb experiments. 

\acknowledgments
We sincerely thank Namit Mahajan and Ketan Patel for their valuable suggestions. We also thank Chayan Majumdar for collaboration during an earlier stage of this work. BD wishes to acknowledge the PRL Theory Group for warm hospitality where this work was initiated. BD also wishes to acknowledge the Center for Theoretical Underground Physics and Related Areas (CETUP*) and the Institute
for Underground Science at SURF for hospitality and for providing a stimulating environment during the completion of this work. The work of BD was partly supported by the U.S. Department of Energy under grant No.~DE-SC0017987, and by a Humboldt Fellowship from the Alexander von Humboldt Foundation. SG acknowledges the J.C. Bose Fellowship (JCB/2020/000011) from the Anusandhan National Research Foundation, Government of India. The work of DP, SKS and SG is supported by the Department of Space (DoS), Government of India. The computations were performed on the Param Vikram-1000 High Performance Computing Cluster of the Physical Research Laboratory (PRL). SKS acknowledges the hospitality offered by the Centre for High Energy Physics at Indian Institute of Science (IISc), Bengaluru during the completion of this project. 

\appendix

\section{Sub-amplitudes, NMEs and PSFs for \nubb}\label{app:sub-amplitudes} 
In the $SU(5)$ model under consideration, the operators that contribute to the \nubb are ${\cal O}_{S+P}^{S+P},\, {\cal O}_{T+T_5}^{T+T_5}$ and ${\cal O}_{V+A}^{V+A}$ and their corresponding Wilson coefficients are $\epsilon_{S+P}^{S+P},\, \epsilon_{T+T_5}^{T+T_5}$ and $\epsilon_{V+A}^{V+A}$. But we find that the dominant contribution to the amplitude comes from ${\cal O}_{S+P}^{S+P}$ due to an enhancement factor $m_N/m_e$ associated with it, where $m_N$ and $m_e$ are the nucleon and electron masses, respectively.  With this Wilson coefficient, we can write Eq.~(\ref{eq:half-life}) as 
\beqa
\left( T_{1/2}^{0\nu}\right)^{-1} &=& g_A^4 \Big[ G_{01} \, \left| {\cal A}_{\nu}\right|^2  \Big], \label{eq:app_half_life}
\eeqa
where the sub-amplitudes ${\cal A}_\nu$ are defined as~\cite{Cirigliano:2018yza}
\beqa
\cal A_{\nu} &=& \frac{m^{\rm std}_{\rm ee}}{m_e}\, {\cal M}_{\nu}^{(3)} + \frac{m_N}{m_e} {\cal M}_{\nu}^{(6)} . \label{eq:amplitude}
\eeqa
Here, ${\cal M}_{\nu}^{(3)} $ denotes the contribution induced by the light Majorana neutrinos and ${\cal M}_{\nu}^{(6)} $ encapsulates the contributions from other dimension-6 $L$-violating operators present in this model; c.f.~Eqs.~\eqref{eq:S+P} and \eqref{eq:TR}. 
These ${\cal M}_\nu$'s can be expressed in terms of Wilson coefficients and NMEs, as follows: 
\beqa
{\cal M}_{\nu}^{(3)} & =& -V_{ud}^2\left(-\frac{M_F}{g_A^2}+M_{GT} + M_T + \frac{2m_{\pi}^2 g_{\nu}^{NN}}{g_A^2}\,M_{F,sd}  \right)\,, \label{eq:Mnu3}\\
 {\cal M}_{\nu}^{(6)} &=& V_{ud}\left( -\frac{B}{m_N}\, \epsilon_{S+P}^{S+P} \,M_{PS} + 
 \epsilon_{T+T_5}^{T+T_5} \,M_{T6} \right)\, , \label{eq:Mnu6}
\eeqa
where $g_{\nu}^{NN} \sim  {\cal O} (f_\pi^{-2}) = -92.9\, \text{GeV}^{-2} \pm 50\%$~\cite{Cirigliano:2018yza,Scholer:2023bnn}  and $B = 2.7 $ GeV  at $\mu=2$ GeV in the $\overline{\rm MS}$ scheme~\cite{Cirigliano:2018yza} . The NMEs can be calculated via~\cite{Cirigliano:2018yza} 
\beqa
M_{GT} &=&  M_{GT}^{AA} + M_{GT}^{AP} + M_{GT}^{PP} +  M_{GT}^{MM}\, ,\\
    M_{T}  &= & M_{T}^{AP} + M_{T}^{PP} + M_{T}^{MM} \, ,\\
    M_{PS}  &= & \frac{1}{2} M_{GT}^{AP} + M_{GT}^{PP} + \frac{1}{2} M_{T}^{AP} + M_{T}^{PP}\, , \\
    M_{T6}  &=&  2\frac{g_T^\prime-g_T^{NN}}{g_A^2} \frac{m_\pi^2}{m_N^2} M_{F,sd} - 8 \frac{g_T}{g_M} (M_{GT}^{MM}+ M_T^{MM}) \nonumber \\
    & & + g_T^{\pi N} \frac{m_\pi^2}{4m_N^2}(M_{GT,sd}^{AP}+ M_{T,sd}^{AP}) +  g_T^{\pi \pi} \frac{m_\pi^2}{4m_N^2}(M_{GT,sd}^{PP}+ M_{T,sd}^{PP}) .
\eeqa

\begin{table}[!t]
\center
\begin{tabular}{ccc||ccc}
\hline
 \multicolumn{3}{c||}{ $n\rightarrow pe\nu$, $\pi \rightarrow e \nu$ } &  \multicolumn{3}{c}{$\pi \pi \rightarrow e e$} \\
 \hline
 $g_A$ & $1.271\pm 0.002$ & \cite{ParticleDataGroup:2016lqr}      & $g^{\pi\pi}_{1}$   		& $  0.36 \pm 0.02 $             & \cite{Nicholson:2018mwc}  \\
 $g_S$ & $0.97\pm 0.13$ & \cite{Bhattacharya:2016zcn} & $g^{\pi\pi}_{2}$   		& $  2.0  \pm 0.2 $  \, GeV$^2$  & \cite{Nicholson:2018mwc}  \\
 $g_M$ &  $ 4.7$ &  \cite{ParticleDataGroup:2016lqr}              & $g^{\pi\pi}_{3}$ 	        & $ -(0.62 \pm 0.06)$  \, GeV$^2$  & \cite{Nicholson:2018mwc}  \\
 $g_T$ & $0.99\pm 0.06$ & \cite{Bhattacharya:2016zcn} & $g^{\pi\pi}_{4}$   		& $ -(1.9  \pm 0.2)$   \, GeV$^2$  & \cite{Nicholson:2018mwc}\\  
 $|g'_{T}|$ & $\mathcal O(1)$  	&		      & $g^{\pi\pi}_{5}$ 		& $ -(8.0  \pm 0.6)$   \, GeV$^2$  & \cite{Nicholson:2018mwc}  \\
 $B$      &    $2.7$~GeV       &     & $|g^{\pi\pi}_{ \rm T}|$     & $\mathcal O(1)$  & \\ 
 \hline\hline
  \multicolumn{3}{c||}{$n \rightarrow p\pi ee$} & \multicolumn{3}{c}{$nn\rightarrow pp\, ee$}   
 \\
  \hline 
   $|g^{\pi N}_{1} |$       & $\mathcal{O}(1)$ &    & $|g^{N N}_1|$         & $\mathcal{O}(1)$ &  \\  
   $|g^{\pi N}_{6,7,8,9}|$ & $\mathcal{O}(1)$ &    & $|g_{6,7}^{N N}|$ & $\mathcal{O}(1)$ & \\
     $|g^{\pi N}_{\rm VL}|$ & $\mathcal{O}(1)$ &    &$ |g^{NN}_{\rm VL}|$ & $ \mathcal{O}(1) $ &     \\ 
           $|g^{\pi N}_{\rm T}|$ 			   & 		$ \Or(1)$     &    &
        $ |g_{\rm T}^{NN}|$ &$\Or(1)$&   
  \\ 
			   & 		      &    & $ |g^{NN}_{\nu}|$ & $ -92.9\,\mathrm{GeV}^{-2} \pm 50\%$ &   \cite{Cirigliano:2020dmx,Cirigliano:2021qko,Wirth:2021pij}  \\ 
			   &&&$ |g_{VL,VR}^{E,m_e}|$ &$\Or(1)$&\\
			   			   &&&            $ |g^{NN}_{2,3,4,5}|$ & $ \mathcal{O}((4\pi)^2) $ &  
  \\\hline
\end{tabular}
\caption{The values of the LECs used in our numerical calculation using \nudobe. The Table is taken from Refs.~\cite{Cirigliano:2018yza, Scholer:2023bnn}. For $g_\nu^{NN}$, we have varied it within the 50\% uncertainty of its value. Moreover, the unknown LECs are varied within their order of magnitude estimates i.e in the range $\pm \left[1/\sqrt{10} :\sqrt{10} \right]\times {\cal O}(|g_i|)$ while all other LECs are kept constant.
}\label{Tab:LECs}
\end{table}

\begin{table}[!t]
\renewcommand{\arraystretch}{1.3}
\centering
\begin{tabular}{c|cccc|cccc}
\hline
NMEs & \multicolumn{4}{c|}{$^{76}$Ge} & \multicolumn{4}{c}{$^{136}$Xe} \\ 
\cline{2-9}
 & IBM2 & QRPA & SM & CDFT  & IBM2 & QRPA & SM & CDFT \\
 \hline\hline 
 $M_F$ & $-0.78$ & $-1.74$ & $-0.59$ & $-1.924$ & $-0.522$ & $-0.89$ &  $-0.54$ & $-1.184$  \\
 $M_{GT}^{AA}$ & 6.062 & 5.477 & 3.15 & 5.743 & 3.203 & 3.165 & 2.45 & 4.003\\
 $M_{GT}^{AP}$ & $-0.857$ & $-2.016$ & $-0.94$  & $-1.462$  & $-0.452$ & $-1.192$ & $-0.79$ & $-1.059$ \\
 $M_{GT}^{PP}$ & 0.174 & 0.664 & 0.3  &  0.423 & 0.0935 & 0.395 & 0.25 & 0.308  \\
 $M_{GT}^{MM}$ & $0.203$ & $0.51$ & $0.22 $&$ 0.326$ & $0.1094$ & $0.303$ & $0.19$ & $0.24$  \\
 $M_{T}^{AA}$ & $0$ & $0$ & $0$ &$ 0$ & $0$  & $0 $& $0$ & $0$  \\
 $M_{T}^{AP}$ &  $0.238$ & $-0.353$ & $-0.01$  & $0.02$ &$ 0.119$ & $-0.278$ & $0.01$ & $0.053$ \\
 $M_{T}^{PP}$ & $-0.06$ & $0.103$ &$ 0$ & $0.174$  & $-0.0268$ & $0.09$ & $0.01$ & $0.108$   \\
 $M_{T}^{MM}$ & $ 0.04$ & $-0.037$ & $0 $& $0.012$ & $0.018$ & $-0.034$ &$ 0$  & $0.014$ \\
 $M_{F,sd}$ & $ -1.261 $& $-3.599$ & $-1.522$ &  $-2.33$ & $-0.77$ & $-1.594$ & $-1.342$ & $-1.684$  \\
 $M_{GT,sd}^{AA}$ & $4.389$ & $11.564$ & $5.057$ & $7.485$ & $2.43$ & $5.937$ & $4.412$ & $5.65$ \\
 $M_{GT,sd}^{AP}$ & $-1.296$ &$-5.558$ & $-2.348$ &$ -3.270$ & $-0.723$ & $-2.905$ & $-2.064$ & $-2.494$  \\
 $M_{T,sd}^{PP}$ &  $0.307$ & $2.067$ & $0.851$ & $1.143$  & $0.176$ & $1.097$ & $0.774$ & $0.882$ \\
 $M_{T,sd}^{AP}$ &  $-0.805$ & $-0.8849$ & $-0.052$ & $0.15$  & $-0.381$ & $-0.955$  & $0.052$ & $0.232$ \\
 $M_{T,sd}^{PP}$ &  $0.241$  &  $0.328$ & $0.026$ &  $-0.237$ & $0.121$ & $0.372$ & $-0.026$ & $-0.238$ \\
 \hline
\end{tabular}
\caption{Values of different NMEs of $^{76}$Ge and $^{136}{\rm Xe}$ using IBM-2~\cite{Deppisch:2020ztt}, QRPA~\cite{Hyvarinen:2015bda}, Shell Model~\cite{Menendez:2017fdf} and CDFT~\cite{Ding:2024obt}.}
\label{tab:NMEs}
\end{table}
\begin{table}[!t]
\renewcommand{\arraystretch}{1.3}
\centering
\begin{tabular}{c|c|c}
\hline
PSFs & $^{76}$Ge & $^{136}$Xe \\
\hline\hline 
$G_{01}$ & $0.22$ & $1.5$\\
$G_{02}$ & $0.35$ & $3.2$ \\
$G_{03}$ & $0.12$ & $0.86$ \\
$G_{04}$ & $0.19$ & $1.2 $ \\
$G_{06}$ & $ 0.33 $ &  $1.8$ \\
$G_{09}$ & $ 0.48$ &  $2.8$\\ \hline
\end{tabular}
\caption{ Phase space factors of  $^{76}$Ge and $^{136}{\rm Xe}$ are given in units of $10^{-14} {\, \rm yr}^{-1}$~\cite{Cirigliano:2018yza}. }
\label{tab:PSFs}
\end{table}

The values of the LECs ($g_i$'s) are given in Tab.~\ref{Tab:LECs}.
The same values are also employed in the $\tt \nu DoBe$ package~\cite{Scholer:2023bnn} which we use to calculate the half-lives. The values of the NMEs are given in Tab.~\ref{tab:NMEs} and the PSFs are given  in Tab.~\ref{tab:PSFs} for two representative nuclei $^{136}$Xe and $^{76}$Ge, which are used in the experiments being considered here, namely, KamLAND-Zen~\cite{KamLAND-Zen:2024eml} and LEGEND-1000~\cite{LEGEND:2021bnm}.

Typically, the NMEs turn out to be ${\cal O}(1)- {\cal O}(10)$ numbers. Then the half-life in Eq.~\eqref{eq:app_half_life} can be simplified as 
\beqa
\left( T_{1/2}^{0\nu}\right)^{-1} &=& g_A^4 G_{01} \left|{\cal M}_{\nu}^{(3)} \right|^2\,  \Big[ \frac{m^{\rm std}_{\rm ee}}{m_e} + \frac{m_N}{m_e} \, \epsilon_{S+P}^{S+P} \, \frac{{\cal M}_{\nu}^{\prime (6)}}{{\cal M}_{\nu}^{(3)}}  \Big]^2 , \nonumber \\
& = & g_A^4 G_{01} \left|{\cal M}_{\nu}^{(3)} \right|^2\,  \Big[ \frac{m^{\rm std}_{\rm ee} + m_{ee}^{\rm nstd}}{m_e}\Big]^2, \nonumber \\
& =& g_A^4 G_{01} \left|{\cal M}_{\nu}^{(3)} \right|^2\,  \Big[ \frac{m^{\rm eff}_{\rm ee}}{m_e}\Big]^2, \label{eq:app_meff}
\eeqa
 where ${\cal M}_{\nu}^{(3)} = -4.832~(-8.635)$,  ${\cal M}_{\nu}^{\prime (6)} = -V_{ud}\,\left(\frac{B}{m_N}\,  \,M_{PS} + \frac 14 \,M_{T6}\right) = 0.327~(0.627)$ for $^{136} \rm Xe$ $(^{76}\rm Ge)$ nucleus and $m_{ee}^{\rm nstd}$ is defined as, $m_{ee}^{\rm nstd} = {m_N} \, \epsilon_{S+P}^{S+P} \, \left( {\cal M}_{\nu}^{\prime (6)}/ {\cal M}_{\nu}^{(3)}\right)$. It should be noted here that the standard and non-standard contributions can  interfere either constructively or destructively depending on the sign of $\epsilon_{S+P}^{S+P}$. For example, when $\epsilon_{S+P}^{S+P}>0$, $m_{ee}^{\rm nstd} < 0 $ because $\left( {\cal M}_{\nu}^{\prime (6)}/ {\cal M}_{\nu}^{(3)}\right) = -0.068~(-0.073)$ for $^{136} \rm Xe$ $(^{76}\rm Ge)$. This leads to destructive inference between the standard and non-standard contributions, as in the case of BP2 in Fig.~\ref{fig:onbb_enhanced}. Conversely, when $\epsilon_{S+P}^{S+P}<0 $, they interfere constructively, as for BP1 in Fig.~\ref{fig:onbb_enhanced}.

\section{\nubb process in the canonical $SU(5)$}\label{app:nubbSU5}
 \begin{figure}[t!]
    \centering
    \includegraphics[scale=0.45]{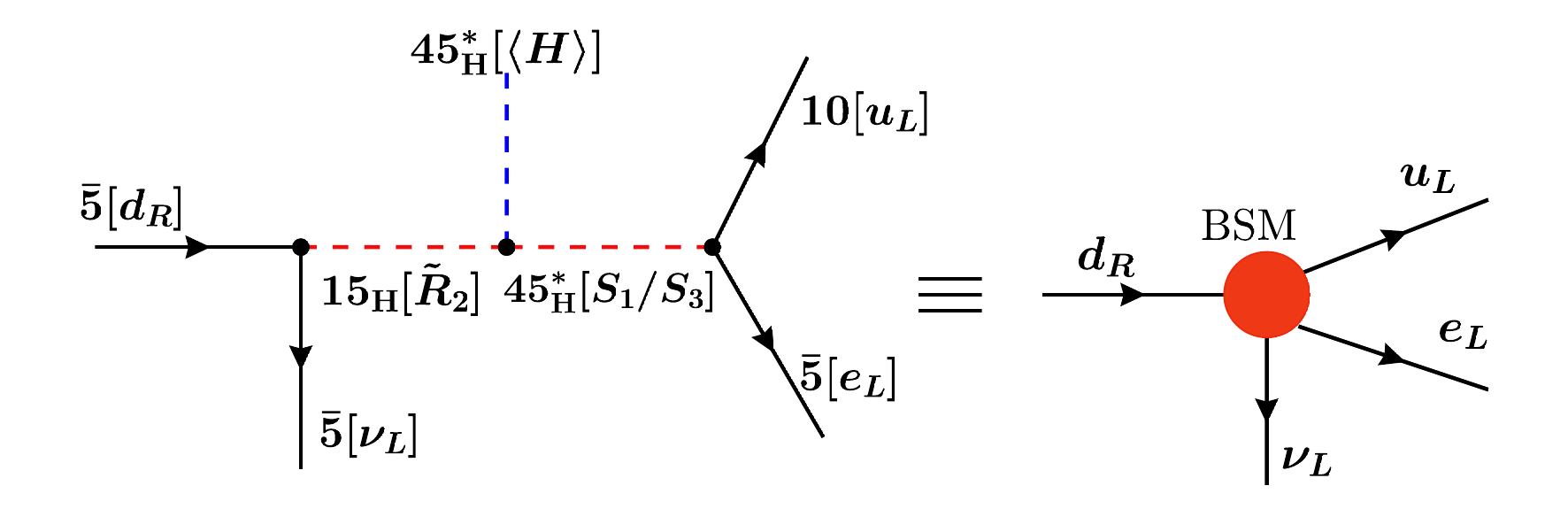}    \caption{\textbf{Left:}  Diagram contributing to \nubb non-standard effective operator in the $SU(5)$ model. \textbf{Right:} The LEFT operator corresponding to the left panel diagram. In $SU(5)$ model leptoquarks from $\mathbf{15_\hh}$ and $\mathbf{45_\hh}$ mix and generate a dimension-6 operator that contribute to $0\nu\beta\beta$. }
    \label{fig:onbb_SU5}
    \end{figure}
In the minimal $SU(5)$ scenario (without the ${\cal Z}_3$ charge assignments discussed in the main text), apart from the canonical long range mechanism shown in Fig.~\ref{fig:SM_operator}, additional non-standard contributions arise through the mixing between different leptoquarks from $\mathbf{45}_\hh$ and $\mathbf{15}_\hh$. Specially, the mixing of $Q_{\rm em} =1/3$ components of $S_{1'} /S_{3}~ (\mathbf{45}_\hh)$ and $\tilde{R}_2~(\mathbf{15}_\hh)$ through the $\mathbf{45}_\hh$ generates a dimension-six operator ${\cal O}^{S+P}_{S+P} = G_F\,\left( \bar u_L d_R \right) \left( \bar e_L \nu_L^c \right)$ which can contribute to \nubb process, as shown in Fig.~\ref{fig:onbb_SU5}. An analogous  operator can be formed through mixing of the $Q_{\rm em}=2/3$ components of $S_3 $ and $\tilde{R}_2$ leptoquarks. The Wilson coefficient of the operators can be written as 
\beqa
\epsilon_{S+P}^{S+P} & =& \frac{\eta \, \left(\sqrt{2} \,Y_{45}\right)_{11}\, \left(\sqrt{2}\,Y_{15}\right)_{11}\, v^3}{M_{S_3}^2 \, M_{\tilde{R}_2}^2} \, . \label{eq:SU5_amplitude}
\eeqa   

Following Eqs.~\eqref{eq:amplitude} and \eqref{eq:Mnu6}, the amplitude of the leptoquark mediated diagram is 
\beqa
{\cal A}_{\rm LQ} & = & V_{ud}\left( \frac{m_N}{m_e}\right)  \epsilon_{S+P}^{S+P} \, M_{PS} \, . \label{eq:ALQ_amplitude}
\eeqa
However, as shown in Eq.~\eqref{eq:nubb_tree}, this contribution is negligible, once the proton decay constraints are imposed. 

\section{Loop integration factors}
\label{app:LF}
The loop integration factors appearing in Eqs.~\eqref{eq:deltaf}, \eqref{eq:kfs} and \eqref{eq:SU5NM}  are given by
\beqa \label{eq:LF_f}
f[M_1^2,M_2^2] &=& -\frac{1}{16 \pi^2} \left(\frac{M_1^2 \log \frac{M_1^2}{\mu^2}-M_2^2\log \frac{M_2^2}{\mu^2}}{M_1^2-M_2^2} - 1 \right)\,, \\ \label{eq:LF_h}
h[M_1^2,M_2^2] &=& \frac{1}{16 \pi^2} \left(\frac{1}{2} \log\frac{M_1^2}{\mu^2} +  \frac{\frac{1}{2} q^2 \log q -\frac{3}{4}q^2 + q - \frac{1}{4}}{(1-q)^2}\right)\,, \\
\label{eq:LF_g}
g[M_1^2,M_2^2] & = & \frac{1}{16 \pi^2} \frac{\frac{q^3}{6}-q^2+\frac{q}{2}+ q \log q + \frac{1}{3}}{(1 -q)^3}\,,\eeqa
where $q=M_2^2/M_1^2$ in the last two equations, and 
\beqa{\label{eq:LF_i}}
p\left(M_1^2,M_2^2\right) \eq \frac{1}{16\pi^2}\,\frac{1}{M_1^2 - M_2^2} \log\left(\frac{M_1^2}{M_2^2}\right) \, .
\eeqa

\section{Charged Lepton flavor Violation}\label{app:clfvs}
 
 As the SU(5) framework naturally inherits leptoquarks, it can significantly contribute to the charged lepton flavor violating (cLFV) processes. For TeV scale leptoquarks these processes may pose severe constraints. In the main text, we mentioned that $R_2, S_3 \in \mathbf{45_\hh}$ and $\tilde{R}_2 \in \mathbf{15_\hh}$ are considered to be light. However, $R_2 $ and $S_3$ can not mediate cLFV as $Y_{45}$ is chosen to be diagonal\footnote{It is to be noted that the non-diagonal Yukawa couplings can be non-zero at low energies due to the RG effects. We have checked that even in those cases the cLFV contributions of these leptoquarks are negligible.}. Therefore, only the contribution stemming from $\tilde{R}_2$ leptoquark needs to considered. For our analysis, we only considered $\mu \to e $ conversion in  nuclei\footnote{One could have considered  $\mu \to e \gamma$ also but in our scenario this decay process is loop suppressed.} i.e. $\mu\, N \rightarrow \, e\, N$. 
This process occurs at tree level, therefore provides the most stringent constraint among the flavor observables on the Yukawa couplings involving first and second generation leptons~\cite{Fajfer:2024uut, Dev:2024tto}. The conversion ratio  is denoted as 
\beqa
\left.\mathcal{R} \right|_{\mu \rightarrow \, e}^{N} &=& \frac{\Gamma^N \left(\mu \rightarrow \, e \right)}{\Gamma^N_{\rm capture}} \, , \label{eq:mutoeconversion}
 \eeqa
  where $N$ denotes a particular nucleus and $\Gamma^N_{\rm capture}$ implies the muon capture rate of that nucleus. 
  The low energy effective Lagrangian describing $\mu \, N \rightarrow \, e\, N$ can be written as~\cite{Kitano:2002mt}
  \beqa
  \mathcal{L}_{\bar{q}q\bar{e}\mu} &\supset & - \sqrt{2} G_F \sum_{q=u,d,s} \sum_{X,Y=L,R} \left[  C_{V_{XY}}^{q} \left(\bar{e} \, \gamma^{\mu} \, P_X \,  \mu \right)\, \left(\bar{q}  \, \gamma_{\mu} \, \, P_Y q\right) \, + C_{S_{XY}}^{q} \left(\bar{e} \, P_X \,  \mu \right)\, \left(\bar{q}  \, P_Y q\right) \, \right. \nonumber \\
  &  &  +  \left. C_{T_{XY}}^{q} \left(\bar{e} \, \sigma^{\mu\nu} \, P_X \,  \mu \right)\, \left(\bar{q}  \, \sigma_{\mu\nu} \, \, P_Y q\right) \, \right]  \hc \, .
  \eeqa
\begin{figure}[t!]
    \centering
    \includegraphics[width=0.5\linewidth]{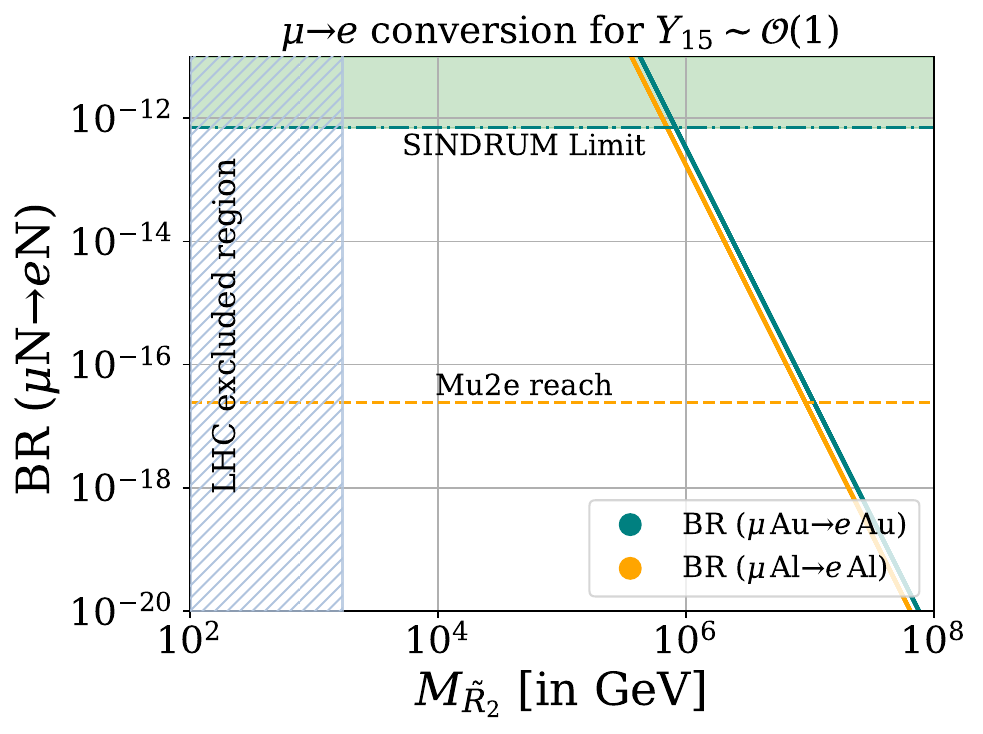}
    \caption{Variation of the branching fraction for $\mu \to e$ conversion in nuclei as a function of the $\tilde{R}_2$ leptoquark mass.
The red dashed horizontal line indicates the current upper limit from the SINDRUM experiment.
The vertical shaded region represents the exclusion bounds on the leptoquark mass from LHC searches.  }
    \label{fig:LFV}
\end{figure}
Here, the scalar operators and operators with the gluon field strength tensor are neglected because they are suppressed in this model~\cite{Plakias:2023esq}. The relevant Wilson coefficients in our framework (only $\tilde{R}_2$ contribution) are given by  
  \beqa
  C_{V_{LR}}^d &=& \frac{1}{2\sqrt{2}G_F} \frac{\left(Y_{15} \right)_{12} \left(Y_{15} \right)_{11}^{\ast}}{m^2_{\tilde{R}_2^{2/3}} }  \, . \label{eq:cVLR}
  \eeqa
For the coherent conversion process in which the initial and final nucleus are the same, the total conversion rate can be written as ~\cite{Kitano:2002mt} 
  \beqa
 \left. \mathcal{R}\right|_{\mu \rightarrow \, e}^{N} &=& \frac{2\, G_F^2 \, m_{\mu}^5}{\Gamma_{\rm capture}^N \, } \, \left|C_{VL}^{p} \, V_p + C_{VL}^n \, V_n  \, \right|^2 , \label{eq:mueconversion}
\eeqa
where the $V_p$ and $V_n$ are called the overlap integrals whose values along with $\Gamma_{\rm{capture}}^N$ are given in Tab.~\ref{tab:conv} for two nuclei. Here the coefficients are defined as:
  \beqa
  C_{VL}^p &=&  \frac 12 C_{V_{LR}}^d \, , \qquad  
   C_{VL}^n =  C_{V_{LR}}^d \, . 
  \eeqa
  Using these values, Eq.~\eqref{eq:mueconversion} can be approximated as
\beqa
\left. \mathcal{R}\right|_{\mu \rightarrow \, e}^{N} &\approx & \frac{2\, G_F^2 \, m_{\mu}^5}{\Gamma_{\rm capture}^N \, } \, V_p^2 \, \frac 94 \, \left|C_{V_{LR}}^d\right|^2 \, \nonumber \\
& \simeq & 5.6 \times 10^{-13}\, \left(\frac{\left(Y_{15} \right)_{12}}{\mathcal{O}\left( 1\right)}\right)^2\, \left(\frac{\left(Y_{15} \right)_{11}^{\ast}}{\mathcal{O}\left( 1\right)}\right)^2 \, \left(\frac{ 10^6\, \rm GeV}{m_{\tilde{R}_2}}\right)^4,
\eeqa
where $V_p \approx V_n$ is assumed and also can be seen in Tab.~\ref{tab:conv}. 

Currently, the most stringent limit on $\mu \rightarrow \,e $ conversion comes from the SINDRUM experiment using $^{197}$Au nucleus: $\left.\mathcal{R} \right|_{\mu \rightarrow \, e}^{\rm Au} \, < 7.0 \times 10^{-13}$~\cite{SINDRUMII:2006dvw}. In the future, the Mu2e experiment at Fermilab using $^{27}$Al aims at improving the sensitivity down to $2.4\times 10^{-17}$~\cite{Mu2e:2014fns}. Similarly, the COMET experiment at J-PARC is projected to reach a sensitivity of $3.1\times 10^{-15}$ in Phase-I~\cite{COMET:2018auw}.
As shown in Fig.~\ref{fig:LFV}, the current constraint on $\mu\to e$ conversion requires the $\tilde{R}_2$ leptoquark mass to be above $8.4 \times 10^5$ GeV, while the future Mu2e reach can probe it up to $9.3 \times 10^6$ GeV for ${\cal O}$(1) Yukawa couplings. In our analysis, we choose $m_{\tilde{R}_2} = 10^6$ GeV for the fermion mass fittings in order to be consistent with the cLFV constraint.   
\begin{table*}[t!]
	\begin{center} 
		\begin{math} 
			\begin{tabular}{cccc}
				\hline
 $N$ & $\Gamma^N_{\rm capture}$ (MeV) & $V_p$ & $V_n$ \\
 \hline
 \hline
 $^{197}$Au & $8.7\times 10^{-15}$ & 0.08059 & 0.108 \\
 $^{27}$Al & $4.6\times 10^{-16}$ & 0.0159 & 0.0169 \\ 
\hline
\end{tabular}
\end{math}
	\end{center}
\caption{Muon capture rates~\cite{Suzuki:1987jf} and overlap integral values~\cite{Kitano:2002mt} relevant for $\mu\to e$ conversion in $^{197}$Au and $^{27}$Al.}
\label{tab:conv}
\end{table*}

 \section{Benchmark Solutions} 
\label{app:bestfit}
We consider two benchmark points (BP1 and BP2) from the solution space fitting the fermion mass spectra. The Yukawa couplings for the benchmark points are given as follows\footnote{The Yukawa entries have been specified with very high precision. Any deviation in these Yukawa values or the heavier scalar masses given in Tab.~\ref{tab:sol} would significantly alter the $\chi^2$ shown in Fig.~\ref{fig:chisquare&Yukawa}.}:
\beqa\label{eq:fittedy1y2}
\text{BP1:} & & \nonumber \\
Y_5 \eq \begin{footnotesize} \begin{pmatrix}
    \left(8.09002 + i\,\, 3.26336\right) \times 10^{-5} &  \left( -1.9705 + i\,\, 0.667620 \right)\times 10^{-4} &
   \left( -3.80612 - i\,\, 1.91173\right) \times 10^{-3}\\
  \left( -1.9705 + i\,\, 0.667620 \right)\times 10^{-4} & \left( 3.05963 + i\,\,-0.02150 \right) \times 10^{-3} &
   \left( -2.83391 - i\,\,0.84882 \right) \times 10^{-2}\\
  \left( -3.80612 - i\,\, 1.91173\right) \times 10^{-3} & \left( -2.83391 - i\,\,0.84882 \right) \times 10^{-2} &
  \left( 7.22552 +  i\,\,1.89986 \right)\times 10^{-1} 
\end{pmatrix}\,\end{footnotesize} ,\nl
Y_{45} \eq \begin{pmatrix}
   1.64251 \times 10^{-2} & 0 & 0 \\
   0 & -9.38426\times 10^{-5} & 0 \\
   0 & 0 &  -3.78806\times10^{-1} 
\end{pmatrix}\,, \text{and}\nl
Y_{15} \eq \begin{pmatrix}
    -1.41343 &   0.00564 &  0.46266\\
     0.00564  & 3.49915 &   -0.00987 \\
    0.46266  &  -0.00987 & 1.30712
    \end{pmatrix}\, ,\\
\text{BP2:} & & \nonumber \\
Y_5 \eq \begin{footnotesize} \begin{pmatrix}
    \left(-0.160022 + i\,\, 1.78568\right) \times 10^{-4} &  \left( 2.37783 - i\,\, 2.28594 \right)\times 10^{-4} &
   \left( 1.07006 - i\,\, 6.42993\right) \times 10^{-3}\\
  \left( 2.37783 - i\,\, 2.28594 \right)\times 10^{-4} & \left( -1.7098 + i\,\, 1.34955 \right) \times 10^{-3} &
   \left( -1.35222 - i\,\,2.13648 \right) \times 10^{-2}\\
  \left( 1.07006 - i\,\, 6.42993\right) \times 10^{-3} & \left( -1.35222 - i\,\,2.13648 \right) \times 10^{-2} &
  \left( 0.28407 +  i\,\,7.72693 \right)\times 10^{-1} 
\end{pmatrix}\,\end{footnotesize} ,\nl
Y_{45} \eq \begin{pmatrix}
   -1.34103 \times 10^{-2} & 0 & 0 \\
   0 & 1.08804\times 10^{-4} & 0 \\
   0 & 0 &  3.87003\times10^{-1} 
\end{pmatrix}\,, \text{and}\nl
Y_{15} \eq \begin{pmatrix}
    -1.39680 &   0.001433 &  -0.51452\\
    0.001433  & 3.42342 &   0.00113 \\
    -0.51452  & 0.00113 & 1.29648
    \end{pmatrix}\,.    
\eeqa

\begin{table*}[t!]
	\begin{center} 
		\begin{math} 
			\begin{tabular}{cccc}
				\hline
				\hline 
      \multirow{2}{*}{~~Observable~~} & \multirow{2}{*}{ \hspace*{0.3cm}  $O_{\rm exp}$ \hspace*{0.3cm} } & 
      \multicolumn{1}{c}{\bf \hspace*{0.3cm}  BP1  \hspace*{0.3cm} } & 
      \multicolumn{1}{c}{\bf \hspace*{0.3cm}  BP2  \hspace*{0.3cm} } \\
	   &   & $O_{\rm th}$ & $O_{\rm th}$ \\
	\hline
	$y_u/10^{-6}$  & $4.087 $ & $4.086 $ & $4.083$ \\
	$y_c/10^{-3}$   & $2.060$ & $ 2.060$ & $2.061$ \\
	$y_t$  & $0.6580  $ & $0.6845$ & $0.6826$ \\
	$y_d/10^{-6}$  & $ 8.777  $ & $8.776$ & $8.789$ \\
	$y_s/10^{-4}$   & $1.792 $ & $1.791$ & $1.800$ \\
	$y_b/10^{-3}$   & $9.992 $ & $9.9921$ & $9.513$ \\
	$y_e/10^{-6}$   & $3.921 $ & $3.921$ & $3.9179$ \\
	$y_{\mu}/10^{-4}$  &$8.172 $ & $8.174$ & $8.0957$ \\
	$y_{\tau}/10^{-3}$   & $13.82 $ & $13.78$ & $14.535$ \\
	$|V_{us}|$ & $0.2286$ & $0.2286$ & $0.2289$ \\
	$|V_{cb}|$ & $0.03794$ & $0.03791$ & $0.03798$ \\
	$|V_{ub}|$ & $0.003518$ & $0.003518$ & $0.003516$ \\
	$\sin\delta_{\rm CKM}$ & $0.78$ & $0.78$ & $0.78$ \\
	$\Delta m^2_{\text{sol}}~ [{\rm eV}^2]/10^{-5}$ & $7.49$& $7.49$ & $7.49$ \\
	$\Delta m^2_{\text{atm}}~ [{\rm eV}^2]/10^{-3}$ & $2.534$& $2.532$ & $2.532$ \\
    $\sin^2 \theta _{12}$ & $0.307$ & $0.307$ & $0.2940$ \\
	$\sin^2 \theta _{23}$ & $0.561$ & $0.5602$ & $0.5685$ \\
	$\sin^2 \theta _{13}$ & $0.02195$ & $0.02194$ & $0.0229$ \\
    \hline
     $\chi^2_{\rm min}$    & &  $  10^{-3} $ &  $  0.7$ \\
    \hline
    $M_{S_1}$ [GeV]& & $ 5.45405 \times 10^{15}$  & $ 4.39036 \times 10^{16}$\\
    $M_{S^{'}_1}$ [GeV]& & $8.55261 \times 10^{15}$ & $ 9.30636 \times 10^{16}$\\
    $M_{\tilde{S}}$ [GeV]& & $  1.75281 \times 10^{15}$ & $ 2.28598 \times 10^{15}$\\
    $M_{{\mathbb{S}}}$ [GeV] & & $ 2.64956 \times 10^{13}$ & $  10^{3}$ \\
    $M_{O}$ [GeV] & & $  5.98515 \times 10^{15}$ & $ 10^{15}$ \\
    $M_{\Sigma}$ [GeV] & & $6.62310 \times 10^{10} $ & $ 3.12650 \times 10^{10}$ \\
    $M_{\Delta}$ [GeV] & & $ 2.70915 \times 10^{16}$ & $ 2.69230 \times 10^{16}$ \\
    $\eta$ [GeV] & & $ 9.54343 \times 10^{16}$ & $ 9.68322 \times 10^{16}$ \\
    \hline
    $M_{R_2}$ [GeV] & & $ 2500$ & $ 2500$ \\
    $M_{S_3}$ [GeV] & & $ 2004$ & $ 2004$ \\
    $M_{\tilde{R}_2}$ [GeV] & & $ 10^{6}$ & $ 10^{6}$ \\
	\hline \hline
			\end{tabular}
		\end{math}
	\end{center}
	\caption{The best-fit values of the predicted theoretical observables corresponding to BP1 and BP2, and the minimum $\chi^2$ values are given. The definition of $\chi^2$ includes charged and neutral fermion sector observables. The extrapolated values of the experimental observables at the scale of $\mu=10^{16}$ GeV are provided along with the values reproduced through $\chi^2$ minimization. The fitted values of the masses of various scalars and the cubic coupling $\eta$ are also given, where the masses of $R_2$ and $S_3$ are fixed close to 2 TeV (just above the LHC constraint) and $\tilde{R}_{2}$ is $10^{6}$ GeV (to satisfy cLFV).} 
	\label{tab:sol} 
\end{table*}

The fitted values of different observables corresponding to the above Yukawa matrices are shown in Tab.~\ref{tab:sol}. The obtained values of the optimized $\chi^2$ for the given benchmark points are $10^{-3}$ and $0.7$ respectively, indicating that the considered $SU(5)$ model appropriately reproduces the observed charged and neutral fermion mass spectra. In addition to the fitted Yukawa coupling, the absolute values of obtained neutrino masses and $m_{ee}^{\rm std}$ are also provided;\\
\beqa\label{eq:NMprediction}
\text{BP1:}\,\, {\rm Diag}(M_{\nu}) \eq \begin{pmatrix}
    0.0021666 & 0 & 0\\
    0 & 0.0023311 & 0\\
    0 & 0 & 0.0547934\\
\end{pmatrix}~{\rm eV}\,,\quad m_{ee}^{\rm std} = 0.0229\,\rm eV ,\\
\text{BP2:}\,{\rm Diag}\, (M_{\nu}) \eq
\begin{pmatrix}
    0.0022384 & 0 & 0\\
    0 & 0.002400 & 0\\
    0 & 0 & 0.0547934\\
\end{pmatrix}~{\rm eV}\,, \quad m_{ee}^{\rm std} = 0.0236\, \rm eV .
\eeqa 
\newpage

\bibliographystyle{JHEP}
\bibliography{SU5}

@article{Buccella:1984ft,
    author = "Buccella, F. and Cocco, L. and Wetterich, C.",
    title = "{An SO(10) Model With 54 + 126 + 10 Higgs}",
    reportNumber = "Print-84-0510 (NAPLES)",
    doi = "10.1016/0550-3213(84)90029-4",
    journal = "Nucl. Phys. B",
    volume = "243",
    pages = "273--284",
    year = "1984"
}

@article{Super-Kamiokande:2013rwg,
    author = "Abe, K. and others",
    collaboration = "Super-Kamiokande",
    title = "{Search for Nucleon Decay via $n \to \bar{\nu} \pi^{0}$ and $p \to \bar{\nu} \pi^{+}$ in Super-Kamiokande}",
    eprint = "1305.4391",
    archivePrefix = "arXiv",
    primaryClass = "hep-ex",
    doi = "10.1103/PhysRevLett.113.121802",
    journal = "Phys. Rev. Lett.",
    volume = "113",
    number = "12",
    pages = "121802",
    year = "2014"
}

@article{Fritzsch:1974nn,
    author = "Fritzsch, Harald and Minkowski, Peter",
    title = "{Unified Interactions of Leptons and Hadrons}",
    reportNumber = "CALT-68-467",
    doi = "10.1016/0003-4916(75)90211-0",
    journal = "Annals Phys.",
    volume = "93",
    pages = "193--266",
    year = "1975"
}

@article{Georgi:1974sy,
    author = "Georgi, H. and Glashow, S. L.",
    title = "{Unity of All Elementary Particle Forces}",
    doi = "10.1103/PhysRevLett.32.438",
    journal = "Phys. Rev. Lett.",
    volume = "32",
    pages = "438--441",
    year = "1974"
}

@article{Langacker:1980js,
    author = "Langacker, Paul",
    title = "{Grand Unified Theories and Proton Decay}",
    reportNumber = "SLAC-PUB-2544",
    doi = "10.1016/0370-1573(81)90059-4",
    journal = "Phys. Rept.",
    volume = "72",
    pages = "185",
    year = "1981"
}

@article{Dorsner:2016wpm,
    author = "Dor\v{s}ner, I. and Fajfer, S. and Greljo, A. and Kamenik, J. F. and Ko\v{s}nik, N.",
    title = "{Physics of leptoquarks in precision experiments and at particle colliders}",
    eprint = "1603.04993",
    archivePrefix = "arXiv",
    primaryClass = "hep-ph",
    doi = "10.1016/j.physrep.2016.06.001",
    journal = "Phys. Rept.",
    volume = "641",
    pages = "1--68",
    year = "2016"
}

@article{Nath:2006ut,
    author = "Nath, Pran and Fileviez Perez, Pavel",
    title = "{Proton stability in grand unified theories, in strings and in branes}",
    eprint = "hep-ph/0601023",
    archivePrefix = "arXiv",
    doi = "10.1016/j.physrep.2007.02.010",
    journal = "Phys. Rept.",
    volume = "441",
    pages = "191--317",
    year = "2007"
}

@article{Dueck:2013gca,
    author = "Dueck, Alexander and Rodejohann, Werner",
    title = "{Fits to SO(10) Grand Unified Models}",
    eprint = "1306.4468",
    archivePrefix = "arXiv",
    primaryClass = "hep-ph",
    doi = "10.1007/JHEP09(2013)024",
    journal = "JHEP",
    volume = "09",
    pages = "024",
    year = "2013"
}

@article{Babu:2016bmy,
    author = "Babu, K. S. and Bajc, Borut and Saad, Shaikh",
    title = "{Yukawa Sector of Minimal SO(10) Unification}",
    eprint = "1612.04329",
    archivePrefix = "arXiv",
    primaryClass = "hep-ph",
    reportNumber = "OSU-HEP-16-08",
    doi = "10.1007/JHEP02(2017)136",
    journal = "JHEP",
    volume = "02",
    pages = "136",
    year = "2017"
}

@article{Slansky:1981yr,
    author = "Slansky, R.",
    title = "{Group Theory for Unified Model Building}",
    reportNumber = "LA-UR-80-3495",
    doi = "10.1016/0370-1573(81)90092-2",
    journal = "Phys. Rept.",
    volume = "79",
    pages = "1--128",
    year = "1981"
}

@article{Patel:2022wya,
    author = "Patel, Ketan M. and Shukla, Saurabh K.",
    title = "{Anatomy of scalar mediated proton decays in SO(10) models}",
    eprint = "2203.07748",
    archivePrefix = "arXiv",
    primaryClass = "hep-ph",
    doi = "10.1007/JHEP08(2022)042",
    journal = "JHEP",
    volume = "08",
    pages = "042",
    year = "2022"
}

@article{Schechter:1980gr,
    author = "Schechter, J. and Valle, J. W. F.",
    title = "{Neutrino Masses in SU(2) x U(1) Theories}",
    reportNumber = "SU-4217-167, COO-3533-167",
    doi = "10.1103/PhysRevD.22.2227",
    journal = "Phys. Rev. D",
    volume = "22",
    pages = "2227",
    year = "1980"
}

@article{Wright:1994qb,
    author = "Wright, Brian D.",
    title = "{Yukawa coupling thresholds: Application to the MSSM and the minimal supersymmetric SU(5) GUT}",
    eprint = "hep-ph/9404217",
    archivePrefix = "arXiv",
    reportNumber = "MAD-PH-812",
    month = "3",
    year = "1994"
}

@article{Georgi:1979df,
    author = "Georgi, Howard and Jarlskog, C.",
    title = "{A New Lepton - Quark Mass Relation in a Unified Theory}",
    reportNumber = "HUTP-79-A026",
    doi = "10.1016/0370-2693(79)90842-6",
    journal = "Phys. Lett. B",
    volume = "86",
    pages = "297--300",
    year = "1979"
}

@article{Ellis:1979fg,
    author = "Ellis, John R. and Gaillard, Mary K.",
    title = "{Fermion Masses and Higgs Representations in SU(5)}",
    reportNumber = "CERN-TH-2746, LAPP-TH-05",
    doi = "10.1016/0370-2693(79)90476-3",
    journal = "Phys. Lett. B",
    volume = "88",
    pages = "315--319",
    year = "1979"
}

@article{Bajc:2006ia,
    author = "Bajc, Borut and Senjanovic, Goran",
    title = "{Seesaw at LHC}",
    eprint = "hep-ph/0612029",
    archivePrefix = "arXiv",
    doi = "10.1088/1126-6708/2007/08/014",
    journal = "JHEP",
    volume = "08",
    pages = "014",
    year = "2007"
}

@article{Helo:2015ffa,
    author = "Helo, J. C. and Hirsch, M.",
    title = "{LHC dijet constraints on double beta decay}",
    eprint = "1509.00423",
    archivePrefix = "arXiv",
    primaryClass = "hep-ph",
    reportNumber = "IFIC-15-51",
    doi = "10.1103/PhysRevD.92.073017",
    journal = "Phys. Rev. D",
    volume = "92",
    number = "7",
    pages = "073017",
    year = "2015"
}

@article{Sartore:2020gou,
    author = "Sartore, Lohan and Schienbein, Ingo",
    title = "{PyR@TE 3}",
    eprint = "2007.12700",
    archivePrefix = "arXiv",
    primaryClass = "hep-ph",
    doi = "10.1016/j.cpc.2020.107819",
    journal = "Comput. Phys. Commun.",
    volume = "261",
    pages = "107819",
    year = "2021"
}

@article{Kane:1993td,
    author = "Kane, Gordon L. and Kolda, Christopher F. and Roszkowski, Leszek and Wells, James D.",
    title = "{Study of constrained minimal supersymmetry}",
    eprint = "hep-ph/9312272",
    archivePrefix = "arXiv",
    reportNumber = "UM-TH-93-24",
    doi = "10.1103/PhysRevD.49.6173",
    journal = "Phys. Rev. D",
    volume = "49",
    pages = "6173--6210",
    year = "1994"
}

@article{Oliensis:1982sd,
    author = "Oliensis, J. and Fischler, M.",
    title = "{Two Loop Calculations of $M_b/M_{\tau}$ and Heavy Fermion Masses in the SU(5) Model}",
    reportNumber = "FERMILAB-PUB-82-063-THY, FERMILAB-PUB-82-063-T",
    doi = "10.1103/PhysRevD.28.194",
    journal = "Phys. Rev. D",
    volume = "28",
    pages = "194",
    year = "1983"
}

@article{Allwicher:2021rtd,
    author = "Allwicher, Lukas and Arnan, Pere and Barducci, Daniele and Nardecchia, Marco",
    title = "{Perturbative unitarity constraints on generic Yukawa interactions}",
    eprint = "2108.00013",
    archivePrefix = "arXiv",
    primaryClass = "hep-ph",
    doi = "10.1007/JHEP10(2021)129",
    journal = "JHEP",
    volume = "10",
    pages = "129",
    year = "2021"
}

@article{Chankowski:1993tx,
    author = "Chankowski, Piotr H. and Pluciennik, Zbigniew",
    title = "{Renormalization group equations for seesaw neutrino masses}",
    eprint = "hep-ph/9306333",
    archivePrefix = "arXiv",
    reportNumber = "ZU-TH-20-93, DFPD-93-TH-44",
    doi = "10.1016/0370-2693(93)90330-K",
    journal = "Phys. Lett. B",
    volume = "316",
    pages = "312--317",
    year = "1993"
}

@article{Babu:1993qv,
    author = "Babu, K. S. and Leung, Chung Ngoc and Pantaleone, James T.",
    title = "{Renormalization of the neutrino mass operator}",
    eprint = "hep-ph/9309223",
    archivePrefix = "arXiv",
    reportNumber = "IUHET-252, UDHEP-93-03, BA-93-44",
    doi = "10.1016/0370-2693(93)90801-N",
    journal = "Phys. Lett. B",
    volume = "319",
    pages = "191--198",
    year = "1993"
}

@article{Mei:2005qp,
    author = "Mei, Jian-wei",
    title = "{Running neutrino masses, leptonic mixing angles and CP-violating phases: From M(Z) to Lambda(GUT)}",
    eprint = "hep-ph/0502015",
    archivePrefix = "arXiv",
    doi = "10.1103/PhysRevD.71.073012",
    journal = "Phys. Rev. D",
    volume = "71",
    pages = "073012",
    year = "2005"
}

@article{Antusch:2005gp,
    author = {Antusch, Stefan and Kersten, J\"orn and Lindner, Manfred and Ratz, Michael and Schmidt, Michael Andreas},
    title = "{Running neutrino mass parameters in see-saw scenarios}",
    eprint = "hep-ph/0501272",
    archivePrefix = "arXiv",
    reportNumber = "DESY-05-013, TUM-HEP-576-05, SHEP-0504",
    doi = "10.1088/1126-6708/2005/03/024",
    journal = "JHEP",
    volume = "03",
    pages = "024",
    year = "2005"
}

@article{Weinberg:1980wa,
    author = "Weinberg, Steven",
    title = "{Effective Gauge Theories}",
    reportNumber = "HUTP-80/A001",
    doi = "10.1016/0370-2693(80)90660-7",
    journal = "Phys. Lett. B",
    volume = "91",
    pages = "51--55",
    year = "1980"
}

@article{Hall:1980kf,
    author = "Hall, Lawrence J.",
    title = "{Grand Unification of Effective Gauge Theories}",
    reportNumber = "HUTP-80/A024",
    doi = "10.1016/0550-3213(81)90498-3",
    journal = "Nucl. Phys. B",
    volume = "178",
    pages = "75--124",
    year = "1981"
}

@article{Hempfling:1993kv,
    author = "Hempfling, Ralf",
    title = "{Yukawa coupling unification with supersymmetric threshold corrections}",
    reportNumber = "DESY-93-092",
    doi = "10.1103/PhysRevD.49.6168",
    journal = "Phys. Rev. D",
    volume = "49",
    pages = "6168--6172",
    year = "1994"
}

@article{Dorsner:2012nq,
    author = "Dorsner, Ilja and Fajfer, Svjetlana and Kosnik, Nejc",
    title = "{Heavy and light scalar leptoquarks in proton decay}",
    eprint = "1204.0674",
    archivePrefix = "arXiv",
    primaryClass = "hep-ph",
    reportNumber = "LAL-12-111",
    doi = "10.1103/PhysRevD.86.015013",
    journal = "Phys. Rev. D",
    volume = "86",
    pages = "015013",
    year = "2012"
}

@article{delAguila:1980qag,
    author = "del Aguila, F. and Ibanez, Luis E.",
    title = "{Higgs Bosons in SO(10) and Partial Unification}",
    reportNumber = "OXFORD-TP 41/80",
    doi = "10.1016/0550-3213(81)90266-2",
    journal = "Nucl. Phys. B",
    volume = "177",
    pages = "60--86",
    year = "1981"
}

@article{DESI:2025zgx,
    author = "Abdul Karim, M. and others",
    collaboration = "DESI",
    title = "{DESI DR2 Results II: Measurements of Baryon Acoustic Oscillations and Cosmological Constraints}",
    eprint = "2503.14738",
    archivePrefix = "arXiv",
    primaryClass = "astro-ph.CO",
    reportNumber = "FERMILAB-PUB-25-0169-PPD",
    month = "3",
    year = "2025"
}

@article{Planck:2018vyg,
    author = "Aghanim, N. and others",
    collaboration = "Planck",
    title = "{Planck 2018 results. VI. Cosmological parameters}",
    eprint = "1807.06209",
    archivePrefix = "arXiv",
    primaryClass = "astro-ph.CO",
    doi = "10.1051/0004-6361/201833910",
    journal = "Astron. Astrophys.",
    volume = "641",
    pages = "A6",
    year = "2020",
    note = "[Erratum: Astron.Astrophys. 652, C4 (2021)]"
}

@article{Dimopoulos:1981zb,
    author = "Dimopoulos, Savas and Georgi, Howard",
    title = "{Softly Broken Supersymmetry and SU(5)}",
    reportNumber = "HUTP-81/A022",
    doi = "10.1016/0550-3213(81)90522-8",
    journal = "Nucl. Phys. B",
    volume = "193",
    pages = "150--162",
    year = "1981"
}

@article{KATRIN:2024cdt,
    author = "Aker, Max and others",
    collaboration = "KATRIN",
    title = "{Direct neutrino-mass measurement based on 259 days of KATRIN data}",
    eprint = "2406.13516",
    archivePrefix = "arXiv",
    primaryClass = "nucl-ex",
    doi = "10.1126/science.adq9592",
    journal = "Science",
    volume = "388",
    number = "6743",
    pages = "adq9592",
    year = "2025"
}

@article{Sakai:1981gr,
    author = "Sakai, N.",
    title = "{Naturalness in Supersymmetric Guts}",
    reportNumber = "TU/81/225",
    doi = "10.1007/BF01573998",
    journal = "Z. Phys. C",
    volume = "11",
    pages = "153",
    year = "1981"
}

@article{Dorsner:2024seb,
    author = "Dor{\v{s}}ner, Ilja and Saad, Shaikh",
    title = "{Is doublet-triplet splitting necessary?}",
    eprint = "2404.09021",
    archivePrefix = "arXiv",
    primaryClass = "hep-ph",
    doi = "10.1103/PhysRevD.110.075025",
    journal = "Phys. Rev. D",
    volume = "110",
    number = "7",
    pages = "075025",
    year = "2024"
}

@article{Dorsner:2007fy,
    author = "Dorsner, Ilja and Mocioiu, Irina",
    title = "{Predictions from type II see-saw mechanism in SU(5)}",
    eprint = "0708.3332",
    archivePrefix = "arXiv",
    primaryClass = "hep-ph",
    doi = "10.1016/j.nuclphysb.2007.12.004",
    journal = "Nucl. Phys. B",
    volume = "796",
    pages = "123--136",
    year = "2008"
}

@article{Wilczek:1979hc,
    author = "Wilczek, Frank and Zee, A.",
    title = "{Operator Analysis of Nucleon Decay}",
    reportNumber = "Print-79-0709 (PRINCETON)",
    doi = "10.1103/PhysRevLett.43.1571",
    journal = "Phys. Rev. Lett.",
    volume = "43",
    pages = "1571--1573",
    year = "1979"
}

@article{Patel:2023gwt,
    author = "Patel, Ketan M. and Shukla, Saurabh K.",
    title = "{Quantum corrections and the minimal Yukawa sector of SU(5)}",
    eprint = "2310.16563",
    archivePrefix = "arXiv",
    primaryClass = "hep-ph",
    doi = "10.1103/PhysRevD.109.015007",
    journal = "Phys. Rev. D",
    volume = "109",
    number = "1",
    pages = "015007",
    year = "2024"
}

@article{Mohapatra:1982aq,
    author = "Mohapatra, Rabindra N. and Senjanovic, Goran",
    title = "{Higgs Boson Effects in Grand Unified Theories}",
    reportNumber = "CCNY-HEP-82/8a",
    doi = "10.1103/PhysRevD.27.1601",
    journal = "Phys. Rev. D",
    volume = "27",
    pages = "1601",
    year = "1983"
}

@article{Dimopoulos:1984ha,
    author = "Dimopoulos, S. and Georgi, H. M.",
    title = "{Extended Survival Hypothesis and Fermion Masses}",
    reportNumber = "ITP-759-STANFORD",
    doi = "10.1016/0370-2693(84)91049-9",
    journal = "Phys. Lett. B",
    volume = "140",
    pages = "67--70",
    year = "1984"
}

@article{Klein:2019jgb,
    author = "Klein, Christiane and Lindner, Manfred and Vogl, Stefan",
    title = "{Radiative neutrino masses and successful $SU(5)$ unification}",
    eprint = "1907.05328",
    archivePrefix = "arXiv",
    primaryClass = "hep-ph",
    doi = "10.1103/PhysRevD.100.075024",
    journal = "Phys. Rev. D",
    volume = "100",
    number = "7",
    pages = "075024",
    year = "2019"
}

@article{Magg:1980ut,
    author = "Magg, M. and Wetterich, C.",
    title = "{Neutrino Mass Problem and Gauge Hierarchy}",
    reportNumber = "CERN-TH-2829",
    doi = "10.1016/0370-2693(80)90825-4",
    journal = "Phys. Lett. B",
    volume = "94",
    pages = "61--64",
    year = "1980"
}

@article{Lazarides:1980nt,
    author = "Lazarides, George and Shafi, Q. and Wetterich, C.",
    title = "{Proton Lifetime and Fermion Masses in an SO(10) Model}",
    reportNumber = "FREIBURG-THEP-80-2",
    doi = "10.1016/0550-3213(81)90354-0",
    journal = "Nucl. Phys. B",
    volume = "181",
    pages = "287--300",
    year = "1981"
}

@article{Dorsner:2005fq,
    author = "Dorsner, Ilja and Fileviez Perez, Pavel",
    title = "{Unification without supersymmetry: Neutrino mass, proton decay and light leptoquarks}",
    eprint = "hep-ph/0504276",
    archivePrefix = "arXiv",
    doi = "10.1016/j.nuclphysb.2005.06.016",
    journal = "Nucl. Phys. B",
    volume = "723",
    pages = "53--76",
    year = "2005"
}

@article{Esteban:2024eli,
    author = "Esteban, Ivan and Gonzalez-Garcia, M. C. and Maltoni, Michele and Martinez-Soler, Ivan and Pinheiro, Jo{\~a}o Paulo and Schwetz, Thomas",
    title = "{NuFit-6.0: updated global analysis of three-flavor neutrino oscillations}",
    eprint = "2410.05380",
    archivePrefix = "arXiv",
    primaryClass = "hep-ph",
    reportNumber = "IFT-UAM/CSIC-24-140, YITP-SB-2024-24, IPPP/24/64, IPPP/24/64, IFT-UAM/CSIC-24-140, YITP-SB-2024-24",
    doi = "10.1007/JHEP12(2024)216",
    journal = "JHEP",
    volume = "12",
    pages = "216",
    year = "2024"
}

@article{Dorsner:2017wwn,
    author = "Dor\v{s}ner, Ilja and Fajfer, Svjetlana and Ko\v{s}nik, Nejc",
    title = "{Leptoquark mechanism of neutrino masses within the grand unification framework}",
    eprint = "1701.08322",
    archivePrefix = "arXiv",
    primaryClass = "hep-ph",
    doi = "10.1140/epjc/s10052-017-4987-2",
    journal = "Eur. Phys. J. C",
    volume = "77",
    number = "6",
    pages = "417",
    year = "2017"
}

@article{Mohapatra:1980yp,
    author = "Mohapatra, Rabindra N. and Senjanovic, Goran",
    title = "{Neutrino Masses and Mixings in Gauge Models with Spontaneous Parity Violation}",
    reportNumber = "FERMILAB-PUB-80-061-THY, FERMILAB-PUB-80-061-T",
    doi = "10.1103/PhysRevD.23.165",
    journal = "Phys. Rev. D",
    volume = "23",
    pages = "165",
    year = "1981"
}

@article{Schechter:1981bd,
    author = "Schechter, J. and Valle, J. W. F.",
    title = "{Neutrinoless Double beta Decay in SU(2) x U(1) Theories}",
    reportNumber = "SU-4217-213, COO-3533-213",
    doi = "10.1103/PhysRevD.25.2951",
    journal = "Phys. Rev. D",
    volume = "25",
    pages = "2951",
    year = "1982"
}

@inproceedings{Pas:1997fx,
    author = "Pas, H. and Hirsch, M. and Kovalenko, S. G. and Klapdor-Kleingrothaus, H. V.",
    title = "{Towards a superformula for neutrinoless double beta decay}",
    booktitle = "{Workshop on Physics Beyond the Standard Model: Beyond the Desert: Accelerator and Nonaccelerator Approaches}",
    eprint = "hep-ph/9804374",
    archivePrefix = "arXiv",
    pages = "884--890",
    month = "6",
    year = "1997"
}

@article{Cirigliano:2018yza,
    author = "Cirigliano, V. and Dekens, W. and de Vries, J. and Graesser, M. L. and Mereghetti, E.",
    title = "{A neutrinoless double beta decay master formula from effective field theory}",
    eprint = "1806.02780",
    archivePrefix = "arXiv",
    primaryClass = "hep-ph",
    reportNumber = "LA-UR-18-24895, Nikhef 2018-023, NIKHEF-2018-023, DESY-18-072",
    doi = "10.1007/JHEP12(2018)097",
    journal = "JHEP",
    volume = "12",
    pages = "097",
    year = "2018"
}

@article{Deppisch:2020ztt,
    author = "Deppisch, Frank F. and Graf, Lukas and Iachello, Francesco and Kotila, Jenni",
    title = "{Analysis of light neutrino exchange and short-range mechanisms in $0\nu\beta\beta$ decay}",
    eprint = "2009.10119",
    archivePrefix = "arXiv",
    primaryClass = "hep-ph",
    doi = "10.1103/PhysRevD.102.095016",
    journal = "Phys. Rev. D",
    volume = "102",
    number = "9",
    pages = "095016",
    year = "2020"
}

@article{Ali:2007ec,
    author = "Ali, A. and Borisov, A. V. and Zhuridov, D. V.",
    title = "{Probing new physics in the neutrinoless double beta decay using electron angular correlation}",
    eprint = "0706.4165",
    archivePrefix = "arXiv",
    primaryClass = "hep-ph",
    reportNumber = "DESY-07-097, DESY 07-097",
    doi = "10.1103/PhysRevD.76.093009",
    journal = "Phys. Rev. D",
    volume = "76",
    pages = "093009",
    year = "2007",
    note = "[Erratum: Phys.Rev.D 105, 099902 (2022)]"
}

@article{Kotila:2021xgw, author = "Kotila, Jenni and Ferretti, Jacopo and Iachello, Francesco", title = "{Long-range neutrinoless double beta decay mechanisms}", eprint = "2110.09141", archivePrefix = "arXiv", primaryClass = "hep-ph", month = "10", year = "2021" }

@article{LEGEND:2021bnm, author = "Abgrall, N. and others", collaboration = "LEGEND", title = "{The Large Enriched Germanium Experiment for Neutrinoless $\beta\beta$ Decay}: {LEGEND-1000 Preconceptual Design Report}", eprint = "2107.11462", archivePrefix = "arXiv", primaryClass = "physics.ins-det", month = "7", year = "2021" }

@article{nEXO:2021ujk, author = "Adhikari, G. and others", collaboration = "nEXO", title = "{nEXO: neutrinoless double beta decay search beyond 10$^{28}$ year half-life sensitivity}", eprint = "2106.16243", archivePrefix = "arXiv", primaryClass = "nucl-ex", doi = "10.1088/1361-6471/ac3631", journal = "J. Phys. G", volume = "49", number = "1", pages = "015104", year = "2022" }

@article{KamLAND-Zen:2024eml,
    author = "Abe, S. and others",
    collaboration = "KamLAND-Zen",
    title = "{Search for Majorana Neutrinos with the Complete KamLAND-Zen Dataset}",
    eprint = "2406.11438",
    archivePrefix = "arXiv",
    primaryClass = "hep-ex",
    month = "6",
    year = "2024"
}

@article{Dev:2024tto,
    author = "Dev, P. S. Bhupal and Goswami, Srubabati and Majumdar, Chayan and Pachhar, Debashis",
    title = "{Neutrinoless double beta decay from scalar leptoquarks: interplay with neutrino mass and flavor physics}",
    eprint = "2407.04670",
    archivePrefix = "arXiv",
    primaryClass = "hep-ph",
    doi = "10.1007/JHEP01(2025)004",
    journal = "JHEP",
    volume = "01",
    pages = "004",
    year = "2025"
}

@article{Fajfer:2024uut,
    author = "Fajfer, S. and Leal, L. P. S. and Sumensari, O. and Funchal, R. Zukanovich",
    title = "{Correlating 0\ensuremath{\nu}\ensuremath{\beta}\ensuremath{\beta} decays and flavor observables in leptoquark models}",
    eprint = "2406.20050",
    archivePrefix = "arXiv",
    primaryClass = "hep-ph",
    doi = "10.1007/JHEP01(2025)147",
    journal = "JHEP",
    volume = "01",
    pages = "147",
    year = "2025"
}

@article{Mummidi:2018myd,
    author = "Mummidi, V. Suryanarayana and Patel, Ketan M.",
    title = "{Pseudo-Dirac Higgsino dark matter in GUT scale supersymmetry}",
    eprint = "1811.06297",
    archivePrefix = "arXiv",
    primaryClass = "hep-ph",
    doi = "10.1007/JHEP01(2019)224",
    journal = "JHEP",
    volume = "01",
    pages = "224",
    year = "2019"
}

@article{Scholer:2023bnn,
    author = "Scholer, Oliver and de Vries, Jordy and Gr\'af, Luk\'a\v{s}",
    title = "{\ensuremath{\nu}DoBe \textemdash{} A Python tool for neutrinoless double beta decay}",
    eprint = "2304.05415",
    archivePrefix = "arXiv",
    primaryClass = "hep-ph",
    doi = "10.1007/JHEP08(2023)043",
    journal = "JHEP",
    volume = "08",
    pages = "043",
    year = "2023"
}

@article{ATLAS:2021jol,
    author = "Aad, Georges and others",
    collaboration = "ATLAS",
    title = "{Search for doubly and singly charged Higgs bosons decaying into vector bosons in multi-lepton final states with the ATLAS detector using proton-proton collisions at $ \sqrt{\mathrm{s}} $ = 13 TeV}",
    eprint = "2101.11961",
    archivePrefix = "arXiv",
    primaryClass = "hep-ex",
    reportNumber = "CERN-EP-2020-240",
    doi = "10.1007/JHEP06(2021)146",
    journal = "JHEP",
    volume = "06",
    pages = "146",
    year = "2021"
}

@article{Fonseca:2015ena,
    author = "Fonseca, Renato M. and Hirsch, Martin",
    title = "{SU(5)-inspired double beta decay}",
    eprint = "1505.06121",
    archivePrefix = "arXiv",
    primaryClass = "hep-ph",
    reportNumber = "IFIC-15-30",
    doi = "10.1103/PhysRevD.92.015014",
    journal = "Phys. Rev. D",
    volume = "92",
    number = "1",
    pages = "015014",
    year = "2015"
}

@book{Mohapatra:1986uf,
    author = "Mohapatra, R. N.",
    title = "Unification and Supersymmetry: The Frontiers of Quark-Lepton Physics",
    doi = "10.1007/978-1-4757-1928-4",
    isbn = "978-1-4757-1930-7, 978-1-4757-1928-4",
    publisher = "Springer",
    address = "Berlin",
    year = "1986"
}

@article{Das:2000uk,
    author = "Das, C. R. and Parida, M. K.",
    title = "{New formulas and predictions for running fermion masses at higher scales in SM, 2 HDM, and MSSM}",
    eprint = "hep-ph/0010004",
    archivePrefix = "arXiv",
    reportNumber = "NEHU-PHYS-MP-03-2000",
    doi = "10.1007/s100520100628",
    journal = "Eur. Phys. J. C",
    volume = "20",
    pages = "121--137",
    year = "2001"
}

@article{Branco:2011iw,
    author = "Branco, G. C. and Ferreira, P. M. and Lavoura, L. and Rebelo, M. N. and Sher, Marc and Silva, Joao P.",
    title = "{Theory and phenomenology of two-Higgs-doublet models}",
    eprint = "1106.0034",
    archivePrefix = "arXiv",
    primaryClass = "hep-ph",
    doi = "10.1016/j.physrep.2012.02.002",
    journal = "Phys. Rept.",
    volume = "516",
    pages = "1--102",
    year = "2012"
}

@article{ParticleDataGroup:2024cfk,
    author = "Navas, S. and others",
    collaboration = "Particle Data Group",
    title = "{Review of particle physics}",
    doi = "10.1103/PhysRevD.110.030001",
    journal = "Phys. Rev. D",
    volume = "110",
    number = "3",
    pages = "030001",
    year = "2024"
}

@article{Shukla:2024bwf,
    author = "Shukla, Saurabh K.",
    title = "{Revisiting $SU(5)$ Yukawa Sectors Through Quantum Corrections}",
    eprint = "2411.06906",
    archivePrefix = "arXiv",
    primaryClass = "hep-ph",
    month = "11",
    year = "2024"
}

@article{CMS:2018lab,
    author = "Sirunyan, Albert M and others",
    collaboration = "CMS",
    title = "{Search for pair production of second-generation leptoquarks at $\sqrt{s}=$ 13 TeV}",
    eprint = "1808.05082",
    archivePrefix = "arXiv",
    primaryClass = "hep-ex",
    reportNumber = "CMS-EXO-17-003, CERN-EP-2018-218",
    doi = "10.1103/PhysRevD.99.032014",
    journal = "Phys. Rev. D",
    volume = "99",
    number = "3",
    pages = "032014",
    year = "2019"
}

@article{CMS:2022nty,
    author = "Tumasyan, Armen and others",
    collaboration = "CMS",
    title = "{Inclusive nonresonant multilepton probes of new phenomena at $\sqrt s$=13\,\,TeV}",
    eprint = "2202.08676",
    archivePrefix = "arXiv",
    primaryClass = "hep-ex",
    reportNumber = "CMS-EXO-21-002, CERN-EP-2022-008",
    doi = "10.1103/PhysRevD.105.112007",
    journal = "Phys. Rev. D",
    volume = "105",
    number = "11",
    pages = "112007",
    year = "2022"
}

@article{ATLAS:2021oiz,
    author = "Aad, Georges and others",
    collaboration = "ATLAS",
    title = "{Search for pair production of third-generation scalar leptoquarks decaying into a top quark and a $\tau$-lepton in $pp$ collisions at $ \sqrt{s} $ = 13 TeV with the ATLAS detector}",
    eprint = "2101.11582",
    archivePrefix = "arXiv",
    primaryClass = "hep-ex",
    reportNumber = "CERN-EP-2020-241",
    doi = "10.1007/JHEP06(2021)179",
    journal = "JHEP",
    volume = "06",
    pages = "179",
    year = "2021"
}

@article{CMS:2018iye,
    author = "Sirunyan, Albert M and others",
    collaboration = "CMS",
    title = "{Search for heavy neutrinos and third-generation leptoquarks in hadronic states of two $\tau$ leptons and two jets in proton-proton collisions at $\sqrt{s} =$ 13 TeV}",
    eprint = "1811.00806",
    archivePrefix = "arXiv",
    primaryClass = "hep-ex",
    reportNumber = "CMS-EXO-17-016, CERN-EP-2018-272",
    doi = "10.1007/JHEP03(2019)170",
    journal = "JHEP",
    volume = "03",
    pages = "170",
    year = "2019"
}

@article{CMS:2020wzx,
    author = "Sirunyan, Albert M and others",
    collaboration = "CMS",
    title = "{Search for singly and pair-produced leptoquarks coupling to third-generation fermions in proton-proton collisions at s=13~TeV}",
    eprint = "2012.04178",
    archivePrefix = "arXiv",
    primaryClass = "hep-ex",
    reportNumber = "CMS-EXO-19-015, CERN-EP-2020-216",
    doi = "10.1016/j.physletb.2021.136446",
    journal = "Phys. Lett. B",
    volume = "819",
    pages = "136446",
    year = "2021"
}

@article{ATLAS:2023prb,
    author = "Aad, Georges and others",
    collaboration = "ATLAS",
    title = "{Search for leptoquark pair production decaying into $te^- \bar{t}e^+$ or $t\mu ^- \bar{t}\mu ^+$ in multi-lepton final states in pp collisions at $\sqrt{s} = 13\,\textrm{TeV}$ with the ATLAS detector}",
    eprint = "2306.17642",
    archivePrefix = "arXiv",
    primaryClass = "hep-ex",
    reportNumber = "CERN-EP-2023-087",
    doi = "10.1140/epjc/s10052-024-12975-4",
    journal = "Eur. Phys. J. C",
    volume = "84",
    number = "8",
    pages = "818",
    year = "2024"
}

@article{Babu:2019mfe,
    author = "Babu, K. S. and Dev, P. S. Bhupal and Jana, Sudip and Thapa, Anil",
    title = "{Non-Standard Interactions in Radiative Neutrino Mass Models}",
    eprint = "1907.09498",
    archivePrefix = "arXiv",
    primaryClass = "hep-ph",
    reportNumber = "FERMILAB-PUB-19-304-T, OSU-HEP-19-04",
    doi = "10.1007/JHEP03(2020)006",
    journal = "JHEP",
    volume = "03",
    pages = "006",
    year = "2020"
}

@article{Babu:2020hun,
    author = "Babu, K. S. and Dev, P. S. Bhupal and Jana, Sudip and Thapa, Anil",
    title = "{Unified framework for $B$-anomalies, muon $g-2$ and neutrino masses}",
    eprint = "2009.01771",
    archivePrefix = "arXiv",
    primaryClass = "hep-ph",
    reportNumber = "OSU-HEP-20-12",
    doi = "10.1007/JHEP03(2021)179",
    journal = "JHEP",
    volume = "03",
    pages = "179",
    year = "2021"
}

@article{Lee:2007qx,
    author = "Lee, Hye-Sung and Luhn, Christoph and Matchev, Konstantin T.",
    title = "{Discrete gauge symmetries and proton stability in the U(1)-prime - extended MSSM}",
    eprint = "0712.3505",
    archivePrefix = "arXiv",
    primaryClass = "hep-ph",
    doi = "10.1088/1126-6708/2008/07/065",
    journal = "JHEP",
    volume = "07",
    pages = "065",
    year = "2008"
}

@article{Hur:2008sy,
    author = "Hur, Taeil and Lee, Hye-Sung and Luhn, Christoph",
    title = "{Common gauge origin of discrete symmetries in observable sector and hidden sector}",
    eprint = "0811.0812",
    archivePrefix = "arXiv",
    primaryClass = "hep-ph",
    reportNumber = "UFIFT-HEP-08-17, UCRHEP-T457, SHEP-08-32",
    doi = "10.1088/1126-6708/2009/01/081",
    journal = "JHEP",
    volume = "01",
    pages = "081",
    year = "2009"
}

@article{Mohapatra:2007vd,
    author = "Mohapatra, Rabindra N. and Ratz, Michael",
    title = "{Gauged Discrete Symmetries and Proton Stability}",
    eprint = "0707.4070",
    archivePrefix = "arXiv",
    primaryClass = "hep-ph",
    reportNumber = "UMD-PP-07-006, TUM-HEP-672-07",
    doi = "10.1103/PhysRevD.76.095003",
    journal = "Phys. Rev. D",
    volume = "76",
    pages = "095003",
    year = "2007"
}

@article{Berasaluce-Gonzalez:2011gos,
    author = "Berasaluce-Gonzalez, Mikel and Ibanez, Luis E. and Soler, Pablo and Uranga, Angel M.",
    title = "{Discrete gauge symmetries in D-brane models}",
    eprint = "1106.4169",
    archivePrefix = "arXiv",
    primaryClass = "hep-th",
    reportNumber = "IFT-UAM-CSIC-11-42",
    doi = "10.1007/JHEP12(2011)113",
    journal = "JHEP",
    volume = "12",
    pages = "113",
    year = "2011"
}

@article{Farrar:1978xj,
    author = "Farrar, Glennys R. and Fayet, Pierre",
    title = "{Phenomenology of the Production, Decay, and Detection of New Hadronic States Associated with Supersymmetry}",
    reportNumber = "CALT-68-648",
    doi = "10.1016/0370-2693(78)90858-4",
    journal = "Phys. Lett. B",
    volume = "76",
    pages = "575--579",
    year = "1978"
}

@article{Dimopoulos:1981dw,
    author = "Dimopoulos, Savas and Raby, Stuart and Wilczek, Frank",
    title = "{Proton Decay in Supersymmetric Models}",
    reportNumber = "NSF-ITP-82-08, UM HE 81-64",
    doi = "10.1016/0370-2693(82)90313-6",
    journal = "Phys. Lett. B",
    volume = "112",
    pages = "133",
    year = "1982"
}

@article{Ibanez:1991hv,
    author = "Ibanez, Luis E. and Ross, Graham G.",
    title = "{Discrete gauge symmetry anomalies}",
    doi = "10.1016/0370-2693(91)91614-2",
    journal = "Phys. Lett. B",
    volume = "260",
    pages = "291--295",
    year = "1991"
}

@article{Ibanez:1991pr,
    author = "Ibanez, Luis E. and Ross, Graham G.",
    title = "{Discrete gauge symmetries and the origin of baryon and lepton number conservation in supersymmetric versions of the standard model}",
    reportNumber = "CERN-TH-6111-91",
    doi = "10.1016/0550-3213(92)90195-H",
    journal = "Nucl. Phys. B",
    volume = "368",
    pages = "3--37",
    year = "1992"
}

@article{Dreiner:2005rd,
    author = "Dreiner, Herbi K. and Luhn, Christoph and Thormeier, Marc",
    title = "{What is the discrete gauge symmetry of the MSSM?}",
    eprint = "hep-ph/0512163",
    archivePrefix = "arXiv",
    doi = "10.1103/PhysRevD.73.075007",
    journal = "Phys. Rev. D",
    volume = "73",
    pages = "075007",
    year = "2006"
}

@article{Forste:2010pf,
    author = "Forste, Stefan and Nilles, Hans Peter and Ramos-Sanchez, Saul and Vaudrevange, Patrick K. S.",
    title = "{Proton Hexality in Local Grand Unification}",
    eprint = "1007.3915",
    archivePrefix = "arXiv",
    primaryClass = "hep-ph",
    reportNumber = "DESY-10-111, LMU-ASC-55-10",
    doi = "10.1016/j.physletb.2010.08.057",
    journal = "Phys. Lett. B",
    volume = "693",
    pages = "386--392",
    year = "2010"
}

@article{Emmanuel-Costa:2013gia,
    author = "Emmanuel-Costa, D. and Simoes, C. and Tortola, M.",
    title = "{The minimal adjoint-SU(5) x $Z_{4}$ GUT model}",
    eprint = "1303.5699",
    archivePrefix = "arXiv",
    primaryClass = "hep-ph",
    reportNumber = "CFTP-13-008, IFIC-13-10",
    doi = "10.1007/JHEP10(2013)054",
    journal = "JHEP",
    volume = "10",
    pages = "054",
    year = "2013"
}

@article{Emmanuel-Costa:2011xdu,
    author = "Emmanuel-Costa, D. and Simoes, C.",
    title = "{Nearest-Neighbour-Interactions from a minimal discrete flavour symmetry within SU(5) Grand Unification}",
    eprint = "1102.3729",
    archivePrefix = "arXiv",
    primaryClass = "hep-ph",
    reportNumber = "CFTP-11-004",
    doi = "10.1103/PhysRevD.85.016003",
    journal = "Phys. Rev. D",
    volume = "85",
    pages = "016003",
    year = "2012"
}

@article{Azatov:2008vu,
    author = "Azatov, A. T. and Mohapatra, R. N.",
    title = "{Flavor Physics in SO(10) GUTs with Suppressed Proton decay Due to Gauged Discrete Symmetry}",
    eprint = "0802.3906",
    archivePrefix = "arXiv",
    primaryClass = "hep-ph",
    reportNumber = "UMD-PP-08-003",
    doi = "10.1103/PhysRevD.78.015002",
    journal = "Phys. Rev. D",
    volume = "78",
    pages = "015002",
    year = "2008"
}

@article{Dev:2012nm,
    author = "Dev, P. S. Bhupal and Dutta, Bhaskar and Mohapatra, R. N. and Severson, Matthew",
    title = "{$\theta_{13}$ and Proton Decay in a Minimal $SO(10) \times S_4$ model of Flavor}",
    eprint = "1202.4012",
    archivePrefix = "arXiv",
    primaryClass = "hep-ph",
    reportNumber = "UMD-PP-012-002, MIFPA-12-05",
    doi = "10.1103/PhysRevD.86.035002",
    journal = "Phys. Rev. D",
    volume = "86",
    pages = "035002",
    year = "2012"
}

@article{Dutta:2004zh,
    author = "Dutta, Bhaskar and Mimura, Yukihiro and Mohapatra, R. N.",
    title = "{Suppressing proton decay in the minimal SO(10) model}",
    eprint = "hep-ph/0412105",
    archivePrefix = "arXiv",
    reportNumber = "UMD-PP-05-026",
    doi = "10.1103/PhysRevLett.94.091804",
    journal = "Phys. Rev. Lett.",
    volume = "94",
    pages = "091804",
    year = "2005"
}

@article{Wu:2022stu,
    author = "Wu, Yongcheng and Xie, Ke-Pan and Zhou, Ye-Ling",
    title = "{Collapsing domain walls beyond Z2}",
    eprint = "2204.04374",
    archivePrefix = "arXiv",
    primaryClass = "hep-ph",
    doi = "10.1103/PhysRevD.105.095013",
    journal = "Phys. Rev. D",
    volume = "105",
    number = "9",
    pages = "095013",
    year = "2022"
}

@article{Wu:2022tpe,
    author = "Wu, Yongcheng and Xie, Ke-Pan and Zhou, Ye-Ling",
    title = "{Classification of Abelian domain walls}",
    eprint = "2205.11529",
    archivePrefix = "arXiv",
    primaryClass = "hep-ph",
    doi = "10.1103/PhysRevD.106.075019",
    journal = "Phys. Rev. D",
    volume = "106",
    number = "7",
    pages = "075019",
    year = "2022"
}

@article{Gelmini:1988sf,
    author = "Gelmini, Graciela B. and Gleiser, Marcelo and Kolb, Edward W.",
    title = "{Cosmology of Biased Discrete Symmetry Breaking}",
    reportNumber = "NSF-ITP-88-148, FERMILAB-PUB-88-151-A",
    doi = "10.1103/PhysRevD.39.1558",
    journal = "Phys. Rev. D",
    volume = "39",
    pages = "1558",
    year = "1989"
}

@article{Kajiyama:2005rk,
    author = "Kajiyama, Yuji and Itou, Etsuko and Kubo, Jisuke",
    title = "{NonAbelian discrete family symmetry to soften the SUSY flavor problem and to suppress proton decay}",
    eprint = "hep-ph/0511268",
    archivePrefix = "arXiv",
    reportNumber = "KANAZAWA-05-15",
    doi = "10.1016/j.nuclphysb.2006.02.042",
    journal = "Nucl. Phys. B",
    volume = "743",
    pages = "74--103",
    year = "2006"
}

@article{ATLAS:2023vxj,
    author = "Aad, Georges and others",
    collaboration = "ATLAS",
    title = "{Search for leptoquarks decaying into the b\ensuremath{\tau} final state in $pp$ collisions at $ \sqrt{\textrm{s}} $ = 13 TeV with the ATLAS detector}",
    eprint = "2305.15962",
    archivePrefix = "arXiv",
    primaryClass = "hep-ex",
    reportNumber = "CERN-EP-2023-033",
    doi = "10.1007/JHEP10(2023)001",
    journal = "JHEP",
    volume = "10",
    pages = "001",
    year = "2023"
}

@article{CMS:2018ncu,
    author = "Sirunyan, Albert M and others",
    collaboration = "CMS",
    title = "{Search for pair production of first-generation scalar leptoquarks at $\sqrt{s} =$ 13  TeV}",
    eprint = "1811.01197",
    archivePrefix = "arXiv",
    primaryClass = "hep-ex",
    reportNumber = "CMS-EXO-17-009, CERN-EP-2018-265",
    doi = "10.1103/PhysRevD.99.052002",
    journal = "Phys. Rev. D",
    volume = "99",
    number = "5",
    pages = "052002",
    year = "2019"
}

@article{ATLAS:2019ebv,
    author = "Aaboud, Morad and others",
    collaboration = "ATLAS",
    title = "{Searches for scalar leptoquarks and differential cross-section measurements in dilepton-dijet events in proton-proton collisions at a centre-of-mass energy of $\sqrt{s}$ = 13 TeV with the ATLAS experiment}",
    eprint = "1902.00377",
    archivePrefix = "arXiv",
    primaryClass = "hep-ex",
    reportNumber = "CERN-EP-2018-262",
    doi = "10.1140/epjc/s10052-019-7181-x",
    journal = "Eur. Phys. J. C",
    volume = "79",
    number = "9",
    pages = "733",
    year = "2019"
}

@article{ATLAS:2020dsk,
    author = "Aad, Georges and others",
    collaboration = "ATLAS",
    title = "{Search for pairs of scalar leptoquarks decaying into quarks and electrons or muons in $ \sqrt{s} $ = 13 TeV $pp$ collisions with the ATLAS detector}",
    eprint = "2006.05872",
    archivePrefix = "arXiv",
    primaryClass = "hep-ex",
    reportNumber = "CERN-EP-2020-084",
    doi = "10.1007/JHEP10(2020)112",
    journal = "JHEP",
    volume = "10",
    pages = "112",
    year = "2020"
}

@article{Parida:2018apw,
    author = "Parida, M. K. and Satpathy, Rajesh",
    title = "{High Scale Type-II Seesaw, Dominant Double Beta Decay within Cosmological Bound and LFV Decays in SU(5)}",
    eprint = "1809.06612",
    archivePrefix = "arXiv",
    primaryClass = "hep-ph",
    doi = "10.1155/2019/3572862",
    journal = "Adv. High Energy Phys.",
    volume = "2019",
    pages = "3572862",
    year = "2019"
}

@article{Kitano:2002mt,
    author = "Kitano, Ryuichiro and Koike, Masafumi and Okada, Yasuhiro",
    title = "{Detailed calculation of lepton flavor violating muon electron conversion rate for various nuclei}",
    eprint = "hep-ph/0203110",
    archivePrefix = "arXiv",
    reportNumber = "KEK-TH-808",
    doi = "10.1103/PhysRevD.76.059902",
    journal = "Phys. Rev. D",
    volume = "66",
    pages = "096002",
    year = "2002",
    note = "[Erratum: Phys.Rev.D 76, 059902 (2007)]"
}

@article{Bonnet:2012kh,
    author = "Bonnet, Florian and Hirsch, Martin and Ota, Toshihiko and Winter, Walter",
    title = "{Systematic decomposition of the neutrinoless double beta decay operator}",
    eprint = "1212.3045",
    archivePrefix = "arXiv",
    primaryClass = "hep-ph",
    reportNumber = "IFIC-12-079, MPP-2012-134, STUPP-12-212",
    doi = "10.1007/JHEP03(2013)055",
    journal = "JHEP",
    volume = "03",
    pages = "055",
    year = "2013",
    note = "[Erratum: JHEP 04, 090 (2014)]"
}

@article{Gonzalez:2016ztm,
    author = "Gonzalez, L. and Helo, J. C. and Hirsch, M. and Kovalenko, S. G.",
    title = "{Scalar-mediated double beta decay and LHC}",
    eprint = "1606.09555",
    archivePrefix = "arXiv",
    primaryClass = "hep-ph",
    doi = "10.1007/JHEP12(2016)130",
    journal = "JHEP",
    volume = "12",
    pages = "130",
    year = "2016"
}

@article{Graesser:2022nkv,
    author = "Graesser, Michael L. and Li, Gang and Ramsey-Musolf, Michael J. and Shen, Tianyang and Urrutia-Quiroga, Sebasti\'an",
    title = "{Uncovering a chirally suppressed mechanism of 0\ensuremath{\nu}\ensuremath{\beta}\ensuremath{\beta} decay with LHC searches}",
    eprint = "2202.01237",
    archivePrefix = "arXiv",
    primaryClass = "hep-ph",
    reportNumber = "ACFI-T22-02, LA-UR-21-32454",
    doi = "10.1007/JHEP10(2022)034",
    journal = "JHEP",
    volume = "10",
    pages = "034",
    year = "2022"
}

@article{Graf:2022lhj,
    author = "Gr\'af, Luk\'a\v{s} and Lindner, Manfred and Scholer, Oliver",
    title = "{Unraveling the 0\ensuremath{\nu}\ensuremath{\beta}\ensuremath{\beta} decay mechanisms}",
    eprint = "2204.10845",
    archivePrefix = "arXiv",
    primaryClass = "hep-ph",
    doi = "10.1103/PhysRevD.106.035022",
    journal = "Phys. Rev. D",
    volume = "106",
    number = "3",
    pages = "035022",
    year = "2022"
}

@article{Hirsch:1996ye,
    author = "Hirsch, M. and Klapdor-Kleingrothaus, H. V. and Kovalenko, S. G.",
    title = "{New leptoquark mechanism of neutrinoless double beta decay}",
    eprint = "hep-ph/9603213",
    archivePrefix = "arXiv",
    doi = "10.1103/PhysRevD.54.R4207",
    journal = "Phys. Rev. D",
    volume = "54",
    pages = "R4207--R4210",
    year = "1996"
}

@article{Plakias:2023esq,
    author = "Plakias, I. and Sumensari, O.",
    title = "{Lepton flavor violation in semileptonic observables}",
    eprint = "2312.14070",
    archivePrefix = "arXiv",
    primaryClass = "hep-ph",
    doi = "10.1103/PhysRevD.110.035016",
    journal = "Phys. Rev. D",
    volume = "110",
    number = "3",
    pages = "035016",
    year = "2024"
}

@article{SINDRUMII:2006dvw,
    author = "Bertl, Wilhelm H. and others",
    collaboration = "SINDRUM II",
    title = "{A Search for muon to electron conversion in muonic gold}",
    doi = "10.1140/epjc/s2006-02582-x",
    journal = "Eur. Phys. J. C",
    volume = "47",
    pages = "337--346",
    year = "2006"
}

@article{Mu2e:2014fns,
    author = "Bartoszek, L. and others",
    collaboration = "Mu2e",
    title = "{Mu2e Technical Design Report}",
    eprint = "1501.05241",
    archivePrefix = "arXiv",
    primaryClass = "physics.ins-det",
    reportNumber = "FERMILAB-TM-2594, FERMILAB-DESIGN-2014-01",
    doi = "10.2172/1172555",
    month = "10",
    year = "2014"
}

@article{Suzuki:1987jf,
    author = "Suzuki, T. and Measday, David F. and Roalsvig, J. P.",
    title = "{Total Nuclear Capture Rates for Negative Muons}",
    reportNumber = "TRI-PP-87-5",
    doi = "10.1103/PhysRevC.35.2212",
    journal = "Phys. Rev. C",
    volume = "35",
    pages = "2212",
    year = "1987"
}

@article{Li:2023wfi,
    author = "Li, Gang and Yu, Jiang-Hao and Zhao, Xiang",
    title = "{Complementary LHC searches for UV resonances of the 0\ensuremath{\nu}\ensuremath{\beta}\ensuremath{\beta} decay operators}",
    eprint = "2311.10079",
    archivePrefix = "arXiv",
    primaryClass = "hep-ph",
    doi = "10.1103/PhysRevD.109.055012",
    journal = "Phys. Rev. D",
    volume = "109",
    number = "5",
    pages = "055012",
    year = "2024"
}

@article{Helo:2013ika,
    author = {Helo, J. C. and Hirsch, M. and P\"as, H. and Kovalenko, S. G.},
    title = "{Short-range mechanisms of neutrinoless double beta decay at the LHC}",
    eprint = "1307.4849",
    archivePrefix = "arXiv",
    primaryClass = "hep-ph",
    reportNumber = "IFIC-13-34",
    doi = "10.1103/PhysRevD.88.073011",
    journal = "Phys. Rev. D",
    volume = "88",
    pages = "073011",
    year = "2013"
}

@article{Pas:2015hca,
    author = {P\"as, Heinrich and Schumacher, Erik},
    title = "{Common origin of $R_K$ and neutrino masses}",
    eprint = "1510.08757",
    archivePrefix = "arXiv",
    primaryClass = "hep-ph",
    doi = "10.1103/PhysRevD.92.114025",
    journal = "Phys. Rev. D",
    volume = "92",
    number = "11",
    pages = "114025",
    year = "2015"
}

@article{COMET:2018auw,
    author = "Abramishvili, R. and others",
    collaboration = "COMET",
    title = "{COMET Phase-I Technical Design Report}",
    eprint = "1812.09018",
    archivePrefix = "arXiv",
    primaryClass = "physics.ins-det",
    doi = "10.1093/ptep/ptz125",
    journal = "PTEP",
    volume = "2020",
    number = "3",
    pages = "033C01",
    year = "2020"
}

@article{Graf:2018ozy,
    author = "Graf, Lukas and Deppisch, Frank F. and Iachello, Francesco and Kotila, Jenni",
    title = "{Short-Range Neutrinoless Double Beta Decay Mechanisms}",
    eprint = "1806.06058",
    archivePrefix = "arXiv",
    primaryClass = "hep-ph",
    doi = "10.1103/PhysRevD.98.095023",
    journal = "Phys. Rev. D",
    volume = "98",
    number = "9",
    pages = "095023",
    year = "2018"
}

@article{Stojkovic:2005zh,
    author = "Stojkovic, Dejan and Freese, Katherine and Starkman, Glenn D.",
    title = "{Holes in the walls: Primordial black holes as a solution to the cosmological domain wall problem}",
    eprint = "hep-ph/0505026",
    archivePrefix = "arXiv",
    doi = "10.1103/PhysRevD.72.045012",
    journal = "Phys. Rev. D",
    volume = "72",
    pages = "045012",
    year = "2005"
}

@article{King:2018fke,
    author = "King, Stephen F. and Zhou, Ye-Ling",
    title = "{Spontaneous breaking of $SO(3)$ to finite family symmetries with supersymmetry - an $A_4$ model}",
    eprint = "1809.10292",
    archivePrefix = "arXiv",
    primaryClass = "hep-ph",
    reportNumber = "IPPP/18/83",
    doi = "10.1007/JHEP11(2018)173",
    journal = "JHEP",
    volume = "11",
    pages = "173",
    year = "2018"
}

@article{Kibble:1976sj,
    author = "Kibble, T. W. B.",
    title = "{Topology of Cosmic Domains and Strings}",
    reportNumber = "ICTP/75/5",
    doi = "10.1088/0305-4470/9/8/029",
    journal = "J. Phys. A",
    volume = "9",
    pages = "1387--1398",
    year = "1976"
}

@article{Zeldovich:1974uw,
    author = "Zeldovich, Ya. B. and Kobzarev, I. Yu. and Okun, L. B.",
    title = "{Cosmological Consequences of the Spontaneous Breakdown of Discrete Symmetry}",
    reportNumber = "SLAC-TRANS-0165, IPM-MOSCOW-15",
    journal = "Zh. Eksp. Teor. Fiz.",
    volume = "67",
    pages = "3--11",
    year = "1974"
}

@article{Vilenkin:1984ib,
    author = "Vilenkin, Alexander",
    title = "{Cosmic Strings and Domain Walls}",
    reportNumber = "PRINT-84-0840 (TUFTS)",
    doi = "10.1016/0370-1573(85)90033-X",
    journal = "Phys. Rept.",
    volume = "121",
    pages = "263--315",
    year = "1985"
}

@article{Larsson:1996sp,
    author = "Larsson, Sebastian E. and Sarkar, Subir and White, Peter L.",
    title = "{Evading the cosmological domain wall problem}",
    eprint = "hep-ph/9608319",
    archivePrefix = "arXiv",
    reportNumber = "OUTP-96-11-P",
    doi = "10.1103/PhysRevD.55.5129",
    journal = "Phys. Rev. D",
    volume = "55",
    pages = "5129--5135",
    year = "1997"
}

@book{Kolb:1990vq,
    author = "Kolb, Edward W. and Turner, Michael S.",
    title = "{The Early Universe}",
    reportNumber = "FERMILAB-BOOK-1990-01",
    doi = "10.1201/9780429492860",
    isbn = "978-0-429-49286-0, 978-0-201-62674-2",
    publisher = "Taylor and Francis",
    volume = "69",
    month = "5",
    year = "2019"
}

@article{Dvali:1995cc,
    author = "Dvali, G. R. and Senjanovic, Goran",
    title = "{Is there a domain wall problem?}",
    eprint = "hep-ph/9501387",
    archivePrefix = "arXiv",
    reportNumber = "IFUP-TH-61-94",
    doi = "10.1103/PhysRevLett.74.5178",
    journal = "Phys. Rev. Lett.",
    volume = "74",
    pages = "5178--5181",
    year = "1995"
}

@article{Croon:2019kpe,
    author = "Croon, Djuna and Gonzalo, Tom{\'a}s E. and Graf, Lukas and Ko{\v{s}}nik, Nejc and White, Graham",
    title = "{GUT Physics in the era of the LHC}",
    eprint = "1903.04977",
    archivePrefix = "arXiv",
    primaryClass = "hep-ph",
    reportNumber = "CoEPP-MN-19-01",
    doi = "10.3389/fphy.2019.00076",
    journal = "Front. in Phys.",
    volume = "7",
    pages = "76",
    year = "2019"
}

@article{Pati:1974yy,
    author = "Pati, Jogesh C. and Salam, Abdus",
    title = "{Lepton Number as the Fourth Color}",
    reportNumber = "IC-74-7",
    doi = "10.1103/PhysRevD.10.275",
    journal = "Phys. Rev. D",
    volume = "10",
    pages = "275--289",
    year = "1974",
    note = "[Erratum: Phys.Rev.D 11, 703--703 (1975)]"
}

@article{Saad:2019vjo,
    author = "Saad, Shaikh",
    title = "{Origin of a two-loop neutrino mass from SU(5) grand unification}",
    eprint = "1902.11254",
    archivePrefix = "arXiv",
    primaryClass = "hep-ph",
    reportNumber = "OSU-HEP-19-02",
    doi = "10.1103/PhysRevD.99.115016",
    journal = "Phys. Rev. D",
    volume = "99",
    number = "11",
    pages = "115016",
    year = "2019"
}

@article{Dorsner:2019vgf,
    author = "Dor{\v{s}}ner, Ilja and Saad, Shaikh",
    title = "{Towards Minimal $SU(5)$}",
    eprint = "1910.09008",
    archivePrefix = "arXiv",
    primaryClass = "hep-ph",
    reportNumber = "OSU-HEP-19-06, OSU-HEP-19-08",
    doi = "10.1103/PhysRevD.101.015009",
    journal = "Phys. Rev. D",
    volume = "101",
    number = "1",
    pages = "015009",
    year = "2020"
}

@article{Dorsner:2025rzj,
    author = "Dor{\v{s}}ner, Ilja and Matkovi{\'c}, Mijo and Saad, Shaikh",
    title = "{Nonrenormalizable SU(5) GUTs: Leptoquark-induced neutrino masses}",
    eprint = "2504.16022",
    archivePrefix = "arXiv",
    primaryClass = "hep-ph",
    doi = "10.1103/3plx-vwrb",
    journal = "Phys. Rev. D",
    volume = "111",
    number = "11",
    pages = "115039",
    year = "2025"
}

@article{Hyvarinen:2015bda,
    author = {Hyv{\"a}rinen, Juhani and Suhonen, Jouni},
    title = "{Nuclear matrix elements for $0\nu\beta\beta$ decays with light or heavy Majorana-neutrino exchange}",
    doi = "10.1103/PhysRevC.91.024613",
    journal = "Phys. Rev. C",
    volume = "91",
    number = "2",
    pages = "024613",
    year = "2015"
}

@article{Menendez:2017fdf,
    author = "Men{\'e}ndez, J.",
    title = "{Neutrinoless $\beta\beta$ decay mediated by the exchange of light and heavy neutrinos: The role of nuclear structure correlations}",
    eprint = "1804.02105",
    archivePrefix = "arXiv",
    primaryClass = "nucl-th",
    doi = "10.1088/1361-6471/aa9bd4",
    journal = "J. Phys. G",
    volume = "45",
    number = "1",
    pages = "014003",
    year = "2018"
}

@article{Ding:2024obt,
    author = "Ding, Chen-rong and Li, Gang and Yao, Jiang-ming",
    title = "{Nuclear matrix elements of neutrinoless double-beta decay in covariant density functional theory with different mechanisms}",
    eprint = "2403.17722",
    archivePrefix = "arXiv",
    primaryClass = "nucl-th",
    doi = "10.1016/j.physletb.2024.138896",
    journal = "Phys. Lett. B",
    volume = "856",
    pages = "138896",
    year = "2024"
}

@article{Lindestam:2021dyk,
    author = "Lindestam, Malte and Ohlsson, Tommy and Pernow, Marcus",
    title = "{Flavor symmetries in an SU(5) model of grand unification}",
    eprint = "2110.09533",
    archivePrefix = "arXiv",
    primaryClass = "hep-ph",
    doi = "10.1007/JHEP01(2022)009",
    journal = "JHEP",
    volume = "01",
    pages = "009",
    year = "2022"
}

@article{ParticleDataGroup:2016lqr,
    author = "Patrignani, C. and others",
    collaboration = "Particle Data Group",
    title = "{Review of Particle Physics}",
    doi = "10.1088/1674-1137/40/10/100001",
    journal = "Chin. Phys. C",
    volume = "40",
    number = "10",
    pages = "100001",
    year = "2016"
}

@article{Nicholson:2018mwc,
    author = "Nicholson, A. and others",
    title = "{Heavy physics contributions to neutrinoless double beta decay from QCD}",
    eprint = "1805.02634",
    archivePrefix = "arXiv",
    primaryClass = "nucl-th",
    reportNumber = "LLNL-JRNL-751220, RBRC-1266, RIKEN-ITHEMS-REPORT-18, RIKEN-iTHEMS-Report-18, BNL-209118-2018-JAAM",
    doi = "10.1103/PhysRevLett.121.172501",
    journal = "Phys. Rev. Lett.",
    volume = "121",
    number = "17",
    pages = "172501",
    year = "2018"
}

@article{Bhattacharya:2016zcn,
    author = "Bhattacharya, Tanmoy and Cirigliano, Vincenzo and Cohen, Saul and Gupta, Rajan and Lin, Huey-Wen and Yoon, Boram",
    title = "{Axial, Scalar and Tensor Charges of the Nucleon from 2+1+1-flavor Lattice QCD}",
    eprint = "1606.07049",
    archivePrefix = "arXiv",
    primaryClass = "hep-lat",
    reportNumber = "LA-UR-16-20522",
    doi = "10.1103/PhysRevD.94.054508",
    journal = "Phys. Rev. D",
    volume = "94",
    number = "5",
    pages = "054508",
    year = "2016"
}

@article{Cirigliano:2020dmx,
    author = "Cirigliano, Vincenzo and Dekens, Wouter and de Vries, Jordy and Hoferichter, Martin and Mereghetti, Emanuele",
    title = "{Toward Complete Leading-Order Predictions for Neutrinoless Double $\beta$ Decay}",
    eprint = "2012.11602",
    archivePrefix = "arXiv",
    primaryClass = "nucl-th",
    reportNumber = "LA-UR-20-30355",
    doi = "10.1103/PhysRevLett.126.172002",
    journal = "Phys. Rev. Lett.",
    volume = "126",
    number = "17",
    pages = "172002",
    year = "2021"
}

@article{Cirigliano:2021qko,
    author = "Cirigliano, Vincenzo and Dekens, Wouter and de Vries, Jordy and Hoferichter, Martin and Mereghetti, Emanuele",
    title = "{Determining the leading-order contact term in neutrinoless double $\beta$ decay}",
    eprint = "2102.03371",
    archivePrefix = "arXiv",
    primaryClass = "nucl-th",
    reportNumber = "LA-UR-21-20994",
    doi = "10.1007/JHEP05(2021)289",
    journal = "JHEP",
    volume = "05",
    pages = "289",
    year = "2021"
}

@article{Wirth:2021pij,
    author = "Wirth, R. and Yao, J. M. and Hergert, H.",
    title = "{Ab~Initio Calculation of the Contact Operator Contribution in the Standard Mechanism for Neutrinoless Double Beta Decay}",
    eprint = "2105.05415",
    archivePrefix = "arXiv",
    primaryClass = "nucl-th",
    doi = "10.1103/PhysRevLett.127.242502",
    journal = "Phys. Rev. Lett.",
    volume = "127",
    number = "24",
    pages = "242502",
    year = "2021"
}
\end{document}